\documentclass[12pt,preprint]{emulateapj}
\def\arcs{$''$}

\begin{document}
\title{UV Luminosity Functions from 132 z$\sim$7 and z$\sim$8 Lyman-Break
  Galaxies in the ultra-deep HUDF09 and wide-area ERS WFC3/IR
  Observations \altaffilmark{1}}
\author{R. J. Bouwens\altaffilmark{2,3}, G.D. Illingworth\altaffilmark{2}, P. A. Oesch\altaffilmark{4}, I. Labb\'{e}\altaffilmark{5}, M. Trenti\altaffilmark{6}, P. van Dokkum\altaffilmark{7}, M. Franx\altaffilmark{3}, M. Stiavelli\altaffilmark{8}, C. M. Carollo\altaffilmark{4}, D. Magee\altaffilmark{2}, V. Gonzalez\altaffilmark{2}}

\altaffiltext{1}{Based on observations made with the NASA/ESA Hubble
  Space Telescope, which is operated by the Association of
  Universities for Research in Astronomy, Inc., under NASA contract
  NAS 5-26555. These observations are associated with programs
  \#11563, 9797, 10632.}  
\altaffiltext{2}{UCO/Lick Observatory, University
  of California, Santa Cruz, CA 95064}
\altaffiltext{3}{Leiden Observatory, Leiden University, NL-2300 RA
  Leiden, Netherlands}
\altaffiltext{4}{Institute for Astronomy, ETH Zurich, 8092 Zurich,
  Switzerland; poesch@phys.ethz.ch}
\altaffiltext{5}{Carnegie Observatories, Pasadena, CA 91101, Hubble
  Fellow}
\altaffiltext{6}{University of Colorado, Center for
  Astrophysics and Space Astronomy, 389-UCB, Boulder, CO 80309, USA}
\altaffiltext{7}{Department of Astronomy, Yale University, New Haven,
  CT 06520}
\altaffiltext{8}{Space Telescope Science Institute, Baltimore, MD 
21218, United States}
\begin{abstract}
We identify 73 $z\sim7$ and 59 $z\sim8$ candidate galaxies in the
reionization epoch, and use this large 26-29.4 AB mag sample of
galaxies to derive very deep luminosity functions to $<-18$ AB mag and
the star formation rate density at $z\sim7$ and $z\sim8$ (just 800 Myr
and 650 Myr after recombination, respectively). The galaxy sample is
derived using a sophisticated Lyman-Break technique on the full
two-year WFC3/IR and ACS data available over the HUDF09 ($\sim$29.4 AB
mag, 5$\sigma$), two nearby HUDF09 fields ($\sim$29 AB mag, 5$\sigma$,
14 arcmin$^2$) and the wider area ERS ($\sim$27.5 AB mag, 5$\sigma$,
$\sim$40 arcmin$^2$). The application of strict optical non-detection
criteria ensures the contamination fraction is kept low (just
$\sim$7\% in the HUDF).  This very low value includes a full
assessment of the contamination from lower redshift sources,
photometric scatter, AGN, spurious sources, low mass stars, and
transients (e.g., SNe).  From careful modeling of the selection
volumes for each of our search fields we derive luminosity functions
for galaxies at $z\sim7$ and $z\sim8$ to $<-18$ AB mag.  The faint-end
slopes $\alpha$ at $z\sim7$ and $z\sim8$ are uncertain but very steep
at $\alpha = -2.01\pm0.21$ and $\alpha=-1.91\pm0.32$,
respectively. Such steep slopes contrast to the local $\alpha\sim-1.4$
and may even be steeper than that at $z\sim4$ where
$\alpha=-1.73\pm0.05$.  With such steep slopes ($\alpha\lesssim -1.7$)
lower luminosity galaxies dominate the galaxy luminosity density
during the epoch of reionization. The star formation rate densities
derived from these new $z\sim7$ and $z\sim8$ luminosity functions are
consistent with the trends found at later times (lower redshifts). We
find reasonable consistency, with the SFR densities implied from
reported stellar mass densities, being only $\sim$40\% higher at
$z<7$.  This suggests that (1) the stellar mass densities inferred
from the Spitzer IRAC photometry are reasonably accurate and (2) that
the IMF at very high redshift may not be very different from that at
later times.

\end{abstract}

\section{Introduction}

One of the primary frontiers in extragalactic astronomy is observing
the build-up of the galaxies in the first 1-2 billion years of the
universe.  With the recent installation of WFC3/IR on HST, we are in
position to study galaxy evolution out to $z\sim8$ (e.g., Bouwens et
al.\ 2010b; McLure et al.\ 2010; Bunker et al.\ 2010; Finkelstein et
al.\ 2010; Trenti et al.\ 2011) and even plausibly to $z\sim10$ (e.g.,
Bouwens et al.\ 2011a).  Through the detection and characterization of
galaxies at such high redshifts, we can place important constraints on
the accretion rate of gas onto galaxies at early times, feedback
effects, and perhaps even the dark matter power spectrum.  These
studies also provide essential inputs to reionization discussions
(e.g., Bolton et al.\ 2007; Pawlik et al.\ 2009), galaxy modeling
(e.g., Finlator et al.\ 2011; Lacey et al.\ 2011; Salvaterra et
al.\ 2010; Choi \& Nagamine 2010; Trenti et al.\ 2010) as well as our
understanding of how quickly metals pollute the IGM (e.g., Ryan-Weber
et al.\ 2009; Becker et al.\ 2009; Simcoe 2006).

One very useful way of characterizing the galaxy population over an
extended period of cosmic time is through a careful quantification of
the $UV$ luminosity function (LF) of galaxies with redshift.  The UV
LF is a particularly useful tracer of the build-up of galaxies at
early times, given the observed correlation of UV light with stellar
mass (Stark et al.\ 2009; Gonzalez et al.\ 2010; Labb\'{e} et
al.\ 2010a,b) and halo mass (e.g., Lee et al.\ 2006, 2009; Ouchi et
al.\ 2005; Overzier et al.\ 2006; McLure et al.\ 2009).  Comprehensive
studies of both the $UV$ LF and the differential evolution of the $UV$
LF with cosmic time are available at $z$$\sim$4-5 (Yoshida et
al.\ 2006), $z$$\sim$4-6 (Bouwens et al.\ 2007: see also Beckwith et
al.\ 2006), $z$$\sim$5-6 (McLure et al.\ 2009), $z$$\sim$4-7 (Bouwens
et al.\ 2008), and $z$$\sim$2-4 (Reddy et al.\ 2008; Reddy \& Steidel
2009: see also Sawicki \& Thompson 2006).  The general result of these
studies at $z\lesssim6$ has been that the volume density of galaxies
evolves much more rapidly at the bright end of the $UV$ LF than at the
faint end.

Recently, as a result of the installation of the WFC3/IR camera on
HST, it has been possible to extend these studies to $z\gtrsim6$.  The
first ultra-deep WFC3/IR observations over the Hubble Ultra Deep Field
(Beckwith et al.\ 2006) as part of the HUDF09 program (GO 11563: PI
Illingworth) made it possible to identify samples of $\gtrsim$20
galaxies at $z$$\sim$7-8 (Oesch et al.\ 2010a; Bouwens et al.\ 2010b;
McLure et al.\ 2010; Bunker et al.\ 2010; Yan et al.\ 2010;
Finkelstein et al.\ 2010) and even out to $z$$\sim$8.5.  These early
WFC3/IR samples have already been used to derive
astrophysically-interesting results on the sizes and structure (Oesch
et al.\ 2010a), $UV$-continuum slopes (Bouwens et al.\ 2010a;
Finkelstein et al.\ 2010; Bunker et al.\ 2010), stellar masses
(Labb\'{e} et al. 2010a,b; Finkelstein et al.\ 2010), and specific
star formation rates (sSFRs: Labb\'{e} et al. 2010b) of $z\gtrsim7$
galaxies.

Here we significantly improve upon these early studies by taking
advantage of the full two year observations over the HUDF and two
other ultra-deep WFC3/IR fields taken as part of the HUDF09 program,
as well as the deep wide-area ($\sim$40 arcmin$^2$) WFC3/IR Early
Release Science (ERS) observations (Windhorst et al.\ 2011).  These
observations allow us to significantly expand the sample of $z\sim7$
and $z\sim8$ galaxies and substantially improve LF determinations.
The ultra-deep HUDF09 observations are valuable for extending the
previous results to even fainter levels and adding substantial
statistics -- allowing for significantly improved determinations of
the LF shape and its faint-end slope $\alpha$.  Meanwhile, the
wide-area ERS observations provide important constraints on the rarer,
luminous population of $z$$\sim$7-8 galaxies.  The ERS fields reach
$\sim$1 mag deeper than other recent wide-area surveys in the near-IR
(e.g., Castellano et al.\ 2010a,b; Hickey et al.\ 2010; Bouwens et
al.\ 2010c; Ouchi et al.\ 2009).  Analyses of shallower versions of
this data set are provided by Wilkins et al. (2011), McLure et
al.\ (2011), and Lorenzoni et al.\ (2011).

We begin our manuscript by describing the data sets we use for our
$z\sim7$ and $z\sim8$ selections (\S2).  We then move on to describing
our procedure for constructing catalogs, selecting our Lyman Break
samples, and estimating contamination (\S3).  In \S4, we compare our
samples with those available in the literature and discuss the
robustness of current Lyman Break selections.  In \S5, we use the
present Lyman Break samples to quantify the LF at $z\sim7$ and
$z\sim8$ and then compare our LFs with those in the literature (\S6).
In \S7, we discuss the implications of our results for the evolution
of the LF at high redshift, the SFR density and stellar mass density,
and the reionization of the universe.  Finally, in \S8 we summarize
our results.  The appendices provide a detailed description of how we
utilize the optical data to ensure our selections are as clean as
possible, as well as compiling a number of other useful but secondary
results and tests.  Throughout this work, we find it convenient to
quote results in terms of the luminosity $L_{z=3}^{*}$ Steidel et
al.\ (1999) derived at $z\sim3$, i.e., $M_{1700,AB}=-21.07$.  We will
refer to the HST F435W, F606W, F775W, F850LP, F098M, F105W, F125W, and
F160W bands as $B_{435}$, $V_{606}$, $i_{775}$, $z_{850}$, $Y_{098}$,
$Y_{105}$, $J_{125}$, and $H_{160}$, respectively, for simplicity.  We
assume $\Omega_0 = 0.3$, $\Omega_{\Lambda} = 0.7$, $H_0 =
70\,\textrm{km/s/Mpc}$.  We express all magnitudes in the AB system
(Oke \& Gunn 1983).

\begin{deluxetable}{cccc}
\tablecolumns{4}
\tablecaption{A summary of the observational data used to search for
  $z\sim7$ and $z\sim8$ galaxies.\label{tab:obsdata}}
\tablehead{
\colhead{} & \colhead{Detection Limits\tablenotemark{a,b}} & \colhead{PSF FWHM} & \colhead{Areal Coverage}\\
\colhead{Passband} & \colhead{(5$\sigma$)} & \colhead{(arcsec)} & \colhead{(arcmin$^2$)}}
\startdata
\multicolumn{4}{c}{HUDF09 (WFC3/IR HUDF)} \\
$B_{435}$ & 29.7 & 0.09 & 4.7 \\
$V_{606}$ & 30.1 & 0.09 & 4.7 \\
$i_{775}$ & 29.9 & 0.09 & 4.7 \\
$z_{850}$ & 29.4 & 0.10 & 4.7 \\
$Y_{105}$ & 29.6 & 0.15 & 4.7 \\
$J_{125}$ & 29.9 & 0.16 & 4.7 \\
$H_{160}$ & 29.9 & 0.17 & 4.7 \\
\multicolumn{4}{c}{} \\
\multicolumn{4}{c}{HUDF09-1 (WFC3/IR P12)} \\
$V_{606}$ & 29.0 & 0.09 & 4.7 \\
$i_{775}$ & 29.0 & 0.09 & 4.7 \\
$z_{850}$ & 29.0 & 0.10 & 4.7 \\
$Y_{105}$ & 29.0 & 0.15 & 4.7 \\
$J_{125}$ & 29.3 & 0.16 & 4.7 \\
$H_{160}$ & 29.1 & 0.17 & 4.7 \\
\multicolumn{4}{c}{} \\
\multicolumn{4}{c}{HUDF09-2 (WFC3/IR P34)} \\
$B_{435}$\tablenotemark{c} & 28.8 & 0.09 & 3.3 \\
$V_{606}$\tablenotemark{c} & 29.9 & 0.09 & 4.7 \\
$i_{775}$\tablenotemark{c} & 29.3 & 0.09 & 4.7 \\
$I_{814}$\tablenotemark{c} & 29.0 & 0.09 & 3.3 \\
$z_{850}$\tablenotemark{c} & 29.2 & 0.10 & 4.7 \\
$Y_{105}$ & 29.2 & 0.15 & 4.7 \\
$J_{125}$ & 29.5 & 0.16 & 4.7 \\
$H_{160}$ & 29.3 & 0.17 & 4.7 \\
\multicolumn{4}{c}{} \\
\multicolumn{4}{c}{WFC3/IR ERS Fields (CDF-S GOODS)} \\
$B_{435}$ & 28.2 & 0.09 & 39.2 \\
$V_{606}$ & 28.5 & 0.09 & 39.2 \\
$i_{775}$ & 28.0 & 0.09 & 39.2 \\
$z_{850}$ & 28.0 & 0.10 & 39.2 \\
$Y_{098}$ & 27.9 & 0.15 & 39.2 \\
$J_{125}$ & 28.4 & 0.16 & 39.2 \\
$H_{160}$ & 28.1 & 0.17 & 39.2 \\
\enddata
\tablenotetext{a}{$0.35''$-diameter apertures}
\tablenotetext{b}{No correction is made for the light outside of the
  $0.35''$-diameter aperture.  This is in contrast to several previous
  WFC3/IR studies by our team (e.g., Bouwens et al.\ 2010a,b) where
  our quoted depths were corrected for the missing light (which can
  result in a $\sim$0.3 mag and $\sim$0.5 mag correction to the quoted
  depths for the ACS and WFC3/IR data, respectively, but depend upon
  the profile assumed).}
\tablenotetext{c}{Our reductions of the ACS data over the HUDF09-2
  field include both those observations taken as part of the HUDF05
  (82 orbits: see dark blue region covering the HUDF09-2 field in
  Figure~\ref{fig:obsdata}) and HUDF09 (111 orbits: see cyan ``P1''
  region covering the HUDF09-2 field) programs.  The latter
  observations add $\sim$0.15-0.4 mag to the total optical depths and
  are important for ensuring our Lyman Break selections are largely free
  of contamination to the limits of our selection ($\sim$29 AB mag).}
\end{deluxetable}

\begin{figure}
\epsscale{1.15}
\plotone{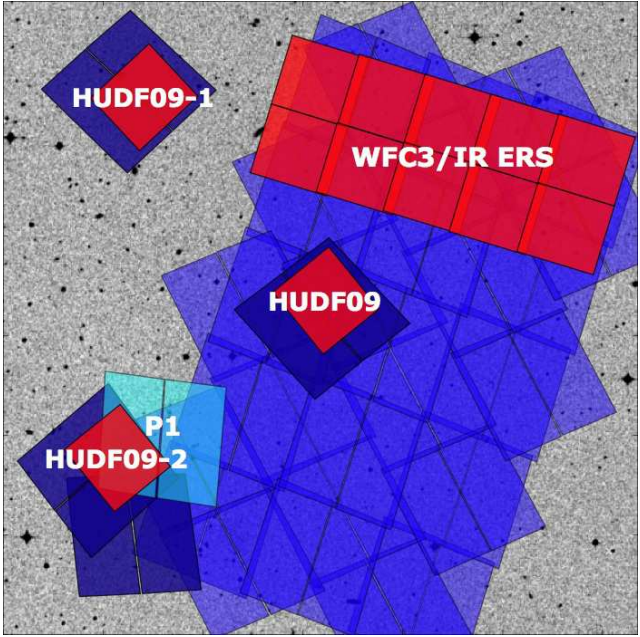}
\caption{Deep WFC3/IR data over the extended CDF-South GOODS field
  used to search for $z\gtrsim7$ Lyman Break galaxies.  The red rectangles
  over the HUDF, HUDF05-1 (P12), and HUDF05-2 (P34) fields (labelled
  HUDF09, HUDF09-1, and HUDF09-2) show the position of the three
  $\sim$4.7 arcmin$^2$ ultra-deep WFC3/IR observations ($\sim$29-29.5
  AB mag at $5\sigma$) that make up the HUDF09 program (\S2.1).  The
  upper ten rectangles show the WFC3/IR pointings that make up the
  Early Release Science observations (GO 11359: PI O'Connell) and
  cover $\sim$40 arcmin$^2$ (\S2.2).  The dark blue regions show the
  position of the ultra-deep $100+$-orbit ACS/WFC data over the HUDF
  (Beckwith et al.\ 2006) and HUDF05-1/HUDF05-2 (P12, P34: Oesch et
  al.\ 2009) fields, while the blue regions show the position of the
  ACS observations over the GOODS fields (Giavalisco et al.\ 2004).
  The cyan region labeled P1 show the position of the deep 111-orbit
  ACS observations we obtained over the HUDF09-2 WFC3/IR field while
  observing the HUDF09 field with the WFC3/IR camera.  A convenient
  summary of the properties of these search fields is given in
  Table~\ref{tab:obsdata}.\label{fig:obsdata}}
\end{figure}

\section{Observational Data}

The primary data set for this analysis is the two-year ultra-deep
WFC3/IR observations that were taken as part of the HUDF09 program (GO
11563: PI Illingworth).  These IR data were obtained over the HUDF and
the two ACS HUDF05 fields.  We also include results from the new wide
area WFC3/IR Early Release Science (ERS) Observations over the
CDF-South GOODS field (GO 11359: PI O'Connell).  A summary of the
observational data we use for the present analysis is provided in
Table~\ref{tab:obsdata}.  The layout of these deep WFC3/IR
observations over and around the CDF-South GOODS field is illustrated
in Figure~\ref{fig:obsdata}.

\subsection{HUDF09 observations}

The full two-year WFC3/IR observations from the HUDF09 program consist
of 192 orbits of ultra-deep WFC3/IR data over the HUDF09 (111 orbits),
HUDF09-1 (33 orbits), and HUDF09-2 (48 orbits) fields.  The
observations from this program are now complete.

All 111 orbits of imaging data over the HUDF were obtained on a single
$\sim$4.7 arcmin$^2$ WFC3/IR pointing and were distributed over three
bands $Y_{105}$ (24 orbits), $J_{125}$ (34 orbits), and $H_{160}$ (53
orbits).  The reductions of those data are similar those already
described in Oesch et al.\ (2010a) and Bouwens et al.\ (2010b), but
now include the full two years of observations of the HUDF.  Great
care was exercised to ensure the image registration to the optical
HUDF was as accurate as possible.  Not only were latest distortion
solutions (10/11/2010) utilized, but each source in our WFC3/IR
reductions was cross-correlated with the corresponding source in the
v1.0 HUDF ACS $z_{850}$-band observations (Beckwith et al.\ 2006).
Small corrections to the distortion solution were required to obtain
excellent registration (i.e., RMS differences of $<$0.01$''$) for the
$Y_{105}$-band data (no WFC3/IR distortion solution were explicitly
derived for this filter by STScI).  Pixels affected by source
persisitence were explicitly masked out.  This masking was performed
by remapping our initial reductions of the data back to the original
frames, subtracting them from the original frames, coadding these
subtracted frames for all exposures within a visit, smoothing, and
then flagging all pixels above a $3\sigma$ threshold.  The final
reduced frames were drizzled onto the v1.0 HUDF ACS reductions
rebinned on a 0.06$''$-pixel frame.

Our HUDF09 WFC3/IR reductions reach to 29.6, 29.9, and 29.9 AB mag
(5$\sigma$: 0.35$''$-diameter apertures) in the $Y_{105}$, $J_{125}$,
and $H_{160}$ bands, respectively.  The optical ACS imaging over the
HUDF reach to 29.7, 30.1, 29.9, and 29.4 AB mag (5$\sigma$:
0.35$''$-diameter apertures) in the $B_{435}$, $V_{606}$, $i_{775}$,
$z_{850}$ bands, respectively.  These depths do not include any
correction for the light outside of the $0.35''$-diameter aperture and
on the wings of the PSF, so that the depths can be readily reproduced.

The availability of the WFC3/IR observations over the HUDF09-1 (33
orbits) and HUDF09-2 (48 orbits) fields allows us to substantially
extend our $z\sim7$ and $z\sim8$ samples.  In both fields, the WFC3/IR
observations were concentrated in single $\sim$4.7 arcmin$^2$ WFC3/IR
pointings and distributed across the three bands $Y_{105}$, $J_{125}$,
and $H_{160}$.  In the HUDF09-1 field, 8 orbits in the $Y_{105}$-band,
12 orbits in the $J_{125}$-band, and 13 orbits in the $H_{160}$-band
were acquired.  In the HUDF09-2 field, 11 orbits in the
$Y_{105}$-band, 18 orbits in the $J_{125}$-band, and 19 orbits in the
$H_{160}$-band were acquired.  These new observations were reduced
using the same procedures as for the HUDF.  The WFC3/IR observations
were registered and drizzled onto the same frame as our reductions of
the ACS data on the two fields rebinned on a 0.06$''$-pixel frame.  As
with our UDF reductions, care was taken to ensure that the
registration to the optical data was as accurate as possible
($<0.01''$) and that even issues such as velocity aberration were
properly treated.  Our reductions of the WFC3/IR over the HUDF09-1
field reach to 29.0, 29.3, and 29.1 AB mag in the $Y_{105}$,
$J_{125}$, and $H_{160}$ bands, respectively.  These depths are 29.1,
29.4, and 29.3 AB mag, respectively, over the HUDF09-2 field (see
Table~\ref{tab:obsdata}).

For the ACS observations over the HUDF09-1 field, we could only take
advantage of those from the HUDF05 (GO10632: PI Stiavelli) program,
and so we used the publicly available v1.0 reductions of Oesch et
al.\ (2007).  For the HUDF09-2 field, on the other hand, deep ACS
observations are available as a result of two programs: the HUDF05 and
HUDF09 programs.  The ACS observations over this field from the HUDF05
program (102 orbits in total: 9 orbits $V_{606}$ band, 23 orbits
$i_{775}$ band, 70 orbits $z_{850}$ band) were obtained over an
$\sim$18 month period in 2005 and 2006 at a number of different
orientations, as well as a $1'$ coordinate shift for 20 orbits from
the primary pointing.  111 additional orbits of ACS data (10 orbits
$B_{435}$ band, 23 orbits $V_{606}$ band, 23 orbits $i_{775}$ band, 16
orbits $I_{814}$ band, 39 orbits $z_{850}$ band) were acquired over
this field in 2009 and 2010 from our HUDF09 program.

Reductions of the recent ACS observations were conducted using the ACS
GTO apsis pipeline (Blakeslee et al.\ 2003).  Special care was
required to cope with the significantly reduced charge transfer
efficiency in the new post-SM4 ACS observations and to correct for
row-by-row banding artifacts.  Our recent ACS observations (from
HUDF09) cover $\sim$70\% of the HUDF09-2 footprint and add
$\sim$0.15-0.5 mag to the depth of the earlier HUDF05 observations
(after coaddition).  Reductions of the HUDF05 ACS observations are
described in Bouwens et al.\ (2007).  The depths of the WFC3/IR and
ACS observations over these two fields are given in
Table~\ref{tab:obsdata} and reach to $\gtrsim$29 AB mag ($5\sigma$).

\subsection{WFC3/IR ERS Observations}

The WFC3/IR ERS observations cover $\sim$40 arcmin$^2$ in the upper
region of the CDF South GOODS field (Windhorst et al.\ 2011).  These
observations include 10 separate $\sim$4.7 arcmin$^2$ WFC3/IR
pointings (see Figure~\ref{fig:obsdata}).  2 orbits of near-IR data
are obtained in the F098M, F125W, and F160W bands, for a total of 6
orbits per field (60 orbits in total).  These near-IR observations are
reduced in a very similar way to the procedure used for the WFC3/IR
observations from our HUDF09 program (Oesch et al.\ 2010a; Bouwens et
al.\ 2010b).  To keep the size of the drizzled WFC3/IR frames
manageable, we split the output mosaic into 8 discrete pieces --
corresponding to different segments in the ACS GOODS mosaic (i.e.,
S14, S24, S25, S34, S35, S43, S44, S45).\footnote{The configuration of
  these segments within the CDF-South GOODS mosaic is illustrated in
  http://archive.stsci.edu/pub/hlsp/goods/v1/h\_cdfs\_v1.0sects\_plt.jpg.}
For each segment, we first aligned the observations against the ACS
data binned on a 0.06$''$-pixel scale and then drizzled the data onto
that frame.  For the ACS data, we made use of our own reductions
(Bouwens et al.\ 2006; Bouwens et al.\ 2007) of the deep GOODS ACS/WFC
data over the GOODS fields (Giavalisco et al.\ 2004).  These
reductions are similar to the GOODS v2.0 reductions, but reach
$\sim$0.1-0.3 mag deeper in the $z_{850}$-band due to our inclusion of
the SNe follow-up data (e.g., Riess et al.\ 2007).  Our reduced
WFC3/IR data reach to 27.9, 28.4, and 28.0 in the $Y_{098}$, $J_{125}$
and $H_{160}$ bands, respectively ($5\sigma$: $0.35''$ apertures).
The ACS observations reach to 28.2, 28.5, 28.0, and 28.0 in the
$B_{435}$, $V_{606}$, $i_{775}$, $z_{850}$, respectively ($5\sigma$:
$0.35''$ apertures).  The FWHMs of the PSFs are $\sim$0.16$''$ for the
WFC3/IR data and $\sim$0.10$''$ for the ACS/WFC data.

\section{Lyman Break Selection}

\begin{figure}
\epsscale{1.15}
\plotone{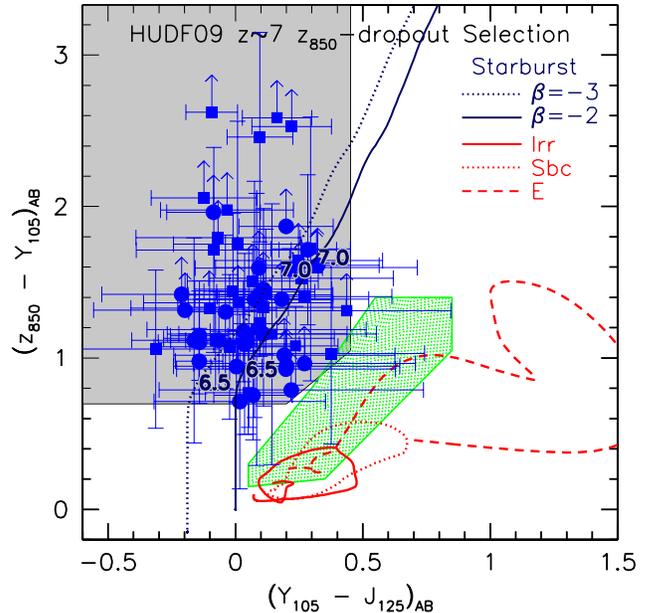}
\caption{(\textit{left}) $z_{850}-Y_{105}$ vs. $Y_{105}-J_{125}$
  two-color diagram we use to identify $z\sim7$ $z_{850}$-dropouts
  over the three ultra-deep WFC3/IR HUDF09 fields (14 arcmin$^2$).
  The $z-Y$/$Y-J$ colors required for our $z_{850}$-dropout selection
  is indicated by the gray region.  $z_{850}$-dropout candidates
  identified in the HUDF09, HUDF09-1, and HUDF09-2 fields are shown
  with the blue solid circles, blue open squares, and blue solid
  squares, respectively.  The error bars and lower limits are
  $1\sigma$.  The blue lines show the expected colors of star-forming
  galaxies with $UV$-continuum slopes $\beta$ of $-3$ and $-2$ while
  the red lines show the expected colors of low-redshift contaminants.
  The colors of low-mass L,T dwarf stars (e.g., Knapp et al.\ 2004)
  are indicated by the green squares.  In addition to the presented
  two-color Lyman-Break selection, we also enforce a very strict
  optical non-detection criterion involving a $\chi_{opt} ^2$ quantity
  (\S3.3; Appendix D).  Through extensive simulations, we have found
  that a two-color LBG selection combined with strong constraints on
  the optical flux allow for the robust selection of star-forming
  galaxies at $z\sim7$.\label{fig:zyyj}}
\end{figure}

In this section, we describe our procedure for selecting star-forming
galaxies at $z\sim7$ and $z\sim8$ using the observational data (\S2).
We begin by detailing our technique for constructing source catalogs
(\S3.1).  We then describe our procedure for selecting galaxies at
$z\sim7$ and $z\sim8$ using Lyman-Break Galaxy selection criteria in
the HUDF09 (\S3.2) and ERS (\S3.4) fields.  In \S3.5, we discuss
possible sources of contamination and attempt to quantify their
importance, and in \S3.6, we provide a brief summary of our final
samples of $z\sim7$ and $z\sim8$ galaxies.

\subsection{Catalog Construction}

We generate separate catalogs for each of our Lyman Break selections,
to obtain more optimal photometry for each selection.  This is done
through the use of square root of $\chi^2$ detection images (Szalay et
al.\ 1999: similar to a coadded inverse noise-weighted image)
constructed from only those bands that are expected to have flux for a
given Lyman Break selection.  Specifically, this image is constructed
from the $Y_{105}$, $J_{125}$ and $H_{160}$ band images for our
$z\sim7$ HUDF09 $z_{850}$-dropout selection, the $J_{125}$ and
$H_{160}$ band images for our $z\sim8$ HUDF09 $Y_{105}$-dropout
selection, and from the $J_{125}$ and $H_{160}$ band images for our
ERS $z_{850}$ and $Y_{098}$-dropout selections; the square root of
$\chi^2$ image therefore includes all deep WFC3/IR observations
redward of the break.

Object detection and photometry is performed using the SExtractor
(Bertin \& Arnouts 1996) software run in dual image mode.  Object
detection is performed off the square root of $\chi^2$ image.  Colors
are measured in small scalable apertures (MAG\_AUTO) defined using a
Kron (1980) factor of 1.2 (where the Kron radius is established from
the square root of $\chi^2$ image).  All of our imaging data
(optical/ACS and near-IR/WFC3) are PSF-matched to the WFC3/IR F160W
imaging data before making these color measurements.  Fluxes in these
small scalable aperture are then corrected to total magnitudes in two
steps.  First a correction is made for the additional flux in a larger
scalable aperture (with Kron factor of 2.5).  Then a correction is
made for the light outside this larger scalable aperture (a
$0.7''$-diameter aperture is typical) and on the wings of the PSF.
The latter correction (typically 0.2 mag) is made based upon the
encircled energy measured outside this aperture (for
stars).\footnote{See
  http://www.stsci.edu/hst/wfc3/documents/handbooks/
  currentIHB/c07\_ir07.html.}  The typical size of the total
correction (including both steps) is $\sim$0.4-0.8 mag.

\begin{figure}
\epsscale{1.15}
\plotone{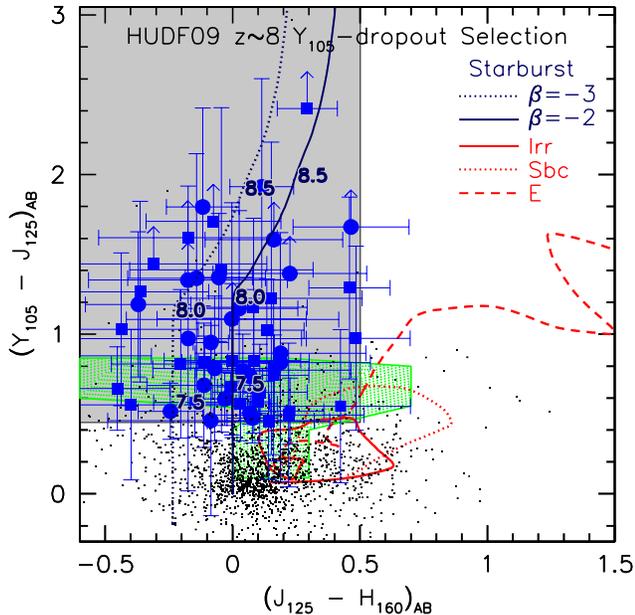}
\caption{(\textit{left}) $Y_{105}-J_{125}$ vs. $J_{125}-H_{160}$
  two-color diagram used to identify $z\sim8$ $Y_{105}$-dropout
  galaxies over the three ultra-deep WFC3/IR HUDF09 fields (14
  arcmin$^2$).  The $Y_{105}-J_{125}$/$J_{125}-H_{160}$ colors
  required for our $Y_{105}$-dropout selection are indicated by the
  gray region.  The other symbols and lines are as in
  Figure~\ref{fig:zyyj}.  The overlap of our color selection region
  with that occupied by L and T dwarfs is not a concern -- given that
  an exceeding small fraction of the sources ($\lesssim$2\%) in our
  selection appear to be unresolved (see \S3.5).  In addition to the
  presented two-color Lyman-Break selection, we also enforce a very
  strict optical non-detection criterion involving a $\chi_{opt} ^2$
  quantity (\S3.3; Appendix D).  As noted in Figure~\ref{fig:zyyj}, our
  extensive simulations show that these two color criteria -- combined
  with strict optical non-detection requirements -- allow for the
  robust selection of $z\sim8$ galaxies.\label{fig:yjjh}}
\end{figure}

\subsection{Selection Procedure (HUDF09 Fields)}

Our primary $z$$\sim$7-8 Lyman-Break samples are based upon the
ultra-deep WFC3/IR observations over the three HUDF09 fields.  The
availability of ultra deep three-band ($Y_{105}$, $J_{125}$, and
$H_{160}$) observations over these fields allow us to use traditional
two-color Lyman-Break Galaxy selection criteria to identify
star-forming galaxies at $z\gtrsim7$.  Spectroscopic follow-up has
shown that these criteria are generally quite robust in identifying
star-forming galaxies at $z$$\sim$3-7 (e.g., Steidel et al.\ 1996,
2003; Popesso et al.\ 2009; Vanzella et al.\ 2009; Stark et al.\ 2010)
though admittedly spectroscopic follow-up of the very faintest sources
at $z\geq4$ has been ambiguous since only those sources with
Ly$\alpha$ emission can be confirmed.

We favor the use of Lyman-Break Galaxy (LBG) ``dropout'' selection
criteria over the use of photometric redshifts for the identification
of star-forming galaxies at $z\sim7$ and $z\sim8$ due to (1) the
simplicity and transparency of LBG criteria, (2) the ease with which
optical non-detection criteria can be engineered to optimally remove
contaminants (e.g., see Appendix D), (3) economy with which these
criteria capture the available redshift information (while minimizing
redshift aliasing effects), and (4) accuracy with which it is possible
to model the effective selection volumes.

\begin{figure}
\epsscale{1.15}
\plotone{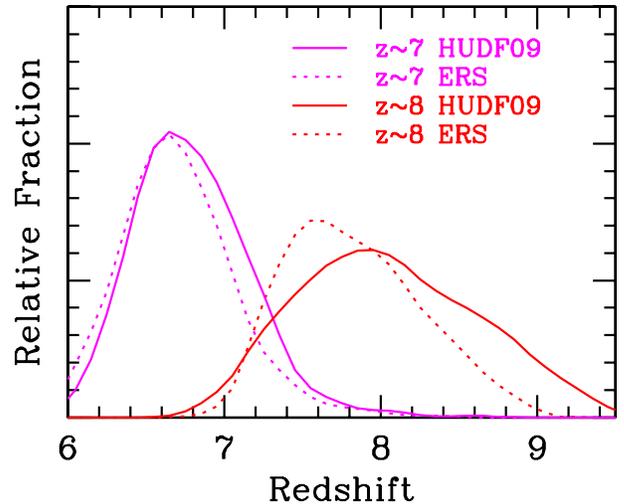}
\caption{Redshift distributions predicted for our $z\sim7$
  $z_{850}$-dropout (solid magenta lines) and $z\sim8$
  $Y_{105}$-dropout (solid red lines) selections over our ultra-deep
  HUDF09 fields (see \S3.2).  Also shown are the redshift
  distributions for our $z\sim7$ $z_{850}$-dropout and $z\sim8$
  $Y_{098}$-dropout selection over the wide-area ERS fields (\S3.4:
  dotted magenta and red lines, respectively).  Different WFC3/IR
  $Y$-band filters are employed for the HUDF09 fields ($Y_{105}$) and
  for the ERS search area ($Y_{098}$); the $J_{125}$ and $H_{160}$
  filters are the same.  The selection criteria are carefully chosen
  to allow for a better match between the HUDF09 and ERS redshift
  selection windows.  The result is that the redshift distributions
  are similar at $z\sim7$ and $z\sim8$.  The mean redshift for our
  $z_{850}$-dropout selections is 6.8 and 6.7 for our HUDF09 and ERS
  selections, respectively.  The mean redshift for our
  $Y_{105}/Y_{098}$-dropout selections is 8.0 and 7.8 for our HUDF09
  and ERS selections, respectively.\label{fig:zdist}}
\end{figure}

The color criteria we utilize for our selection are similar to those
already used by Oesch et al.\ (2010a) and Bouwens et al.\ (2010b).  In
detail, these criteria are
\begin{displaymath}
(z_{850} - Y_{105} > 0.7)\wedge(Y_{105}-J_{125} < 0.45)
\wedge 
\end{displaymath}
\begin{displaymath}
(z_{850}-Y_{105}>1.4(Y_{105}-J_{125})+0.42)
\end{displaymath}
for our $z\sim7$ $z_{850}$-dropout sample and
\begin{displaymath}
(Y_{105} - J_{125} > 0.45)\wedge(J_{125}-H_{160} < 0.5)
\end{displaymath}
for our $z\sim8$ $Y_{105}$-dropout selection, where $\wedge$
represents the logical \textbf{AND} symbol.  These two-color criteria
are illustrated in Figures~\ref{fig:zyyj} and \ref{fig:yjjh},
respectively.  The criteria were chosen to allow for a fairly complete
selection of star-forming galaxies at $z\sim7$ and $z\sim8$ and so
that our HUDF09 samples have a very similar redshift distribution to
those galaxies selected with the ERS $Y_{098}J_{125}H_{160}$ bands
(see Figure~\ref{fig:zdist} and \S3.4).  They are also carefully
crafted so that all star-forming galaxies at $z\gtrsim6.5$ are
included and do not fall between our $z\sim7$ and $z\sim8$ selections.
To ensure that sources are real, we require that sources be detected
at $\gtrsim$3.5$\sigma$ in the $J_{125}$ band and $3\sigma$ in the
$Y_{105}$ or $H_{160}$ bands.  This is equivalent to a detection
significance of $4.5\sigma$ (when combining the $J_{125}$ and
$Y_{105}$/$H_{160}$ detections).

\begin{figure*}
\epsscale{0.85}
\plotone{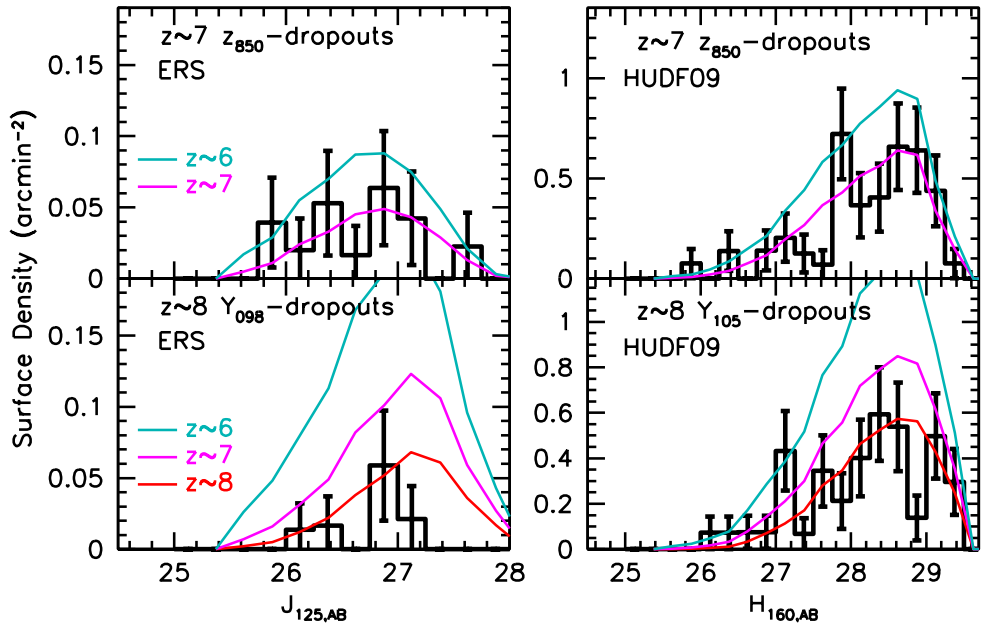}
\caption{Surface densities found in our $z$$\sim$7 $z_{850}$-dropout
  (\textit{upper left}) and $z\sim8$ $Y_{098}$-dropout (\textit{lower
    left}) selections over the wide-area ERS observations
  (\textit{black histograms, with $1\sigma$ errors}) and in our
  $z\sim7$ $z_{850}$-dropout (\textit{upper right}) and $z\sim8$
  $Y_{105}$-dropout (\textit{lower right}) selections over our HUDF09
  observations (\textit{histograms}).  For comparison, we show the
  predicted surface densities based upon the Bouwens et al.\ (2007,
  2008) determinations of the $UV$ LF at $z\sim6$ (\textit{cyan
    lines}) and $z\sim7$ (\textit{magenta lines}).  The solid red
  lines are the surface densities predicted extrapolating the Bouwens
  et al.\ 2008 LF results to $z\sim8$ (based upon the fitting formula
  in that paper).  The two $H_{160,AB}\sim26$ $Y_{105}$-dropouts seen
  in the lower right panel -- while exceeding the predicted surface
  densities for $z\sim8$ star-forming galaxies -- may be part of an
  overdensity (see \S5.3).\label{fig:surfdens}}
\end{figure*}

\textit{For both of our $z\sim7$ $z_{850}$-dropout and $z\sim8$
  $Y_{105}$-dropout samples, we enforce very stringent optical
  non-detection criteria.}  Not only do we reject sources detected at
$2\sigma$ in a single passband or 1.5$\sigma$ in more than one band,
but we also compute a collective $\chi_{opt} ^2$ value from all of the
optical data together and eliminate sources above specific thresholds.
We outline this procedure in detail in \S3.3.

To ensure that our selections do not suffer from significant
contamination from SNe or low mass stars (both of which have
point-like profiles), we examined each of our candidate $z$$\sim$7-8
galaxies with the SExtractor stellarity parameter to identify those
sources consistent with being pointlike.  The only source that we
identified in our fields that was pointlike and satisfied our dropout
criteria was the probable SNe (03:32:34.53, $-$27:47:36.0) previously
identified over the HUDF (Oesch et al.\ 2010a: see also McLure et
al. 2010; Bunker et al.\ 2010; Yan et al.\ 2010).  No other point-like
sources were found (or removed).

We use the above criteria to search for candidate $z\gtrsim7$ galaxies
over all areas of our HUDF09 fields where the WFC3/IR observations are
at least half of their maximum depth in each field (or $\sim$4.7
arcmin$^2$ per field).  In total, 29 $z\sim7$ $z_{850}$-dropout
sources and 24 $z\sim8$ $Y_{105}$-dropouts were identified over our
ultra-deep WFC3/IR HUDF pointing.  17 $z\sim7$ $z_{850}$-dropout
sources and 14 $z\sim8$ $Y_{105}$-dropouts were found over our
ultra-deep WFC3/IR HUDF09-1 pointing, while 14 $z\sim7$
$z_{850}$-dropout sources and 15 $z\sim8$ $Y_{105}$-dropouts were
found over our ultra-deep WFC3/IR HUDF09-2 pointing.  As in the Oesch
et al.\ (2010a) and Bouwens et al.\ (2010b) selections (see also
McLure et al.\ 2010, Bunker et al.\ 2010, Yan et al.\ 2010,
Finkelstein et al.\ 2010), the sources we identified have
$H_{160}$-band magnitudes ranging from $\sim$26 mag to $\sim$29 mag.
A catalog of the $z\sim7$ $z_{850}$-dropouts in the HUDF09 fields is
provided in Tables~\ref{tab:z09candlist}-\ref{tab:z052candlist} of
Appendix E.  A similar catalog of $Y_{105}$-dropouts is provided in
Tables~\ref{tab:y09candlist}-\ref{tab:y052candlist} of Appendix E.
Figures~\ref{fig:zstamp}-\ref{fig:ystamp} of Appendix E show image
cutouts of all of these candidates.  Figure~\ref{fig:surfdens} shows
the approximate surface density of these candidates as a function of
magnitude in our HUDF09 search fields.

As a result of the very limited depths of the ACS data over our
fields, it was necessary to exercise considerable caution in selecting
$z\gtrsim7$ galaxies over our fields.  Our conservative selection
procedure resulted in our eliminating a modest number of sources from
our still very large $z\sim7$ and $z\sim8$ samples.  For comparison
with other studies, we have included a list of these sources (which
have a reasonable probability of being contaminants) in
Tables~\ref{tab:hudf09miss7}-\ref{tab:hudf09miss8} of Appendix E.

\begin{figure*}
\epsscale{1.0}
\plotone{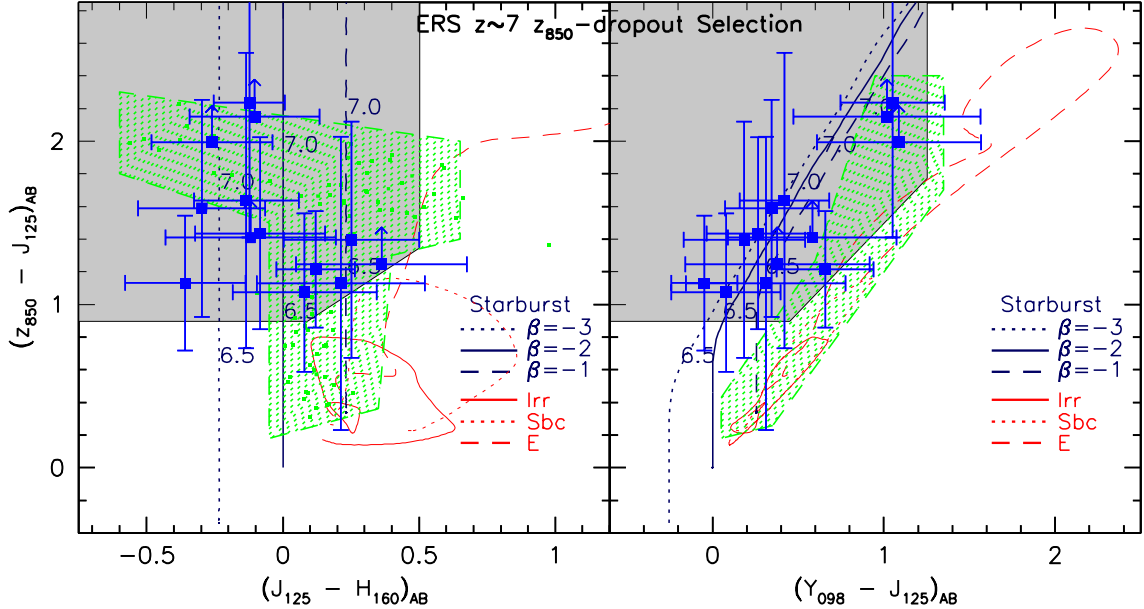}
\caption{(\textit{left}) $z_{850}-J_{125}$ vs. $J_{125}-H_{160}$
  two-color diagram we use to identify $z\sim7$ $z_{850}$-dropouts
  over the $\sim$40 arcmin$^2$ ERS observations (first part of our two
  part $z_{850}$-dropout criterion).  Symbols and lines are the same
  as Figure~\ref{fig:zyyj}.  (\textit{right}) $z_{850}-J_{125}$
  vs. $Y_{098}-J_{125}$ two-color diagram used to identify $z\sim7$
  $z_{850}$-dropout galaxies (second part of our two part
  $z_{850}$-dropout criterion).  Lines and symbols are as in the left
  panel.  The colors required for our $z_{850}$-dropout selection are
  indicated in gray.  We base our $z\sim7$ $z_{850}$-dropout selection
  on a $z_{850}-J_{125}$ dropout criterion rather than a
  $z_{850}-Y_{098}$ dropout criterion (e.g., as used by Wilkins et
  al.\ 2010) to extend our selection over a wider redshift range (see
  \S3.4).  In addition to the presented two-color Lyman-Break
  selection, we also enforce a very strict optical non-detection
  criterion involving a $\chi_{opt} ^2$ quantity (\S3.3; Appendix
  D).\label{fig:zjzj}}
\end{figure*}

\subsection{Controlling for contamination using the measured values of
$\chi_{opt}^2$}

Identifying lower-redshift contaminants in our HUDF09 selections is
very challenging due to the limited depths of the available optical
data.  This is especially true at the faintest magnitudes.  We have
found that the only truly effective way to identify these contaminants
is to make full use of the optical data.

Accordingly, we have developed a rather sophisticated multi-step
procedure to obtain tight controls on contamination.  The first step
is rather standard for high redshift selections: remove all the
sources detected at $2\sigma$ in a single optical band or at
$1.5\sigma$ in more than one optical band.  The second step is more
complex, but very effective.  In this second step we compute a
collective $\chi_{opt} ^2$ from all of the optical data, as
\begin{equation}
\chi_{opt} ^2 = \Sigma_{i} \textrm{SGN}(f_{i}) (f_{i}/\sigma_{i})^2
\end{equation}
We reject all sources with $\chi_{opt} ^2$ greater than a specific
value $\chi_{lim}^2$ that we establish through simulations (Appendix
D).  At brighter $0.5(J_{125,AB}+H_{160,AB})\lesssim28.5$ magnitudes,
we take $\chi_{lim}^2$ equal to 5, 3, and 5 for our HUDF09, HUDF09-1,
and HUDF09-2 selections, respectively, while at fainter
$0.5(J_{125,AB}+H_{160,AB})>29.2$ magnitudes, $\chi_{lim}^2$ is taken
to be half that.  At magnitudes between those limits, the limiting
$\chi_{lim}^2$ is an interpolation between the two extremes.  More
stringent limiting values on $\chi_{lim}^2$ are adopted for the
faintest sources or where the optical data are shallower; more
permissive limiting values on $\chi_{lim}^2$ are adopted for brighter
sources or where the optical data are deeper.  Note that $f_{i}$ is
the flux in band $i$ in our smaller scalable apertures, $\sigma_i$ is
the uncertainty in this flux, and SGN($f_{i}$) is equal to 1 if
$f_{i}>0$ and $-1$ if $f_{i}<0$.  The filters included in the
$\chi_{opt} ^2$ summation are $B_{435}$, $V_{606}$, and $i_{775}$ for
the $z_{850}$-dropout selection and $B_{435}$, $V_{606}$, $i_{775}$,
and $z_{850}$ for the $Y_{105}$-dropout selection.

These particular limits on $\chi_{opt} ^2$ were chosen to minimize
contamination for our Lyman-Break selections while maximizing the
completeness of the $z\sim7$ and $z\sim8$ sources in our samples (see
Appendix D for the relevant simulations).  We determine $\chi_{opt}
^2$ in three different sets of apertures (0.35$''$-diameter apertures,
0.18$''$-diameter apertures, and scalable Kron apertures with typical
and maximum radii of $0.2''$ and $0.4''$) to maximize the information
we have on possible optical flux in our candidates.  Larger aperture
measurements are useful for flagging more extended low-redshift
galaxies in the selection.  Meanwhile, the smaller aperture
measurements provide higher S/N measurements from the higher spatial
resolution ACS data.  Making measurements in such small apertures is
meaningful, given that the alignment we achieve between sources in the
WFC3/IR and ACS images is better than 0.01$''$.

Overall, we have found that a $\chi_{opt} ^2$ criterion is extremely
effective at reducing the contamination rate from low-redshift
galaxies that scatter into our selection.  The reductions in the
contamination rate are very substantial, i.e., factors of $\gtrsim$2-3
over what it would be excluding only those sources detected at
2$\sigma$ in 1 optical band or $>$$1.5\sigma$ in $\geq$2 optical
bands.  Finally, we emphasize that use of the $\chi_{opt} ^2$
statistic and an examination of the $\chi_{opt}^2$ distribution allows
us to verify that our samples are not subject to significant
contamination (since contaminating sources would show up as a tail to
positive $\chi_{opt} ^2$ values: \S4.2 and Appendix D.4).

\subsection{Selection Procedure (ERS Fields)}

To complement our ultra-deep searches for $z$$\sim$7-8 galaxies in the
HUDF09 observations, we also take advantage of the wide area ($\sim$40
arcmin$^2$) ERS observations over the upper portion of the CDF South
GOODS field (Figure~\ref{fig:obsdata}).

Ideally we would use the same selection criteria for identifying
$z$$\sim$7-8 galaxies over the ERS observations as we use over the
three ultra-deep HUDF09 fields.  This is not possible, however since
the $\sim$40 arcmin$^2$ WFC3/IR ERS observations use a different
$Y$-band filter ($Y_{098}$) than in the WFC3/IR HUDF09 observations
($Y_{105}$).  The $J_{125}$ and $H_{160}$ filters remain the same.  We
can nevertheless do quite well in selecting galaxies over a similar
redshift range and with similar properties by judiciously choosing our
selection criteria (as can be seen in Figure~\ref{fig:zdist}).

\begin{figure}
\epsscale{1.15}
\plotone{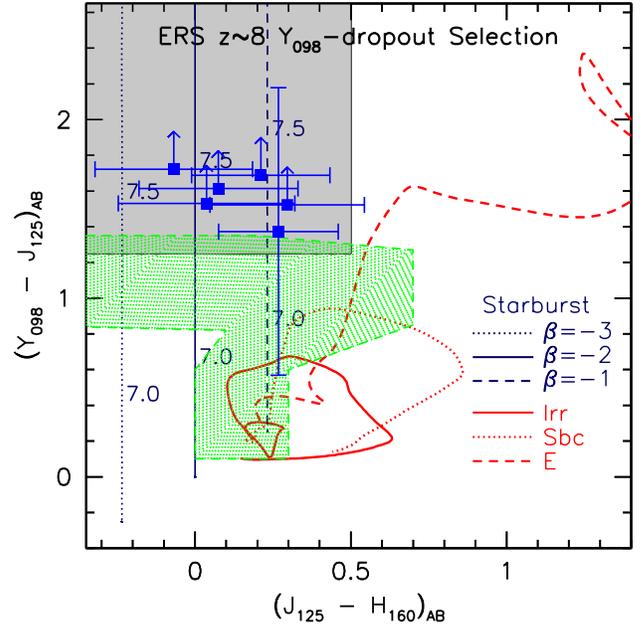}
\caption{$Y_{098}-J_{125}$ vs. $J_{125}-H_{160}$ two-color diagram
  used to identify $z\sim8$ $Y_{098}$-dropout galaxies over the
  $\sim40$ arcmin$^2$ ERS search area.  Regions and lines as per
  Figure~\ref{fig:yjjh}.  In addition to the presented two-color
  Lyman-Break selection, we also enforce a very strict optical
  non-detection criterion involving a $\chi_{opt} ^2$ quantity (\S3.3;
  Appendix D).
\label{fig:zwjjh}}
\end{figure}

The multi-color selection criteria we use to select galaxies at
$z\sim7$ and $z\sim8$ are illustrated in Figures~\ref{fig:zjzj} and
\ref{fig:zwjjh}, respectively, and nominally correspond to $z_{850}$
dropout and $Y_{098}$ dropout selections.  We base our $z\sim7$
$z_{850}$-dropout selection on a $z_{850}-J_{125}$ dropout criterion
rather than a $z_{850}-Y_{098}$ dropout criterion to extend our
selection over a wider redshift range than is possible using a
$z_{850}-Y_{098}$ criterion.  Dropout selections based upon the latter
color criterion (e.g., as used by Wilkins et al.\ 2010) result in
galaxies starting to drop out of the $Y_{098}$ band at $z\gtrsim6.7$.
This makes it more difficult to identify the Lyman Break at
$z\gtrsim7$, resulting in a redshift selection window that is
artificially narrow.

In detail, we require that galaxies show a strong
$z_{850}-J_{125}>0.9$ break and satisfy
$z_{850}-J_{125}>0.8+1.1(J_{125}-H_{160})$ and
$z_{850}-J_{125}>0.4+1.1(Y_{098}-J_{125})$.  The latter two criteria
exclude sources with redshifts $z\lesssim6.5$.  To bound our redshift
window selection on the high end to redshifts $z\lesssim7.4$, we
require that their $Y_{098}-J_{125}$ colors be bluer than 1.25.  In
addition, sources are required to have $J_{125}-H_{160}$ colors
(redward of the break) bluer than 0.5 to exclude intrinsically red
sources (which otherwise might contaminate our selection).  Candidates
are required to be detected at $\geq$4.5$\sigma$ in the $J_{125}$ band
and at $\geq$3.5$\sigma$ in the $H_{160}$ band (to ensure that they
correspond to real sources).  A higher detection significance is
required for our ERS selections than our HUDF09 selections (\S3.2) to
compensate for the fewer exposures that go into each ERS field (and
hence less Gaussian noise).

For our higher $z\sim8$ $Y_{098}$-dropout selections, we require that
galaxies show a strong $Y_{098}-J_{125}>1.25$ break and, similar to
the previous $z\sim7$ $z_{850}$-dropout selection, have
$J_{125}-H_{160}$ colors bluer than 0.5.  As we will later see, our
use of a $Y_{098}-J_{125}>1.25$ criterion selects for star-forming
galaxies at $z\gtrsim7.4$ and is convenient for combining the ERS
dropout samples with the HUDF09 samples (selected using a slightly
different $Y_{105}$ filter).  Candidates are again required to be
detected at $\geq$4.5$\sigma$ in the $J_{125}$-band and $3\sigma$ in
the $H_{160}$ band.

To minimize the contamination from low-redshift galaxies that scatter
into our dropout color windows, we utilize very stringent criteria to
exclude sources that show any evidence for detection in the
optical.  This includes being detected at $2\sigma$ in a single
optical band, $1.5\sigma$ in more than one optical band, or having an
optical $\chi_{opt} ^2$ value $>$2.5 (see \S3.3 and Appendix D for our
definition of $\chi_{opt} ^2$).

To control for potential contamination from low-mass stars or SNe --
both of which have point-like profiles -- we examined each of our
candidate $z\sim7$ $z_{850}$ or $z\sim8$ $Y_{098}$ dropout galaxies
with the SExtractor stellarity parameter to identify those sources
consistent with being pointlike.  One of the sources (03:32:27.91,
$-$27:41:04.2) in our selection had measured stellarity parameters
consistent with being pointlike (with stellarities $>0.8$ in the
$J_{125}$ and $H_{160}$ bands).  This source had a $J-H$$\sim$$-$0.6
color -- which is much bluer than typical $L^*$ galaxies at $z\sim7$
(Bouwens et al.\ 2010a), but consistent with the colors of a T dwarf
(e.g., Knapp et al. 2004).  We therefore identified this source as a
probable T dwarf and removed it from our $z_{850}$-dropout selection.
For all other sources in our selection, the measured stellarity
parameters were substantially less, with typical values $\sim$0.05-0.3
indicative of extended sources.  From a quick inspection of
Figure~\ref{fig:zstamp} and \ref{fig:ystamp}, it is clear that
essentially all the sources in our samples are extended.

The above selection criteria are only applied to those regions of the
WFC3/IR ERS observations over the CDF-South GOODS field where the
optical and near-IR observations reach within 0.2 mag of the typical
ACS GOODS field depth and ERS WFC3/IR depth, i.e., $\gtrsim$3 orbits
(ACS $B_{435}$ band), $\gtrsim$2 orbits (ACS $V_{606}$-band),
$\gtrsim$3 orbits (ACS $i_{775}$-band), $\gtrsim$7 orbits (ACS
$z_{850}$-band), $\gtrsim$2 orbits ($Y_{105}$), $\gtrsim$2 orbits
($J_{125}$), and $\gtrsim$2 orbits ($H_{160}$).  The total effective
area is 39.2 arcmin$^2$ ($\sim$3 arcmin$^2$ of the WFC3/IR ERS fields
extend outside of the deep ACS GOODS data: see
Figure~\ref{fig:obsdata}).

With the above criteria, we selected 13 $z\sim7$ $z_{850}$-dropout
galaxies and 6 $z\sim8$ $z\sim8$ $Y_{098}$-dropout galaxies.  The
$z\sim7$ $z_{850}$-dropout candidates have magnitudes ranging from
25.8 and 27.6 AB mag, while the $z\sim8$ $Y_{098}$-dropout candidates
have magnitudes ranging from 26.0 to 27.2 AB mag.  The properties of
the sources are given in Tables~\ref{tab:zcandlist} and
\ref{tab:z98candlist}.  The surface densities of our $z\sim7$
$z_{850}$ and $z\sim8$ $Y_{098}$-dropouts are shown in
Figure~\ref{fig:surfdens}.

As with our HUDF09 dropout selections, we have compiled a list of
other possible $z\gtrsim7$ candidates that narrowly missed our ERS
selection.  These sources are given in Table~\ref{tab:possiblecand}
along with our reasons for excluding them from our samples.

\subsection{Contamination}

We have carefully crafted our selection criteria to optimally identify
star-forming galaxies at $z\sim7$ and $z\sim8$ while minimizing
contamination.  However, some amount of contamination is almost
inevitable, and so it is important to try to quantify the
contamination rate.  The five most important sources of contamination
for our $z\sim7$ and $z\sim8$ selections are (1) low-mass stars, (2)
SNe and other transient sources, (3) AGN, (4) lower redshift sources
and photometric scatter, and (5) spurious sources.  We consider each
source of contamination in the paragraphs that follow.

\subsubsection{Low-Mass Stars} 

Low-mass stars have $z_{850}-Y_{105}$, $z_{850}-J_{125}$ and
$J_{125}-H_{160}$ colors very similar to that of $z$$\sim$7-8
galaxies, and therefore potentially contaminate our $z\sim7$ and
$z\sim8$ selections.  Fortunately, low mass stars can be identified by
taking advantage of the high-resolution WFC3/IR data, which permit us
to determine which sources are extended and those which are not.
Since essentially all of the sources seem to be extended -- except for
one SNe candidate in the HUDF09 (\S3.2) and one probable T dwarf in
the ERS observations (\S3.4), it seems unlikely that our sample
suffers from significant contamination from low-mass stars (i.e.,
$\lesssim$2\%).  While it is difficult to determine whether our
faintest $z\geq7$ candidates are resolved (due to their low S/N
level), the fact that only 2 out of our brightest $\sim$50 $z\geq7$
candidates are consistent with being pointlike suggests that
contamination from low-mass stars is not particularly important.  Our
conclusions here are similar to those of Finkelstein et al.\ (2010:
see their Figure 5).  Simple arguments based on the surface density of
stars from our galaxy at high latitudes (e.g., Ryan et al.\ 2005) also
suggest that dwarf star contamination must be small.

\subsubsection{Transient Sources}

Since the optical ACS HUDF and GOODS data were taken some 5 years
earlier than the new infrared WFC3/IR data, our samples may suffer
from contamination as a result of transient sources like SNe that
appear in the new IR data but are absent in the old optical data.
That is, they mimic high-redshift dropouts!  Since essentially all
known transient sources likely to show up in our searches as dropouts
are unresolved, we would flag any such contaminants at the same time
as we identify possible T dwarfs in the new observations.  Only one
such source is found in our search fields, and that is an apparent SNe
over the ultra-deep HUDF09 pointing (Oesch et al.\ 2010a; McLure et
al.\ 2010; Bunker et al.\ 2010; Yan et al.\ 2010), and it was indeed
unresolved.  The only other such source found in our selection had
colors consistent with being a T dwarf (see \S3.4).  It therefore
seems unlikely that our samples are subject to significant
contamination from transient sources.

\subsubsection{Active Galactic Nuclei (AGN)} 

Another possible source of contamination for our selections could come
from $z\gtrsim7$ AGN (e.g., Meiksin 2006).  $z\gtrsim7$ AGN would show
similarly strong Lyman break in their optical/near-IR colors and also
very blue near-IR colors.  However, we would also expect AGN to be
unresolved, and as we have already remarked in discussing possible
contamination from low-mass stars and SNe, we only find one such
source in our selection that has a profile consistent with being
unresolved and which is not transient.  It is interesting that obvious
AGN (those that are comparable to, or brighter than, the underlying
galaxy) are not evident (i.e., $\lesssim$2\%) in these $z\sim7$-8
samples.

\subsubsection{Lower Redshift Sources}

Are there lower redshift sources that mimic the colors of $z\sim7$ and
$z\sim8$ galaxies?  They would need to be small but extended, faint
objects that show a distinct and deep spectral break with rather blue
colors at wavelengths longward of the break.  It is not at all obvious
what such objects would be, but one possible contaminating source
might be a relatively-compact, low surface brightness, very faint (and
hence low mass) $z$$\sim$1.5-2.5 galaxy with a Balmer or 4000$\AA$
break.  Yet, whatever they might be, they are unlikely to dominate the
numbers in high redshift samples given (i) the size of the spectral
breaks routinely seen in higher S/N samples and in stacked samples
(e.g., Labbe et al.\ 2010a,b; Gonzalez et al.\ 2010) and (ii) the lack
of large numbers of such sources in spectroscopic samples of
$z$$\sim$4-6 galaxies (e.g., Vanzella et al.\ 2009; Stark et
al.\ 2010).  Instead, they likely contribute to contamination at a low
level and most often as a result of photometric scatter.

\subsubsection{Lower Redshift Sources and Photometric Scatter}

Perhaps the largest source of contamination for our $z\sim7$ and
$z\sim8$ samples is from lower-redshift galaxies that satisfy our
selection criteria due to the effect of noise on their observed
colors.  Clearly, this is a much greater problem for sources within
the faintest 1-2 magnitudes of the sample where the effect of noise is
substantial, and so the contamination rate is expected to be the
highest in the magnitude interval just brightward of the selection
limit.

\textit{The procedure we use to estimate the importance of this effect
  relies upon actual sources from our observed datasets and so
  automatically takes into account the properties of ``real''
  high-redshift galaxies.}  We estimate the number of contaminants in
each of our galaxy samples by the following multi-step process.  For
each faint source in our search fields, we (i) randomly choose a
brighter source from our three ultra-deep HUDF09 fields, and (ii) then
make that bright source match the faint source by scaling the flux of
the brighter source and adding noise.  Having done this for all of the
faint sources in our real catalogs, we then have a "simulated" catalog
with the same number of galaxies, magnitude distribution, and noise
characteristics.  We can use this catalog to test the reliability of
our selections by applying the same $z\sim7$ $z_{850}$-dropout and
$z\sim8$ $Y_{105}$/$Y_{098}$-dropout selection criteria as we applied
on the real data.

The brighter sources are taken from only slightly brighter magnitude
ranges so they have higher S/N but are otherwise expected to be as
similar as practical to the sources being tested for contamination.
Thus, for our dropout samples over our three HUDF09 fields, our
estimated contamination rates are made assuming that the faint
photometric samples ($H_{160,AB} > 27.5$) have the same color
distributions as that found in the higher S/N HUDF09
$26.5<H_{160,AB}<28.0$ subsample.  For our ERS selections, our
contamination estimates are based upon the higher S/N HUDF09
$26.0<H_{160,AB}<27.0$ subsample.  Noise is then added to the
photometry, and our $z_{850}$-dropout or $Y_{105}/Y_{098}$-dropout
selection criteria are applied.  Any source that is selected by our
dropout criteria (after adding noise), but detected in the optical in
the original high S/N observations is counted towards the total
contamination level.  We repeat this simulation 200$\times$ for each
data set.

We find 5.0 $z_{850}$-dropouts and 3.3 $Y_{105}$-dropout contaminants
per simulation from our three HUDF09 fields and 2.9 $z_{850}$-dropouts
and 2.3 $Y_{098}$-dropout contaminants per simulation for our ERS
observations.  We present these contamination rates as a function of
magnitude for each of our search fields in Table~\ref{tab:cont} of
Appendix B.  These simulations argue for an overall contamination rate
of 8\% and 6\% for our $z_{850}$-dropout and $Y_{098}$-dropout
selections, respectively, over our HUDF09 observations and 22\% and
38\% for our $z_{850}$-dropout and $Y_{098}$-dropout selections,
respectively, over the ERS observations.

Our extensive testing has given us some useful insights.  For example,
it is striking how much lower the estimated contamination levels are
in our HUDF09 selections than in our HUDF09-1/HUDF09-2 selections.
The differences are greater than a factor of $\sim$3 typically and are
the direct result of the much greater depth of the optical data over
the HUDF09 (by $\sim$1 mag) than the HUDF09-1/HUDF09-2 selections.
The present tests also show the importance of the ACS parallel
observations obtained as part of the HUDF09 WFC3/IR over the HUDF09-2
field (the cyan ``P1'' region shown on Figure~\ref{fig:obsdata}).
Those regions with the deeper ACS data have an expected contamination
rate that is $2.5\times$ lower than those regions without these data
(see Table~\ref{tab:cont}).

\subsubsection{Spurious Sources}

Spurious sources are unlikely to be a concern for our samples, given
that our $z\geq7$ candidate galaxies show detections in at least two
bands.  One of these bands is always the $J_{125}$-band where a
$\gtrsim$3.5$\sigma$ detection is required.  The second band is either
the $Y_{105}$ or $H_{160}$ band and that detection must be $\gtrsim
3\sigma$.  Obtaining such significant detections in two independent
images is extremely improbable.  One simple technique for estimating
the probable contamination from such sources is to look for similar
$\gtrsim3.5\sigma$ $J_{125}$-band and $\gtrsim3\sigma$ $H_{160}$-band
detections on the negative images.  No such sources were found on the
``negative'' images.

\subsubsection{Summary}

Based upon the above discussion, tests and simulations, it seems clear
that the only meaningful source of contamination for the current
selections are low-redshift sources that enter through photometric
scatter.  Overall, the estimated contamination rate is 5.0
$z_{850}$-dropouts and 3.3 $Y_{105}$-dropouts in our HUDF09 selections
and 2.9 $z_{850}$-dropouts and 2.2 $Y_{098}$-dropouts in our ERS
selections. This works out to an estimated contamination rate of
$\sim$11\% for our $z\sim7$ $z_{850}$-dropout selection and $\sim$9\%
for our $z\sim8$ $Y_{105}$/$Y_{098}$-dropout selection.  We therefore
quote a contamination rate of $\sim$11\% overall.  As expected, most
of this contamination is for the fainter sources.  The estimated
number of contaminants is tabulated as a function of magnitude in
Appendix B in Table~\ref{tab:cont}.

\begin{deluxetable}{lr|rr|rr}
\tablewidth{0pt}
\tablecolumns{11}
\tabletypesize{\footnotesize}
\tablecaption{Summary of $z\sim7$ and $z\sim8$ samples.\tablenotemark{a}\label{tab:dropsamp}}
\tablehead{
\colhead{} & \colhead{Area} &
\multicolumn{2}{c}{$z\sim7$} & \multicolumn{2}{c}{$z\sim8$} \\
\colhead{Sample} & \colhead{(arcmin$^2$)} &
\colhead{\#} & \colhead{Limits\tablenotemark{a}} & \colhead{\#} & \colhead{Limits\tablenotemark{a}}}
\startdata
HUDF09 & 4.7 & 29 & $J\leq29.4$ & 24 & $H\leq29.4$ \\
HUDF09-1 & 4.7 & 17 & $J\leq29.1$ & 14 & $H\leq29.0$ \\
HUDF09-2 & 4.7 & 14 & $J\leq29.2$ & 15 & $H\leq29.0$ \\
ERS & 39.2 & 13 & $J\leq28.0$ & 6 & $H\leq27.5$ \\
\hline
Total & 53.3 & 73 & & 59 & 
\enddata
\tablenotetext{a}{The magnitude limit is the $\sim$5$\sigma$ detection
 limit for objects in a 0.35\arcs-diameter aperture.}
\end{deluxetable}

\subsection{Summary of our $z\sim7$ and $z\sim8$ HUDF09+ERS Samples}

In total, 73 $z\sim7$ $z_{850}$-dropout candidates are identified over
all of our search fields and 59 $z\sim8$ $Y_{105}$/$Y_{098}$-dropout
candidates are found.  A complete catalog of the dropouts in those
samples is provided in
Tables~\ref{tab:z09candlist}-\ref{tab:z98candlist}.  Postage stamps of
these dropout candidates are given in
Figures~\ref{fig:zstamp}-\ref{fig:ystamp}.

A convenient summary of the properties of our $z\sim7$ $z_{850}$ and
$z\sim8$ $Y_{105}$/$Y_{098}$ dropout samples and our search fields is
provided in Table~\ref{tab:dropsamp}.  The redshift distributions
derived for the samples are given in Figure~\ref{fig:zdist}.  The mean
redshift for our $z_{850}$-dropout selections is 6.8 and 6.7 for our
HUDF09 and ERS selections, respectively.  Meanwhile, the mean redshift
for our $Y_{105}/Y_{098}$-dropout selections is 8.0 and 7.8 for our
HUDF09 and ERS selections, respectively.

\section{Assessment of Current LBG Selections}

\subsection{Comparison with earlier LBG selections over our search fields}

Before using the present LBG selections to make inferences about
the rest-frame $UV$ LF at $z\gtrsim7$ and its evolution across cosmic
time, it is instructive to compare these samples with previous samples
over the same fields.

\subsubsection{Previous $z\gtrsim6.5$ Galaxy Selections within the HUDF09 observations over the HUDF}

Nine catalogs of $z\gtrsim6.5$ galaxies have already been published
using the first-year ultra-deep HUDF09 WFC3/IR observations over the
HUDF (Oesch et al.\ 2010a; Bouwens et al.\ 2010b; McLure et al.\ 2010;
Bunker et al.\ 2010; Yan et al.\ 2010; Finkelstein et al.\ 2010;
Wilkins et al.\ 2011; Lorenzoni et al.\ 2011; McLure et al.\ 2011).
The $z\gtrsim6.5$ catalogs from this paper include all 5
$Y_{105}$-dropout candidates in the Bouwens et al.\ (2010b) HUDF09
catalog and 14 of the 16 $z_{850}$-dropout candidates in the Oesch et
al.\ (2010a) HUDF09 catalog.  Two of the Oesch et al.\ (2010a)
$z_{850}$-dropout candidates are blended with nearby neighbors in the
present catalog and are therefore not included here.  In addition,
several of the $z_{850}$-dropouts in our previous catalog are now in
our $Y_{105}$-dropout catalogs, because we have modified our dropout
criteria to better match up with the ERS filter set (see the
discussion in \S3.2 and \S3.4).

In comparison to our previous selection of $z$/$Y$-dropout candidates
over the HUDF09 (Oesch et al.\ 2010a; Bouwens et al.\ 2010a; Bouwens
et al.\ 2011a), the present sample contains 33 additional $z\geq7$
candidates.  The expanded size of our sample is the result of our use
of the full two-year HUDF09 observations and our decision (1) to
extend our selection to lower S/N to include fainter sources in the
HUDF09 observations (as our improved tests for contamination now
allow, as discussed in \S3.5 and Appendix D) and (2) to carefully
match up our $z\sim7$ $z$-dropout and $z\sim8$ $Y$-dropout criteria so
that all credible $z\gtrsim7$ sources fall into one of the two
samples.  Compared to previous $z\gtrsim6$ catalogs, our dropout
sample includes 21 additional candidates (see
Tables~\ref{tab:z09candlist} and \ref{tab:y09candlist}).  Essentially
all of the candidates have $J_{125}$-band magnitudes of $\gtrsim$28.5
AB mag.  The lack of completely new candidates at brighter magnitudes
is not too surprising, given the large number of independent
$z\gtrsim6.5$ selections already performed on this HUDF09 data set.

\subsubsection{Previous $z\gtrsim6.5$ Galaxy Selections within the HUDF09-1/
HUDF09-2 Observations}

Several catalogs of $z\sim7$ $z_{850}$-dropout candidates are
available over the HUDF09-1 and HUDF09-2 fields (Bouwens \&
Illingworth 2006; Bouwens et al.\ 2008; Wilkins et al.\ 2011).  The
first such catalogs were generated based upon the ultra-deep NICMOS
parallels to the HUDF (NICP12 and NICP34) and included two $z\sim7$
$z_{850}$-dropout candidates NICPAR1-3303-4111 and NICPAR2-3308-5229.
Of the two candidates, only the first (NICPAR1-3303-4111) is covered
by the ultra-deep HUDF09 WFC3/IR observations.  The source (originally
reported at 03:33:03.81, $-$27:41:12.1) again makes it into our
$z_{850}$-dropout selection, as UDF091z-03791123.  Oesch et
al.\ (2009) also utilized the ultra-deep NICMOS data in the NICP12
field for a $z\sim7$ search, but did not report any candidates.

More recently, ultra deep WFC3/IR observations of the HUDF09-1 and
HUDF09-2 fields allow for the selection of larger samples of $z\sim7$
$z_{850}$-dropouts.  Wilkins et al.\ (2011) performed exactly such a
selection.  Using a $z_{850}-Y_{105}>1$ criteria, Wilkins et
al.\ (2011) reported 22 candidates over the two fields to
$H_{160,AB}\sim28.2$.  While 8 of the candidates from Wilkins et
al.\ (2011) make it into our selection, strikingly most of the Wilkins
et al.\ (2011) $z\sim7$ candidates do not make it into our sample.
While 4 of their candidates miss our selection as a result of blending
with nearby sources, the most common reason their candidates miss our
selection (9 cases) is as a result of their candidates showing
apparent detections in the optical $B_{435}V_{606}i_{775}$ data.  For
five of the Wilkins et al.\ (2011) candidates (P34.z.4288, P34.z.3053,
P34.z.3990, P34.z.5016,P34.z.2397), the optical detections are quite
prominent, i.e., $>$1.5$\sigma$ in $>$1 band.  McLure et al.\ (2011)
also flag the above sources (from Wilkins et al.\ 2011) as showing
quite significant detections in the optical data.  From the optical
integration times quoted by Wilkins et al.\ (2011) for the HUDF09-2
field, it is not clear they take full advantage of the available ACS
data.

\subsubsection{Previous $z\gtrsim6.5$ Galaxy Selections in the ERS Observations}

The WFC3/IR ERS observations provide another high-quality, though much
shallower dataset, so it is not surprising that catalogs of $z\sim7$
$z_{850}$-dropouts already exist (Wilkins et al.\ 2010, 2011).  Wilkins
et al.\ (2010, 2011) consider $z\sim7$ searches over the ERS observations
using a similar $z_{850}$-dropout criterion to what we use here (but
see \S3.4).  Wilkins et al.\ (2010, 2011) identified 13 $z_{850}$-dropout
candidates over that area.  Strikingly enough, only two of their
thirteen candidates are present in our $z_{850}$-dropout catalogs
(see Table~\ref{tab:zcandlist}).

Most of their sample (11 sources) do not make it into our
$z_{850}$-dropout selection.  What is the reason for this?  Three of
their candidates (zD3, ERS.z.26813, zD1) miss our color selection due
to their having $z_{850}-Y_{098}$ colors that are too blue (e.g.,
ERS.z.26813 is given in Table~\ref{tab:possiblecand}), but seem likely
to correspond to $z\geq6$ galaxies.  One other candidate (zD2) was
blended with a nearby source and therefore not selected.  Another
(ERS.z.87326) was too faint in the $H_{160}$ band to be selected.  The
remaining six $z\sim7$ candidates from Wilkins et al.\ (2010, 2011)
were not included in our selection since they seem much more likely
correspond to low-redshift galaxies or stars.  Five
(zD5,ERS.z.45846,ERS.z.20851,ERS.z.47667,ERS.z.80252) show significant
detections in the optical ($>1.5\sigma$ in $\sim$2-3 bands) and the
sixth (ERS.z.70546: 03:32:27.91, $-$27:41:04.2) is almost certainly a
T-dwarf, given its small size ($\sim$0.1$''$ half-light radius in the
WFC3/IR data), SExtractor stellarity parameter ($\sim$0.95: strongly
suggesting the source is unresolved), and extremely blue
$J_{125}-H_{160}$ color ($\sim-0.6$).

\subsection{Integrity of Our Lyman-Break Galaxy Selections}

One of the most significant challenges in constructing large samples
of $z\geq7$ galaxies to very faint magnitudes is the limited depth of
the optical data.  Without deep optical data, it is very difficult
to discriminate between bona-fide high-redshift galaxies and
$z$$\sim$1-2 galaxies that are simply faint in the optical.  How
confident are we that we have identified all the contaminants in our
selections (or conversely that we have not discarded a substantial
number of bona-fide $z\geq7$ galaxies)?

There are two arguments that suggest the situation is largely under
control.  First, carefully modeling the selection efficiency for each
of our dropout samples and fixing the shape of the LF, we can estimate
$\phi^*$ for the $z\sim7$ and $z\sim8$ LFs on a field-by-field basis
(\S6.2).  If we were misidentifying too many low-redshift galaxies as
high-redshift galaxies -- or vice-versa -- we would expect to observe
huge fluctuations in the value of $\phi^*$ in precisely those fields
with the shallowest optical data.  However, we find essentially the
same value of $\phi^*$ (within 20\%) for each of the three ultra-deep
fields we consider (HUDF09, HUDF09-1, HUDF09-2).  This suggests that
we have a reasonable handle on contamination in our samples.

Second, we can look at the extent to which our $z_{850}$-dropout and
$Y_{105}/Y_{098}$-dropout samples show detections in the optical data
(quantified here using a $\chi_{opt} ^2$ statistic).  If our selections
were largely composed of bona-fide $z$$\sim$7-8 galaxies, then we
would expect there to be no more sources with positive detections in
the optical data than there are negative detections.  Indeed, our
samples show a similar number of sources with negative and positive
detections.  See Appendix D.4 for more details.

It is worthwhile noting that much of this discussion will likely be
highly relevant to future $z\geq7$ selections with the James Webb
Space Telescope -- where the lack of visible imaging of comparable
depths may make the selection of $z\geq7$ galaxies challenging.

\section{Constraints on the rest-frame $UV$ LFs at $z$$\sim$7 and $z$$\sim$8}

The large samples of $z$$\sim$7 and $z$$\sim$8 galaxies we identified
over the wide-area ERS and ultra-deep HUDF09 fields permit us to place
powerful constraints on the shape of the $UV$ LF at $z$$\sim$7 and
$z$$\sim$8.  The available constraints are now much stronger than was
possible using just the early WFC3/IR data over the HUDF (e.g., Oesch
et al.\ 2010a; McLure et al.\ 2010).

We begin this section by establishing the $UV$ LF at $z\sim7$
considering both a stepwise (\S5.1) and Schechter (\S5.2)
representation.  In \S5.3, we move onto a determination of the LF at
$z\sim8$.

\begin{deluxetable}{lcc}
\tablewidth{0pt}
\tabletypesize{\footnotesize}
\tablecaption{Stepwise Constraints on the rest-frame $UV$ LF at $z\sim7$ and $z\sim8$ (\S5.1 and \S5.3).\tablenotemark{a}\label{tab:swlf}}
\tablehead{
\colhead{$M_{UV,AB}$\tablenotemark{c}} & \colhead{$\phi_k$ (Mpc$^{-3}$ mag$^{-1}$)}}
\startdata
\multicolumn{2}{c}{$z$-dropouts ($z\sim7$)}\\
$-$21.36 & $<$0.00002\tablenotemark{b}\\
$-$20.80 & 0.00016$\pm$0.00009\\
$-$20.30 & 0.00015$\pm$0.00009\\
$-$19.80 & 0.00050$\pm$0.00021\\
$-$19.30 & 0.00104$\pm$0.00035\\
$-$18.80 & 0.00234$\pm$0.00068\\
$-$18.30 & 0.00340$\pm$0.00096\\
$-$17.80 & 0.00676$\pm$0.00207\\
\multicolumn{2}{c}{$Y$-dropouts ($z\sim8$)}\\
$-$21.55 & $<$0.00001\tablenotemark{b}\\
$-$20.74 & 0.00011$\pm$0.00007\\
$-$20.14 & 0.00025$\pm$0.00012\\
$-$19.54 & 0.00039$\pm$0.00017\\
$-$18.94 & 0.00103$\pm$0.00035\\
$-$18.34 & 0.00156$\pm$0.00072\\
$-$17.74 & 0.00452$\pm$0.00207\\
\enddata 
\tablenotetext{a}{These stepwise LFs are also shown in Figures~\ref{fig:shapelf7} and \ref{fig:shapelf8}.}
\tablenotetext{b}{Upper limits here are $1\sigma$ (68\% confidence).}
\tablenotetext{c}{The effective rest-frame wavelength is $\sim1600\AA$
  for our $z\sim7$ selection, $\sim1760\AA$ for our $z\sim8$ selection.}
\end{deluxetable}

\subsection{$z\sim7$ LF: Stepwise Determinations} 

We first consider a stepwise determination of the $UV$ LF at $z\sim7$.
Stepwise determinations of the $UV$ LF are valuable since they allow
for a relatively model-independent determination of the shape of the
LF.  This ensures that our determinations are not biased to adhere to
specific functional forms such as a Schechter function or power
law.

Here the stepwise luminosity function $\phi_k$ is derived using a very
similar procedure to the Efstathiou et al.\ (1988) stepwise maximum
likelihood (SWML) method.  With this method, the goal is to find the
shape of the LF which is most likely given the observed distribution
of magnitudes in our search fields.  Since only the shape of the
distribution is considered in this approach (and there is no
dependence on the volume density of sources), we would expect the LF
to be largely insensitive to large-scale structure effects and hence
more robust.  Previously, we employed this procedure to derive the LFs
at $z$$\sim$4-6 (Bouwens et al.\ 2007).

In detail, one can write this likelihood as
\begin{equation}
{\cal L}=\Pi_{field} \Pi_i p(m_i)
\label{eq:ml}
\end{equation}
where
\begin{equation}
p(m_i) = \left (\frac{n_{expected,i}}{\Sigma_j n_{expected,j}} \right )^{n_{observed,i}}
\end{equation}
where $n_{observed,i}$ is the observed number of sources in the
magnitude interval $i$, $n_{expected,j}$ is the expected number of
sources in the magnitude interval $j$, and where $\Pi$ is the product
symbol.  The above expression gives the likelihood that a given
magnitude distribution of sources is observed given a model LF.  The
expected number of sources $n_{expected,i}$ is computed from a model
LF as
\begin{equation}
\Sigma _{k} \phi_k V_{j,k} = n_{expected,j}
\label{eq:numcountg}
\end{equation}
where $n_{expected,j}$ is the surface density of galaxies in some
search with magnitude $j$, $\phi_k$ is the volume density of galaxies
with absolute magnitude $k$, and where $V_{i,j,k}$ is the effective
volume over which galaxies in the absolute magnitude interval $k$ are
selected with a measured magnitude in the interval $j$.  With the
$V_{i,j,k}$ factors, we implicitly account for the effects of
photometric scatter (noise) in the measured magnitudes on the derived
LFs -- allowing us to account for effects like the Malmquist bias on
our LF determinations.  The Malmquist bias is known to boost the
number of sources near the selection limit, simply as a result of the
lower significance sources ($\lesssim$5$\sigma$) scattering into the
selection.  See Appendix B of Bouwens et al. 2007 (and Bouwens et
al. 2008) for a more detailed description of the above maximum
likelihood procedure.  We elected to use magnitude intervals of width
0.5 mag for our LF determination as a compromise between S/N (i.e.,
the number of sources in each interval) and resolution in luminosity.

\begin{deluxetable}{cccccc}
\tablecolumns{6}
\tablecaption{Other Wide-Area Searches for $z\gtrsim7$ Galaxies Considered Here.\label{tab:osearch}}
\tablehead{
\colhead{} & \colhead{Area} & \colhead{Redshift} & \colhead{} & \colhead{Volume} & \colhead{} \\
\colhead{Instrument} & \colhead{(arcmin$^2$)} & \colhead{Range} & \colhead{$<z>$} & \colhead{(Mpc$^3$)} & \colhead{Ref\tablenotemark{a}}}
\startdata
NICMOS$~$+ & $\sim$80 & 6.3-8.5 & 7.1 & 1$\times$10$^{5}$ & \\
MOIRCS/ISAAC & 248 & 6.5-7.5 & 7.0 & 4$\times$10$^{5}$ & [1] \\
Suprime-Cam & 1568 & 6.5-7.1 & 6.8 & 7$\times$10$^{5}$ & [2] \\
HAWK-I & $\sim$151 & 6.5-7.5 & 6.8 & 1$\times$10$^{5}$ & [3] \\
\enddata
\tablenotetext{a}{References: [1] Bouwens et al. 2010c, [2] Ouchi et al.\ 2009, and [3] Castellano et al. 2010a,b}
\end{deluxetable}

We estimate the selection volumes $V_{j,k}$ for our LBG selections
using the same set of simulations described in Appendix C of this
paper.  This procedure is the same as used in our many previous LF
studies (e.g., Bouwens et al.\ 2007, 2008, 2010b) and involves taking
the pixel-by-pixel profiles of actual HUDF $z\sim4$ $B$-dropouts,
inserting them at random positions in the actual observations, and
then attempting to select them using the same procedures as we use on
the real data (including the $\chi_{opt}^2$ non-detection criterion we
describe in \S3.3 and Appendix D).  The $UV$-continuum slopes $\beta$
adopted for the $z\sim7$-8 model galaxies are chosen to match the
observed color distribution (Bouwens et al.\ 2010a).  The selection
volumes $V_{i,j,k}$ we derived from these simulations have a very
similar form to those shown in Figure 8 of Bouwens et al.\ (2006) or
Figure A2 of Bouwens et al.\ (2007).  We would expect these volumes to
be reasonably accurate -- since we reproduce both the observed sizes
of $z\sim7$-8 galaxies (Appendix A) and the $UV$-continuum slopes.

Due to its relative insensitivity to large scale structure effects,
the stepwise maximum likelihood procedure described above only
provides us with constraints on the shape of the LF.  To provide a
normalization for the LF, we require that the total number of
$z\sim7$-8 galaxies we predict from Eq.~\ref{eq:numcountg} in our
search fields matches those we actually find (after correcting for
contamination: see Table~\ref{tab:cont} and Appendix B for details on
how the precise corrections depend on search field and the luminosity
of the sources).

\begin{figure}
\epsscale{1.15}
\plotone{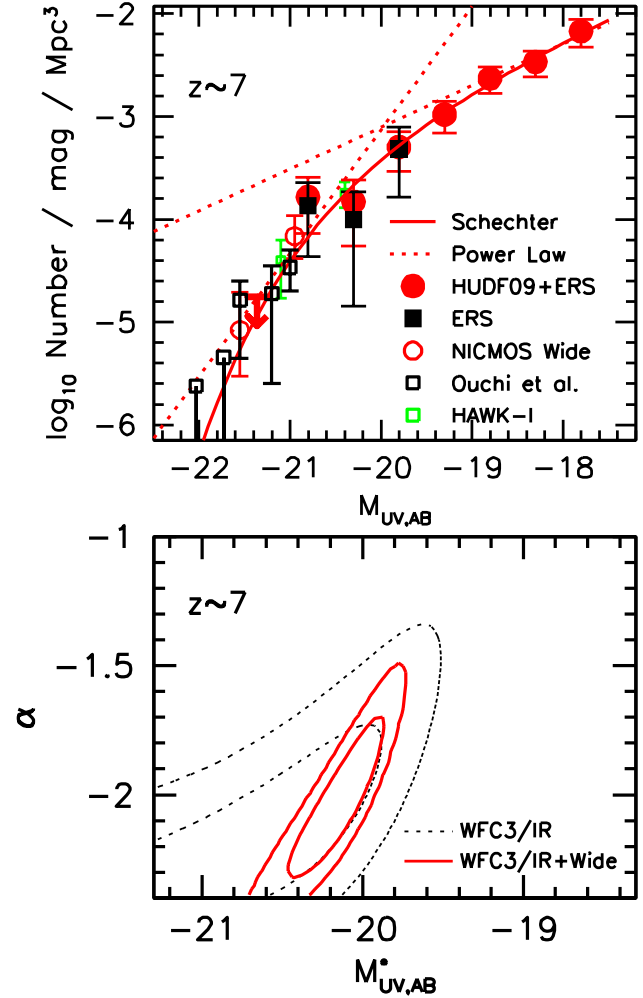}
\caption{(\textit{upper}) Rest-frame $UV$ LF determination at $z\sim7$
  determined from the present WFC3/IR $z\sim7$ samples (\S5.1).  The
  red circles, red downward arrow, and red line show our stepwise LF
  determination, the $1\sigma$ upper limit we have on the bright end
  of the LF from our WFC3/IR $z\sim7$ searches, and best-fit Schechter
  LF, respectively.  Also included in the upper panel are the LFs
  derived from a wide-area Subaru Suprime-Cam search (Ouchi et
  al.\ 2009), a wide-area NICMOS+ISAAC+MOIRCS search (Bouwens et
  al.\ 2010c: see also Mannucci et al.\ 2007; Stanway et al.\ 2008;
  Henry et al.\ 2009), and a wide-area HAWK-I search (Castellano et
  al.\ 2010b).  The black squares show the $z\sim7$ LF constraints
  from the wide-area ERS observations.  These match up well with the
  LF constraints obtained from the deeper HUDF09 observations.  The
  dotted red lines show the approximate asymptotic behavior (linear
  fit) of the $z\sim7$ LF at the bright and faint ends.  Together they
  illustrate the existence of a likely break in the LF at $-20$ AB
  mag.  (\textit{lower}) 68\% and 95\% confidence intervals on the
  characteristic luminosity $M^*$ and faint-end slope $\alpha$ from
  our WFC3/IR samples (\textit{dotted black lines}: \S5.2).  Stronger
  constraints on $M^*$ and $\alpha$ can be obtained by incorporating
  wide-area $z\sim7$ search results (\textit{solid red
    lines}).\label{fig:shapelf7}}
\end{figure}

By applying the above maximum likelihood procedure to our ERS, HUDF09,
HUDF09-1, and HUDF09-2 $z\sim7$ samples, we derive the stepwise LF at
$z\sim7$.  The resulting LF is presented in Table~\ref{tab:swlf} and
Figure~\ref{fig:shapelf7} (red circles).  The $z\sim7$ LFs inferred
from our $z\sim7$ ERS samples alone are shown separately
(\textit{black squares}) to demonstrate the consistency of our derived
LFs in both the wide-area and ultra-deep data.  Also included in
Figure~\ref{fig:shapelf7} are the LF determinations at $z\sim7$ from
two wide-area searches (Ouchi et al.\ 2009; Bouwens et al.\ 2010c: see
Table~\ref{tab:osearch}).  These searches are important in
establishing the shape of the LF at very high luminosities ($\lesssim
-$21 AB mag).

\begin{deluxetable}{ccccc}
\tablewidth{0pt} \tabletypesize{\footnotesize}
\tablecaption{Determinations of the best-fit Schechter Parameters for
  the rest-frame $UV$ LFs at $z\sim7$ and $z\sim8$ and at $z$$\sim$4,
  5, 6 from Bouwens et al.\ (2007).\label{tab:lfparm}} \tablehead{
  \colhead{Dropout} & & & \colhead{$\phi^*$ ($10^{-3}$}
  \\ \colhead{Sample} & \colhead{Redshift} & \colhead{$M_{UV}
    ^{*}$\tablenotemark{a}} & \colhead{ Mpc$^{-3}$)} &
  \colhead{$\alpha$}} \startdata 

$z$ & 6.8 & $-20.14\pm0.26$ & $0.86_{-0.39}^{+0.70}$ & $-2.01\pm0.21$\\ 
$Y$\tablenotemark{b} & 8.0 & $-20.10\pm0.52$ & $0.59_{-0.37}^{+1.01}$ & $-1.91\pm0.32$\\ 
\multicolumn{5}{c}{-----------------------------------------------------------------} \\ 
$B$ & 3.8 & $-20.98\pm0.10$ & $1.3\pm0.2$ & $-1.73\pm0.05$\\ 
$V$ & 5.0 & $-20.64\pm0.13$ & $1.0\pm0.3$ & $-1.66\pm0.09$\\ 
$i$ & 5.9 & $-20.24\pm0.19$ & $1.4_{-0.4}^{+0.6}$ & $-1.74\pm0.16$ 
\enddata
\tablenotetext{a}{Values of $M_{UV}^{*}$ are at $1600\,\AA$ for the
  Bouwens et al.\ (2007) $z\sim4$, $z\sim5$, and $z\sim7$
  LFs, at $\sim1350\,\AA$ for the Bouwens et al.\ (2007)
  $z\sim6$ LF, and at $\sim1750\,\AA$ for our constraints
  on the $z\sim8$ LF.  Since $z$$\sim$6-8 galaxies are
  blue ($\beta\sim-2$: Stanway et al.\ 2005; Bouwens et al.\ 2006,
  2009, 2010a), we expect the value of $M_{UV}^*$ to be very similar
  ($\lesssim0.1$ mag) at $1600\,\AA$ to its value quoted here.}
\tablenotetext{b}{The derived Schechter parameters depend
  significantly on whether we include or exclude the two brightest
  ($\sim$26 AB mag) $z\sim8$ galaxies in the
  HUDF09-2 field.  If we exclude the two brightest galaxies (assuming
  they represent a rare overdensity), we derive a significantly
  fainter characteristic luminosity $M^*$ and larger normalization
  $\phi^*$: $M^*=-19.54\pm0.56$, $\phi^*=1.5_{-1.0}^{+2.9}\times
  10^{-3}$ Mpc$^{-3}$, $\alpha=-1.67\pm0.40$.}
\end{deluxetable}

\begin{figure}
\epsscale{1.15}
\plotone{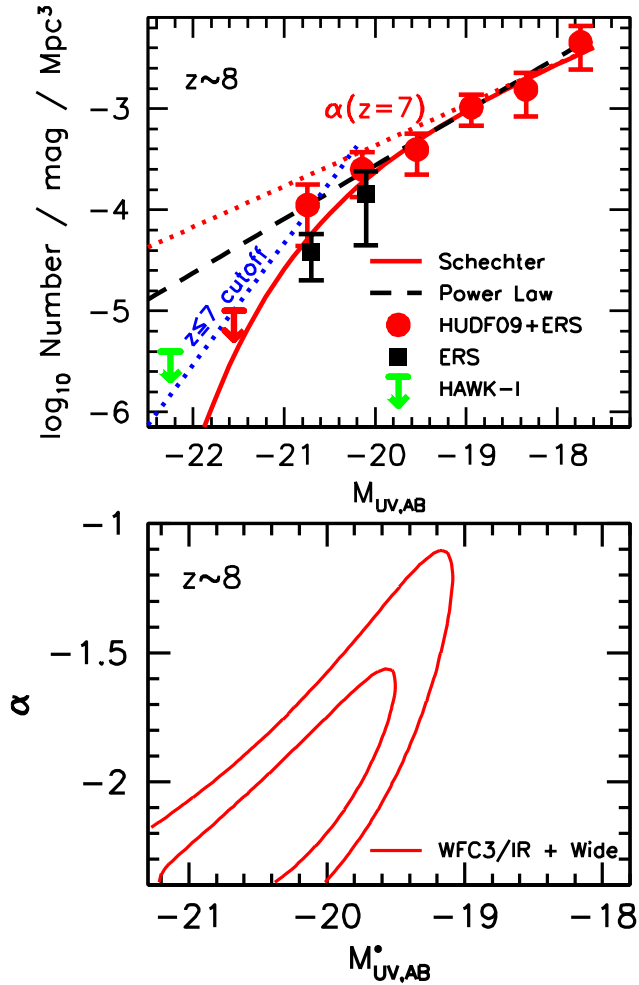}
\caption{(\textit{upper}) Rest-frame $UV$ LF at $z\sim8$ (see \S5.3).
  The red circles show the stepwise LF for the WFC3/IR HUDF09+ERS
  $z\sim8$ sample.  The red downward arrow shows the $1\sigma$ upper
  limit on the bright end of the LF from our WFC3/IR $z\sim8$ search.
  The black points show the stepwise LF at $z\sim8$ derived from the
  $z\sim8$ ERS search.  The dashed black line shows the best-fit
  power-law representation while the red line is for the Schechter
  fit.  The faint-end slope $\alpha$ for the $z\sim7$ LF
  ($\alpha=-2.01$) is indicated on this figure with the dotted red
  line.  The dotted blue line shows the approximate cut-off observed
  at the bright-end of $z\leq7$ LFs.  The green upper limit (at
  $-22.25$ mag) is based upon a recent wide-area $z\sim8$ Y-dropout
  HAWK-I search (Castellano et al.\ 2010b).  See also the caption to
  Figure~\ref{fig:shapelf7}.  Note that the brightest points in our
  $z\sim8$ LF are significantly affected by the three bright sources
  in the HUDF09-2 field (compare the brightest red point with the
  black square: based upon the ERS search).  (\textit{lower}) 68\% and
  95\% maximum likelihood contours on the characteristic luminosity
  $M^*$ and the faint-end slope $\alpha$ inferred for the LF at
  $z\sim8$.\label{fig:shapelf8}}
\end{figure}

\subsection{$z\sim7$ LF: Representation with a Schechter Parameterization}

In the present section, we consider representations of the $z\sim7$ LF
with a Schechter parameterization ($\phi^* (\ln(10)/2.5)
10^{-0.4(M-M^{*})\alpha} e^{-10^{-0.4(M-M^{*})}}$).  This
parameterization features an exponential cut-off at high luminosities
and a power-law shape at fainter luminosities.  Such a parametrization
has been almost universal in characterizing the shape of the galaxy LF
since it works so well.  From the results of the previous section
(e.g., Figure~\ref{fig:shapelf7}), it also appears to be relevant in
describing the LF of $z\sim7$ galaxies.

As with our stepwise LF determinations, we use a maximum likelihood
approach that considers only the shape of the LF in deriving the
best-fit Schechter parameters.  The approach is analogous to that
developed by Sandage et al.\ (1979), but is formulated in terms of the
apparent magnitude distribution (see \S3.1 of Bouwens et al.\ 2007).
The approach has the important advantage that it is almost entirely
insensitive to large-scale structure effects (see Appendix C of
Bouwens et al.\ 2007).

To perform this likelihood analysis, we start with various Schechter
parameter combinations, calculate the equivalent stepwise LF
$\phi_k$'s (adopting 0.1 mag bins), and then make use of
Eq.~\ref{eq:ml} and \ref{eq:numcountg}.  This technique allows us to
set constraints on the characteristic luminosity $M^*$ and faint-end
slope $\alpha$.  The lower panel of Figure~\ref{fig:shapelf7}
(\textit{dotted black lines}) shows the 68\% and 95\% confidence
intervals we are able to obtain on the $z\sim7$ LF.

Inspecting this lower panel, we see that there is a large degree of
freedom in the Schechter parameters $M^*$ and $\alpha$ allowed by our
search results.  The principal reason for this latitude (which occurs
despite the large number of sources and large luminosity range) lies
in the almost featureless power-law shape exhibited by the LF at
$z\sim7$ (top panel: Figure~\ref{fig:shapelf7}).  This restricts us to
characteristic luminosities $M^*$ brighter than $\sim-19.7$ and
approximate power-law slopes $\alpha$ of $-2.0$.

To obtain much tighter constraints on the Schechter parameters
$\alpha$ and $M^*$, we need to incorporate observations which allow us
to resolve more distinct features in the LF (e.g., the expected break
at brighter magnitudes that is not apparent in our small area
HUDF09+ERS data sets).  Such constraints can be obtained by combining
our WFC3/IR results with wide-area searches (e.g., Ouchi et al.\ 2009;
Castellano et al.\ 2010a,b; Bouwens et al.\ 2010c) which more
effectively probe the rarer sources found at the bright end of the LF.
Such searches have consistently found a distinct fall-off in the
volume density of $z\sim7$ sources at bright magnitudes (e.g., upper
panel in Figure~\ref{fig:shapelf7}).  Several of the most notable
wide-area $z\sim7$ searches (Table~\ref{tab:osearch}) include a
$\sim$85 + 248 arcmin$^2$ search with NICMOS and ISAAC/MOIRCS by
Bouwens et al.\ (2010c: see also Mannucci et al.\ 2007; Stanway et
al.\ 2008; Henry et al.\ 2009), a $\sim$150 arcmin$^2$ search with VLT
HAWK-I by Castellano et al.\ (2010b: but see also Castellano et
al.\ 2010a and Hickey et al.\ 2010), and a 1568 arcmin$^2$ search with
Subaru Suprime-Cam by Ouchi et al.\ (2009).  The mean redshift
expected for these samples is typically $\sim$6.8-7.0, which is
comparable to that for the present WFC3/IR $z_{850}$-dropout
selections.

Combining the wide-area $z\sim7$ search results with those available
from the ultra-deep WFC3/IR observations, we arrive at much tighter
constraints on $M^*$ and $\alpha$.  The best results are
$M^*=-20.14\pm0.26$ and $\alpha=-2.01\pm 0.21$ and are presented in
the lower panel to Figure~\ref{fig:shapelf7} with the solid red lines.
In combining the wide-area search results with the WFC3/IR results
(where the overlap in identified $z\sim7$ candidates is minimal), we
find it convenient to modify our LF estimation methodology so that the
likelihood incorporates information on the volume density of the LF
(and not simply its shape) -- comparing the volume density of the
sources found with what is expected from a model LF.  As in our
Bouwens et al.\ (2008) LF analysis, likelihoods are computed assuming
Poissonian statistics.  We verified that the 68\% and 95\% likelihood
contours we compute for the WFC3/IR search results reproduced those
calculated using the STY method.

It is trivial for us to extend the present analysis to constrain the
normalization $\phi^*$ of the Schechter function.  Utilizing the
wide-area and ultra-deep WFC3 search results, we find a $\phi^*$ of
$0.86_{-0.39}^{+0.70}\times10^{-3}$ Mpc$^{-3}$.  Together with the
best-fit values of $\alpha$ and $M^*$, we conveniently include these
parameters in Table~\ref{tab:lfparm}.  The effects of large-scale
structure add somewhat to the uncertainties on $\phi^*$.  Given that
most of the sources which contribute to the faint counts in the LF are
from the CDF South, we estimate that the added uncertainty is 25\% on
$\phi^*$ -- which corresponds to the sample variance expected assuming
a $\sim$100 arcmin$^2$ search area (the approximate area within the
CDF South where our fields are found), a redshift selection volume of
width $\Delta z\sim0.8$ (from Figure~\ref{fig:zdist}), and a bias
factor of $\sim$5 (corresponding to halo volume densities of
$\sim$10$^{-3}$ Mpc$^{-3}$).  These estimates were made using the
Trenti \& Stiavelli (2008) cosmic variance calculator (see also
\S6.2).

Of course, the volume density $\phi^*$ in the Schechter
parameterization shows a significant covariance with both the
characteristic luminosity and faint-end slope $\alpha$, and so there
is considerable freedom in the Schechter parameters that can fit
current search results.  We present the likelihood contours on
$\phi^*$ and the other two Schechter parameters in \S7.1.

\subsection{$z\sim8$ LF}

We now derive the LF at $z\sim8$ based upon the $z\sim8$ selections
presented in \S3.2 and \S3.4.  As expected, we will find that the
$z\sim8$ LF is less well-determined than that at $z\sim7$, but it is
still striking to find very interesting constraints from the combined
HUDF09+ERS datasets.

\begin{figure}
\epsscale{1.15}
\plotone{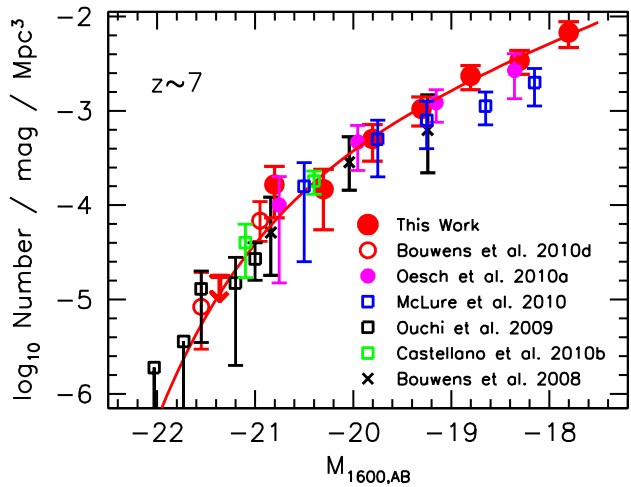}
\caption{Rest-frame $UV$ LF determinations for galaxies at $z\sim7$
  (\S6.1.1).  Shown are the UV LF from this paper (\textit{red
    circles} and \textit{red downward arrow}) and from Bouwens et
  al.\ (2010c: \textit{open red circles}), Oesch et al.\ (2010a:
  \textit{purple circles}), Ouchi et al.\ (2009: \textit{black
    squares}), Castellano et al.\ (2010b: \textit{green squares}),
  McLure et al.\ (2010: \textit{blue squares}) and Bouwens et
  al.\ (2008: \textit{black crosses}).  Constraints from Wilkins et
  al.\ (2011) are similar, but are not shown to reduce confusion.
  See Table~\ref{tab:complf7} for a comparison of the Schechter
  parameters derived here for the $z\sim7$ $UV$ LF with that found in
  other analyses.\label{fig:lf7}}
\end{figure}

\begin{deluxetable*}{lcccc}
\tablewidth{0pt} \tabletypesize{\footnotesize}
\tablecaption{Determinations of the best-fit Schechter Parameters for the
rest-frame $UV$ LFs at $z\sim7$ (\S6.1.1: see also Figure~\ref{fig:lf7}).\label{tab:complf7}}
\tablehead{\colhead{Reference} & \colhead{$M_{UV} ^{*}$} & \colhead{$\phi^*$
($10^{-3}$ Mpc$^{-3}$)} & \colhead{$\alpha$}}
\startdata
This work & $-20.14\pm0.26$ & $0.86_{-0.39}^{+0.70}$ & $-2.01\pm0.21$\\
Castellano et al.\ (2010a) & $-20.24\pm0.45$ & $0.35_{-0.11}^{+0.16}$ & $-1.71$ (fixed) \\
McLure et al.\ (2010) & $-20.04$ & 0.7 & $-1.71$ (fixed) \\
Oesch et al.\ (2010a) & $-19.91\pm0.09$ & 1.4 (fixed) & $-1.77\pm0.20$ \\
Ouchi et al.\ (2009) & $-20.1\pm0.76$ & 0.69$_{-0.55}^{+2.62}$ & $-1.72\pm0.65$ \\
Oesch et al.\ (2009) & $-19.77\pm0.30$ & $1.4$ (fixed) & $-1.74$ (fixed) \\
Bouwens et al.\ (2008) & $-19.8\pm0.4$ & $1.1_{-0.7}^{+1.7}$ & $-1.74$ (fixed) \\
\enddata
\end{deluxetable*}

We use the same procedure for deriving the stepwise LF and Schechter
parameters at $z\sim8$ as we had earlier used at $z\sim7$.  We base
our LF determinations on our combined HUDF09 + ERS $z\sim8$
selections, but also fold in the null results from a shallow,
$\sim$150 arcmin$^2$ $Y$-dropout search over the CDF-South, BDF, and
NTTDF (Castellano et al.\ 2010b).  Finally, to account for the fact
that the mean redshift of our wide-area ERS $z\sim8$ selection is
$\Delta z \sim 0.2$ lower than for our HUDF09 $z\sim8$ selection, we
adjust the volume densities implied by the ERS search downward by a
factor of 1.15 (which we derive from a fit to the stepwise $UV$ LF
evolution from $z\sim6$ to $z\sim4$).  We use 0.6 mag intervals for
the stepwise LFs at $z\sim8$ because of the fewer sources in $z\sim8$
samples.

Our results are presented in Table~\ref{tab:swlf} and
Figure~\ref{fig:shapelf8} (\textit{red points and upper limits}: both
the HUDF09 and the ERS observations are incorporated into the red
circles).  Also included on this figure are our $z\sim8$ LF using only
the ERS observations.  Not surprisingly, given the general form of the
LFs at $z\sim4$-7, the $z\sim8$ LF maintains an approximate power-law
form over a $\sim$3 magnitude range (\textit{red points}), with a
steep faint-end slope.  Brightward of $-21$ AB mag, the LF appears to
cut off somewhat.  This cut-off is clear not only in the upper limits
we derive for the volume density of luminous sources (red and green
downward arrows from WFC3/IR and HAWK-I searches, respectively), but
also in the $z\sim8$ ERS results (black squares).

Nonetheless, sufficiently small area has been probed at present that
the bright end of the $z\sim8$ LF remains fairly uncertain, and indeed
there is significant field-to-field variation across our fields.  The
most notable case is over the HUDF09-2 field where 4 bright
$H_{160,AB}\leq27$ $z\sim8$ galaxy candidates are found, 2 lying
within $3''$ of each other (UDF092y-03781204, UDF092y-03751196) and a
third $H_{160}\sim27$ candidate within $30''$ (UDF092y-04640529: see
Table~\ref{tab:y052candlist}), each of which have similar
$Y_{105}-J_{125}$ colors and presumably redshifts.  As a result of
such variations, the bright end of the $z\sim8$ LF is understandably
still quite uncertain, with some differences between that derived from
the ERS observations alone (black squares) and from the combined
HUDF09 + ERS results (red circles).

Despite the still somewhat sizeable uncertainties, we can model the
$z\sim8$ LF using a Schechter function and derive constraints on the
relevant parameters.  Formal fits yield best-fit Schechter parameters
of $\alpha=-1.91\pm 0.32$, $M^*=-20.10\pm0.52$, and
$\phi^*= 0.59_{-0.37}^{+1.01}\times 10^{-3}$ Mpc$^{-3}$ for the
$z\sim8$ LF.\footnote{The derived Schechter parameters depend
  significantly on whether we include or exclude the two brightest
  ($\sim$26 AB mag) $z\sim8$ galaxies in the HUDF09-2 field.  If we
  exclude the two brightest galaxies (assuming they represent a rare
  overdensity), we derive a significantly fainter characteristic
  luminosity $M^*$ and larger normalization$\phi^*$:
  $M^*=-19.54\pm0.56$, $\phi^*=1.5_{-1.0}^{+2.9}\times 10^{-3}$
  Mpc$^{-3}$, $\alpha=-1.67\pm0.40$.} Though the significance is only
modest, Schechter fits to the $z\sim8$ results (\textit{red line} on
Figure~\ref{fig:shapelf8}) are preferred at 2$\sigma$ over fits to a
power law (\textit{dashed line}).  The implications of such steep
faint-end slopes are discussed below, but we caution against giving
too much weight to these slopes at this time, given the current large
uncertainties.  Table~\ref{tab:lfparm} gives a compilation of these
parameters, along with a comparison with Schechter determinations at
other redshifts.  The trends of the Schechter parameters with redshift
are also discussed more broadly below.

\section{Robustness of LF Results}

\begin{figure}
\epsscale{1.15}
\plotone{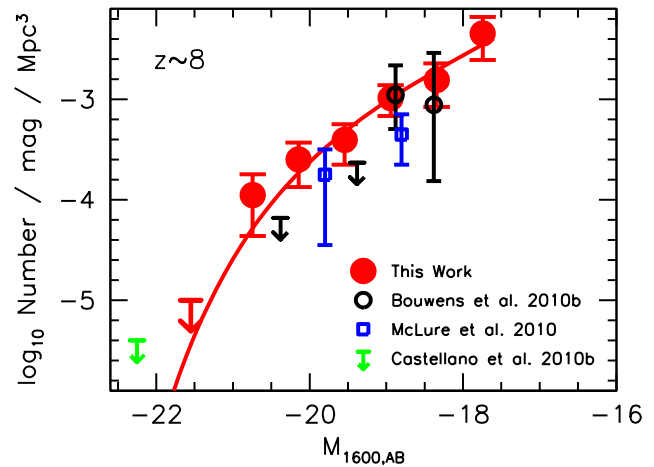}
\caption{Rest-frame $UV$ LF determinations for galaxies at $z\sim8$
  (\S6.1.2).  Shown are the UV LFs from this paper (\textit{solid red
    circles} and \textit{red upper limits}) and from Bouwens et
  al.\ (2010b: \textit{black open circles} and \textit{black upper
    limits}) and McLure et al.\ (2010: \textit{blue squares}).  LF
  estimates from Lorenzoni et al.\ (2011) are similar.  Also presented
  is the $1\sigma$ upper limit from Castellano et al.\ (2010b:
  \textit{green downward arrow}).  See Table~\ref{tab:complf8} for a
  comparison of the Schechter parameters derived here for the $z\sim7$
  $UV$ LF with that found in other analyses.\label{fig:lf8}}
\end{figure}

\begin{deluxetable*}{lcccc}
\tablewidth{0pt} \tabletypesize{\footnotesize}
\tablecaption{Determinations of the best-fit Schechter Parameters for the
rest-frame $UV$ LFs at $z\sim8$ (\S6.1.2: see also Figure~\ref{fig:lf8}).\label{tab:complf8}}
\tablehead{\colhead{Reference} & \colhead{$M_{UV} ^{*}$} & \colhead{$\phi^*$
($10^{-3}$ Mpc$^{-3}$)} & \colhead{$\alpha$}}
\startdata
This work\tablenotemark{a} & $-20.10\pm0.52$ & $0.59_{-0.37}^{+1.01}$ & $-1.91\pm 0.32$\\
McLure et al.\ (2010) & $-20.04$ & 0.35 & $-1.71$ (fixed) \\
Bouwens et al.\ (2010b) & $-19.5\pm0.3$ & $1.1$ (fixed) & $-1.74$ (fixed) \\
Lorenzoni et al.\ (2011) & $-$19.5 & 0.93 & $-1.7$ (fixed) \\
\enddata
\tablenotetext{a}{The derived Schechter parameters depend
  significantly on whether we include or exclude the two brightest
  ($\sim$26 AB mag) $z\sim8$ galaxies in the
  HUDF09-2 field.  If we exclude the two brightest galaxies (assuming
  they represent a rare overdensity), we derive a significantly
  fainter characteristic luminosity $M^*$ and larger normalization $\phi^*$: 
  $M^*=-19.54\pm0.56$, $\phi^*=1.5_{-1.0}^{+2.9}\times 10^{-3}$
  Mpc$^{-3}$, $\alpha=-1.67\pm0.40$ (similar to the values found by 
  Bouwens et al.\ 2010b and Lorenzoni et al.\ 2011).}
\end{deluxetable*}

\subsection{Comparison with previous LF determinations}

\subsubsection{$z\sim7$ LF}

A large number of different determinations of the $UV$ LFs at $z\sim7$
now exist (Bouwens et al.\ 2008; Oesch et al.\ 2009; Ouchi et
al.\ 2009; Oesch et al.\ 2010a; McLure et al.\ 2010; Castellano et
al.\ 2010a,b; Capak et al.\ 2011; Wilkins et al.\ 2011).  Stepwise
determinations of the LF from these studies are presented in
Figure~\ref{fig:lf7}.  The corresponding Schechter parameters are
given in Table~\ref{tab:complf7}.

It is instructive to compare the present LF determination at $z\sim7$
with those previously available.  We begin with those studies where
the primary focus was on the search for luminous, rarer $z\sim7$
galaxies (e.g., Ouchi et al.\ 2009; Castellano et al.\ 2010a,b;
Wilkins et al.\ 2010; Bouwens et al.\ 2010c).  Perhaps the simplest
point of comparison is the volume density of luminous ($\sim-21$ AB
mag) galaxies at $z\sim7$ relative to that at $z\sim6$.  Ouchi et
al.\ (2009), Bouwens et al.\ (2010c), and Wilkins et al.\ (2010) find
a factor of 2 decrease in the volume density of luminous ($-21$ AB
mag) galaxies over this interval while Castellano et al.\ (2010a,b)
find a factor of $\sim$3.5 decrease.\footnote{Note that while Wilkins
  et al.\ (2010) frame the comparison between the $z\sim6$ and
  $z\sim7$ LFs in terms of the $z\sim6$ Bouwens et al.\ (2006) LF,
  here we use the more up-to-date $z\sim6$ LF results of Bouwens et
  al.\ (2007).}  In the present search, we find a similar factor of
$\sim$1.8 decrease at the bright end of the LF (i.e., $-20.5$ AB mag).

At fainter magnitudes, our HUDF09 search fields are essentially unique
in the constraints they provide on the $UV$ LF at $z$$\sim$7.
Nonetheless, it is useful to compare against previous work,
particularly from the early ultra-deep WFC3/IR pointing.  Our results
at $-$19 AB mag are in good agreement with the $UV$ LF of Oesch et
al.\ (2010a), but are somewhat higher than those found by McLure et
al.\ (2010).

Most studies find best-fit values of $M^*$ and $\phi^*$ at $z\sim7$
that are slightly fainter and of lower density than is found at
$z$$\sim$4-6 (Bouwens et al.\ 2008; Oesch et al.\ 2009; Ouchi et
al.\ 2009; Oesch et al.\ 2010a; McLure et al.\ 2010; Castellano et
al.\ 2010a,b).  In the present study, we find something similar.  The
$M^*$ we find at $z\sim7$ is fainter than the $z\sim6$ value (Bouwens
et al.\ 2007) by $\sim 0.1$ mag and the $\phi^*$ we find at $z\sim7$
is lower by a factor of $\sim$1.8.  Of course, $\phi^*$ and $M^*$ are
sufficiently degenerate at $z\sim7$ that there is a modest trade-off
between the evolution found in either parameter.

\begin{figure}
\epsscale{1.18}
\plotone{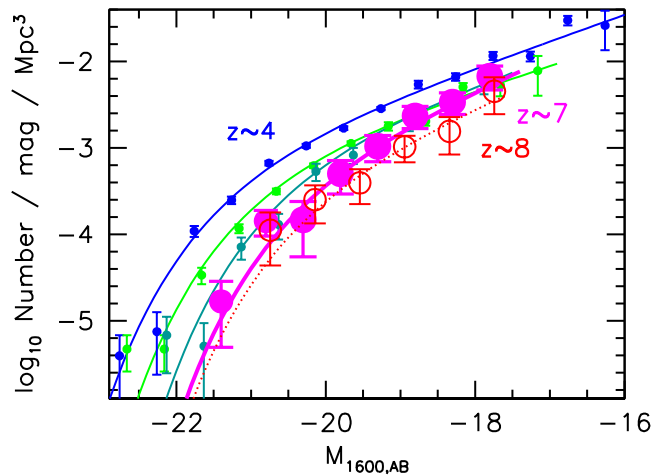}
\caption{Rest-frame $UV$ LFs derived for galaxies at $z\sim7$
  (\textit{magenta circles with $1\sigma$ errors}) at $z\sim8$
  (\textit{open red circles}), compared against similar LF
  determinations at $z\sim4$ (\textit{blue}), $z\sim5$
  (\textit{green}), and $z\sim6$ (\textit{cyan}) from Bouwens et
  al.\ (2007).  The $z\sim7$ LF results incorporate the Bouwens et
  al.\ (2010c) NICMOS + ISAAC + MOIRCS search results (see
  Figure~\ref{fig:lf7}).  The upper limits are $1\sigma$.  The magenta
  and red lines show the best-fit Schechter functions at $z\sim7$ and
  $z\sim8$.  The uniformly steep faint-end slopes $\alpha$ of the $UV$
  LF are quite apparent.  Most of the evolution in the $UV$ LF from
  $z\sim8$ to $z\sim4$ appears to be in the characteristic luminosity
  (by $\sim$1 mag).
\label{fig:lfall}}
\end{figure}

\begin{figure*}
\epsscale{1.15}
\plotone{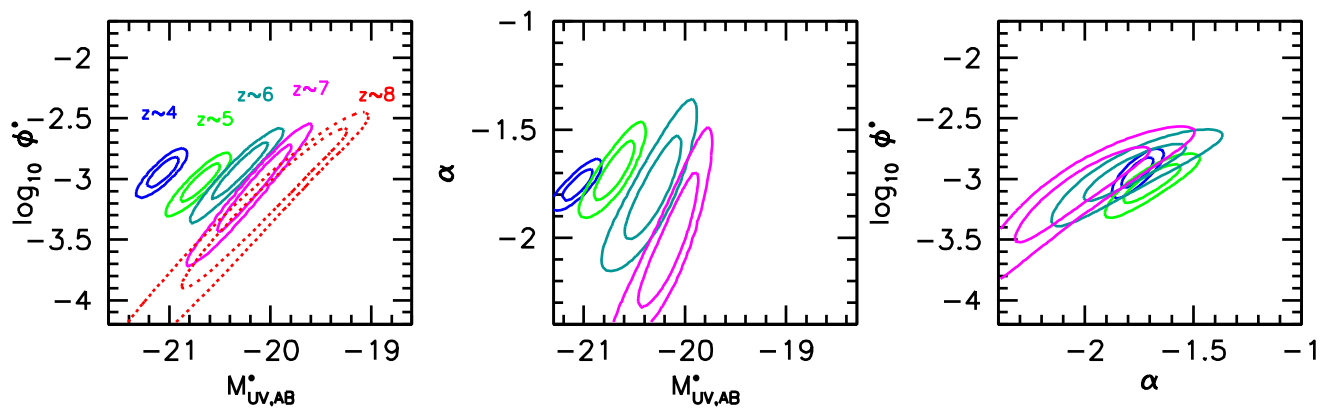}
\caption{68\% and 95\% likelihood contours on the model Schechter
  parameters for our $UV$ (rest-frame $\sim$1700\AA) LF determination
  at $z\sim7$ (\textit{magenta lines}) and $z\sim8$ (\textit{dotted
    red lines}).  For comparison, we also include the LF
  determinations at $z\sim4$ (\textit{blue lines}), $z\sim5$
  (\textit{green lines}), and $z\sim6$ (\textit{cyan lines}) from
  Bouwens et al.\ (2007).  We do not include the $z\sim8$ contours in
  the center and right panels given the large uncertainties on the
  $z\sim8$ Schechter parameters.  It is striking how uniform the rate
  of evolution in the $UV$ LF (left and middle panels) is as a
  function of redshift -- though there is a significant degree of
  degeneracy between $\phi^*$ and $M^*$.  Most of the evolution in the
  LF appears to be in $M^*$ (particularly from $z\sim7$ to $z\sim4$).
  Within the current uncertainties, there is no evidence for evolution
  in $\phi^*$ or $\alpha$ (\textit{rightmost panel}).\label{fig:schlf}}
\end{figure*}

The faint-end slope to our $UV$ LF is steep with
$\alpha=-2.01\pm0.21$.  Previously, it was not possible to set tight
constraints on the faint-end slope.  Ouchi et al.\ (2009) estimated a
slope $\alpha$ of $-1.72\pm0.65$ while Oesch et al.\ (2010a) -- fixing
$\phi^*$ to $0.0014$ Mpc$^{-3}$ -- found $-1.77\pm0.20$.  The present
values of the faint-end slope are consistent with those measured at
$z$$\sim$4-6 where this slope was also found to be steep at
$\alpha\sim-1.7$ (Bouwens et al.\ 2007; Oesch et al.\ 2007; Yoshida et
al.\ 2006; Yan \& Windhorst 2004).  This new result may suggest a
possible steepening towards higher redshifts $z\gtrsim7$, but the
significance is weak, i.e., $\sim$1.5$\sigma$. Such a steepening is
expected in many theoretical models (e.g., Salvaterra et al.\ 2010;
Trenti et al.\ 2010; Jaacks et al.\ 2011).  Clearly we will require
deeper data to confirm this trend.

Capak et al.\ (2011) have suggested that the $z\sim7$ LF may extend to
substantially brighter magnitudes than have been considered by other
groups.  Performing follow-up spectroscopy on three very bright
($H\sim23$ mag) $z\sim7$ candidates over the COSMOS field, Capak et
al.\ (2011) claim to have uncovered statistically significant
detections of line flux.  These lines could represent possible
Ly$\alpha$ emission given the existence of strong continuum breaks in
the sources.  If this is the explanation, these sources would imply
that the $UV$ LF at $z\sim7$ extends to very bright magnitudes indeed,
i.e., $-24$ AB mag, and in fact the LF that such sources would suggest
is essentially power-law in shape.

What should we make of this intriguing possibility?  First, such a LF
would seem to violate the constraints on the bright end of the LF set
by deep, wide-area searches for $z\sim7$ galaxies (Ouchi et al.\ 2009;
Bouwens et al.\ 2010c; Castellano et al.\ 2010a,b).  Secondly, it is
difficult to understand how one might create such bright $-24$ AB mag
sources in the UV.  The challenge is that any source forming stars
rapidly enough to be so bright in the $UV$ (i.e., $\sim$200
$M_{\odot}$/yr) would most likely have built up sufficient quantities
of dust and metals to be quite optically thick in the $UV$.  Under
such circumstances, it is unlikely that the source would be brighter
than $\sim$$-22$ AB mag.  Such is the situation at significantly later
times ($z$$\sim$2-3) where there are exceedingly few system with $UV$
magnitudes brighter than $\sim-23$ (e.g., Reddy et al.\ 2008), but
many ULIRGs (e.g., Chapman et al.\ 2005).  It is hard to imagine
therefore that at $z\sim7$, just 500-600 Myr after the first stars,
that there is a population of galaxies that are substantially brighter
in the $UV$ than at $z\sim3$.  Additional work utilizing deeper
spectroscopic and photometric data is clearly required.

\subsubsection{$z\sim8$ LF}

There are also several published determinations of the $UV$ LF at
$z\sim8$.  These LFs are presented in stepwise form in
Figure~\ref{fig:lf8}.  Schechter parameters for these LFs are given in
Table~\ref{tab:complf8}.  In general, the present $z\sim8$ LF
determination is in good agreement with previous determinations,
within the quoted errors -- though the present $z\sim8$ LF is a little
higher than the McLure et al.\ (2010) result at the lowest luminosities.

For the derived Schechter parameters (see Table~\ref{tab:complf8}), we
find an $M^*$ of $-20.10\pm0.52$, a $\phi^*$ of
$0.59_{-0.37}^{+1.01}\times10^{-3}$ Mpc$^{-3}$, and faint-end slope
$\alpha=-1.91\pm0.32$.  These $M^*$ and $\phi^*$ values are somewhat
brighter and lower, respectively, than what we find at $z$$\sim$6 and
$z$$\sim$7.  This might indicative that some of the evolution in the
UV LF from $z\sim8$ to $z\lesssim6$ can be represented by a change in
the volume density $\phi^*$.  However, the uncertainties on $\phi^*$
and $M^*$ are too large to be sure of these conclusions.  In
particular, the existence of a possible overdensity at $z\sim8$ within
the HUDF09-2 field (see \S5.3) illustrates the challenges of
establishing reliable parameters from these initial datasets.

\subsection{Large-Scale Structure Uncertainties}

One of the most important challenges for LF determinations is
large-scale structure.  Variations in the volume density of galaxies
as a function of position can have a significant impact on both the
shape and overall normalization of the derived LFs.  As a result of
our use of the stepwise maximum likelihood and STY procedures
(Efstathiou et al.\ 1988; Sandage et al.\ 1979) to derive the LF, the
impact of large scale structure on the shape of our LF should be
minimal.  On the other hand, we would expect this structure to have an
effect on the overall normalization of our LFs.  One can estimate the
size of these uncertainties using the Trenti \& Stiavelli (2008)
cosmic variance calculator.  Using the approximate volume densities
$\sim$5$\times$10$^{-3}$ Mpc$^{-4}$ we derived for $z$$\sim$7-8
sources, the width of our redshift selection windows $\Delta
z$$\sim$1, and the $10'$$\times$$10'$ search geometry we use within
the CDF South, we find that the uncertainty in the normalization of
the LF at $z\sim7$ and $z\sim8$ is a very modest $\sim$25\% and
$\sim$20\%, respectively.

Large-scale structure variations can also have an effect on the
faint-end slope $\alpha$ determinations as a result of mass-dependent
(and hence likely luminosity-dependent) bias parameters (e.g.,
Robertson 2010).  We estimate the luminosity-dependent bias parameters
based upon the approximate halo masses that correspond to the volume
density of sources we are probing at specific points in the LF (again
using the Trenti \& Stiavelli 2008 calculator).  We then perturb the
LFs using these bias parameters and the power spectrum at $z\gtrsim7$
and determine the effect on the faint-end slope $\alpha$.  We find
that large-scale structure introduces a $1\sigma$ uncertainty of
$\sim$0.05 in our faint-end slope determinations $\alpha$ at $z\sim7$
and $z\sim8$.

We can also look at the issue of field-to-field variations from a more
empirical standpoint.  Fixing the shape of the $UV$ LFs at $z\sim7$
and $z\sim8$ to the best-fit $\alpha$ and $M^*$, we can look at the
variations in the normalization $\phi^*$ for our three HUDF09 fields.
We find a $1\sigma$ variation in $\phi^*$ of just $\sim$20\%.  This is
similar to what we would expect from the Poissonian uncertainties.
This suggests that large-scale structure uncertainties do not pose an
especially huge problem for LF determinations, in the current
luminosity range ($-21$ AB to $-18$ AB mag).

Thus, the impact of large scale structure on our $z\sim7$ and $z\sim8$
LFs appears to be quite modest.

\begin{figure}
\epsscale{1.15}
\plotone{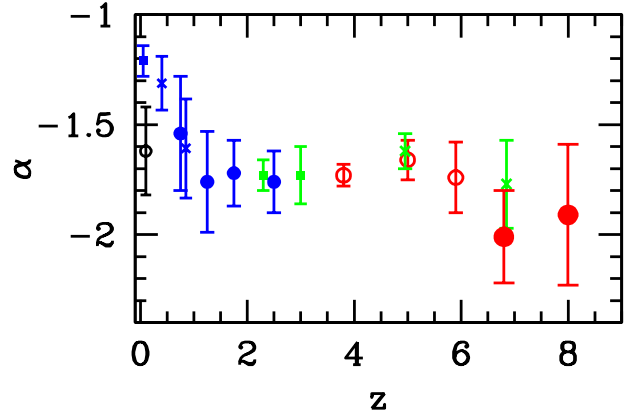}
\caption{The derived faint-end slope $\alpha$ to the $UV$ LF versus
  redshift.  The present constraints on $\alpha$ are shown as the
  large solid red circles, with the $1\sigma$ uncertainties plotted as
  the error bars.  The values of $\alpha$ derived at $z$$\sim$4-6
  (Bouwens et al.\ 2007: but see also Yoshida et al.\ 2006; McLure et
  al.\ 2009) are plotted as the open red circles.  Other
  determinations are from Oesch et al.\ (2010a: \textit{blue circles};
  see also Hathi et al.\ 2010), Reddy \& Steidel (2009: \textit{green
    squares}), Arnouts et al.\ (2005: \textit{blue crosses}), Wyder et
  al.\ (2005: \textit{blue squares}), Treyer et al.\ (1998:
  \textit{black open circle}), and Oesch et al.\ (2007, 2010a:
  \textit{green crosses}).  While the current constraints on the
  faint-end slope $\alpha$ at $z\sim7$ and $z\sim8$ are of limited
  statistical weight (see also Ouchi et al.\ 2009; Oesch et
  al.\ 2010a), they suggest the faint-end slope $\alpha$ is at least as
  steep as $-1.7$ -- and possibly as steep as $\sim-2.0$ (\S5.2;
  \S5.3).  We remark that the derived slopes simply describe the shape
  of the LF to several magnitudes below $L^*$.  Thus slopes that are
  even steeper than $-2$ do not imply a divergent luminosity density
  (since the LF will cut off at some faint value).\label{fig:abmagz}}
\end{figure}

\begin{figure}
\epsscale{1.15}
\plotone{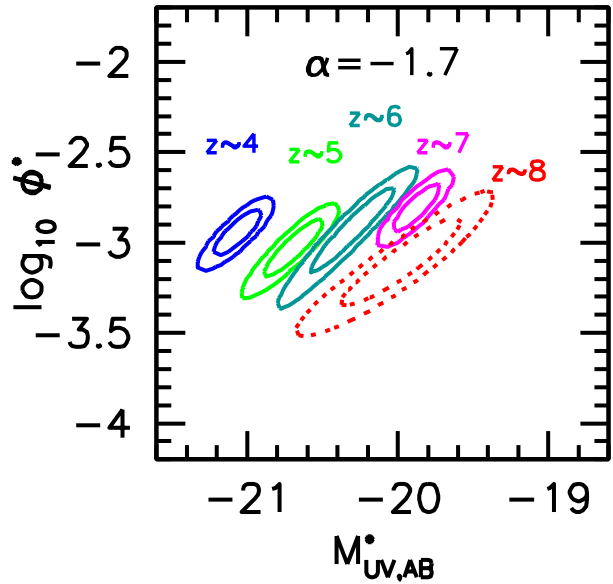}
\caption{68\% and 95\% likelihood contours on the model Schechter
  parameters $M^*$ and $\phi^*$ for our LF determination at $z\sim7$
  (\textit{magenta lines}) and $z\sim8$ (\textit{dotted red lines}).
  For comparison, we also show the constraints on the $UV$ LF at
  $z\sim4$ (\textit{blue lines}), $z\sim5$ (\textit{green lines}), and
  $z\sim6$ (\textit{cyan lines}) from Bouwens et al.\ (2007).  This
  figure is the same as the left panel to Figure~\ref{fig:schlf}, but
  now fixing the faint-end slope $\alpha$ to $-1.7$.  A faint-end
  slope $\alpha$ of $-1.7$ is similar to the values found at
  $z$$\sim$1-6 (Yoshida et al.\ 2006; Bouwens et al.\ 2007; Oesch et
  al.\ 2007; McLure et al.\ 2009; Reddy \& Steidel 2009; Oesch et
  al.\ 2010c; Arnouts et al.\ 2005: see Figure~\ref{fig:abmagz}).  No
  constraint on the faint-end slope at $z\sim4$-6 is imposed, and
  therefore the error contours there are somewhat larger than at
  $z\sim7$.\label{fig:schlff}}
\end{figure}

\section{Discussion}

\subsection{Evolution of the $UV$ LF from $z\sim8$ to $z\sim4$}

In \S5, we derived detailed constraints on the $UV$ LFs at $z\sim7$
and $z\sim8$.  By comparing these LFs with those derived at
$z$$\sim$4-6, we can examine the evolution with cosmic time.  In
Figure~\ref{fig:lfall}, we compare the stepwise LFs derived at
$z\sim7$ and $z\sim8$ with those found at $z$$\sim$4-6.  The $z\sim7$
LF presented in Figure~\ref{fig:lfall} incorporates the wide-area
NICMOS + ISAAC + MOIRCS $z\sim7$ search results from Bouwens et
al.\ (2010c: see the LFs in Figure~\ref{fig:lf7}).  It is remarkable
how uniform the evolution of the $UV$ LF is with cosmic time.  At
higher luminosities ($-20$ AB mag), the UV LF evolves by a factor of
$\sim$6-7 from $z\sim8$ to $z\sim4$ while at lower luminosities ($-18$
AB mag), the decrease is a factor of $\sim$5.

Figure~\ref{fig:schlf} presents the 68\% and 95\% confidence intervals
derived here for the UV LF at $z\sim7$ and $z\sim8$.  For comparison,
the LFs at $z\sim4$, $z\sim5$, and $z\sim6$ are also shown so that the
evolution can readily be inferred.  A clear trend is evident in the
$M^*$/$\phi^*$ LF fit results (\textit{left panel}) from $z\sim8$ to
$z\sim4$.  The characteristic luminosity shows evidence for a
consistent brightening with cosmic time from $z\sim7$ to $z\sim4$.
Ascertaining how the $\phi^*$ and $M^*$ evolves out to $z\sim8$ is not
possible at present, given the large uncertainties and degeneracy
between these two parameters (though there is a hint that some of the
evolution may be in $\phi^*$: see also discussion in McLure et
al.\ 2010).  Tighter constraints on the volume density of bright
$z\sim8$ galaxies are required for progress on this question.

As far as constraints on the shape of the LF (\textit{middle panel}),
again we see a clear trend towards fainter values of $M^*$ from
$z\sim7$ to $z\sim4$.  The faint-end slope $\alpha$ at $z\sim7$ and
$z\sim8$ also is very steep, similar to the values $\alpha\sim-1.7$
found at $z$$\sim$2-6 (Figure~\ref{fig:abmagz}) and possibly even
steeper.  This possible steepening of the LF with redshift is in
qualitative agreement with that predicted from various theoretical
models for the LF (e.g., Trenti et al.\ 2010; Salvaterra et al.\ 2010;
Bouwens et al.\ 2008).  Such steep faint-end slopes are potentially
very important for reionization since it would suggest that low
luminosity galaxies at $z\gtrsim7$ are major contributors to the
needed UV ionizing photon flux.  We discuss this in detail in Bouwens
et al.\ (2011b).

No trend is evident as a function of redshift in $\phi^*$ and $\alpha$
considered together (\textit{right panel}: Figure~\ref{fig:schlf}).
However, the uncertainties remain large at $z\sim7$ and $z\sim8$.
Given these uncertainties, we can potentially gain more insight into
the evolution of the LF at high redshift by fixing the faint-end slope
$\alpha$ to $-1.7$.  The faint-end slope $\alpha$ is found to be
consistent with this value over a very wide redshift range
$z$$\sim$1-8 (Bouwens et al.\ 2007; Reddy \& Steidel 2009; Beckwith et
al.\ 2006; McLure et al.\ 2009; Ouchi et al.\ 2009; Oesch et
al.\ 2007, 2010a, 2010c), so it seems reasonable to fix $\alpha$ to
$-1.7$ at $z\sim7$ and $z\sim8$.  The constraints we derive on $M^*$
and $\phi^*$ fixing $\alpha$ are shown in Figure~\ref{fig:schlff}.
This figure provides further support to the suggestion that the
primary mode of evolution in the LF from $z\sim7$ to $z\sim4$ is in
the characteristic luminosity (see also Bouwens et al.\ 2006; Yoshida
et al.\ 2006; Bouwens et al.\ 2007; McLure et al.\ 2009).  For
$z\sim8$ LF, the normalization is somewhat lower -- suggesting that
some of the evolution in the LF at $z\geq4$ may be in volume density
(e.g., McLure et al.\ 2009; Castellano et al.\ 2010b).

\begin{figure*}
\epsscale{1.20}
\plotone{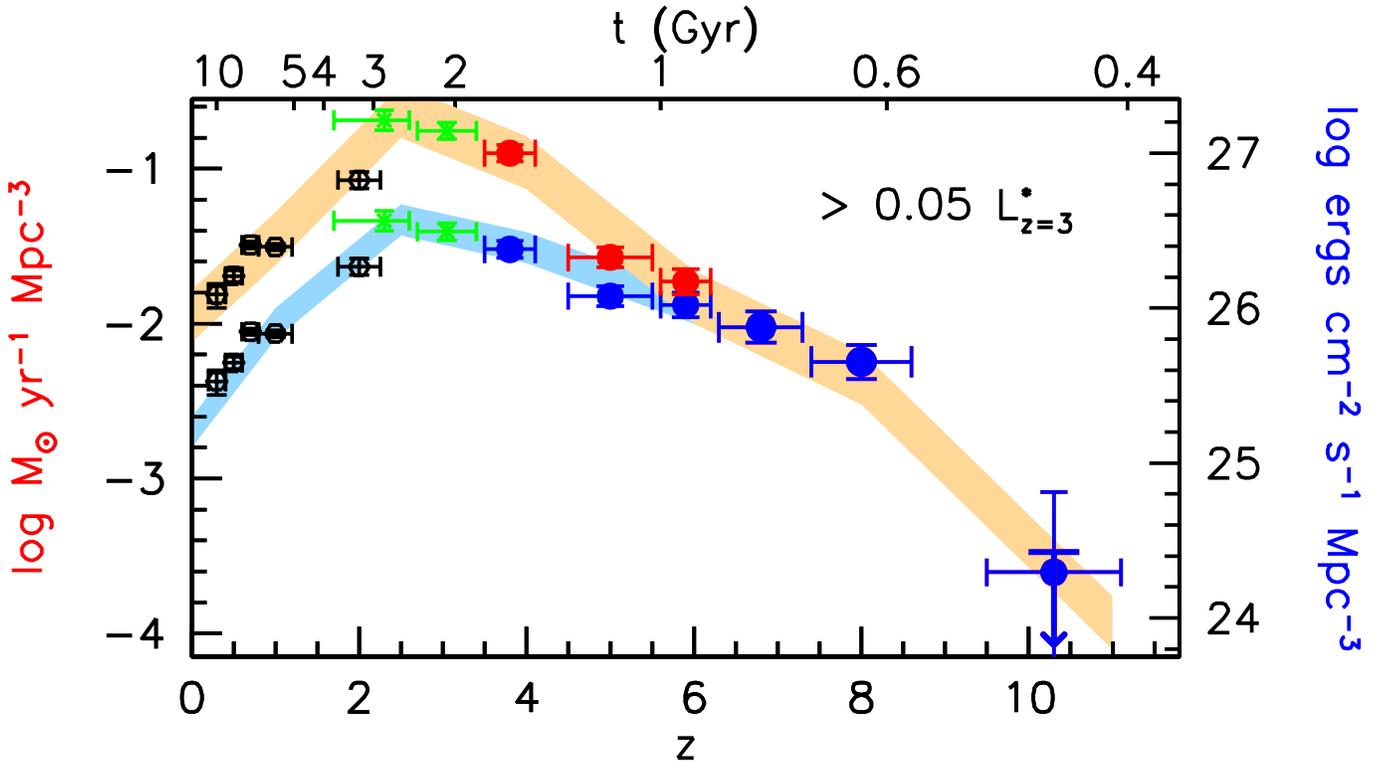}
\caption{Luminosity density and SFR density determinations, as a
  function of redshift (\S7.3).  These determinations are integrated
  to $-17.7$ AB mag ($0.05$ $L_{z=3}^{*}$).  The lower set of points
  (blue for our measurements) and shaded blue region are the observed
  luminosity densities and SFR density estimates before dust
  correction.  The upper set of points (red for our measurements) and
  shaded orange region indicates the SFR density estimates after dust
  correction.  The SFR density estimates assume $\gtrsim100$ Myr
  constant SFR and a Salpeter IMF.  Conversion to a Kroupa (2001) IMF
  would result in a factor of $\sim$1.7 (0.23 dex) decrease in the SFR
  density estimates given here.  At $z\lesssim2$ dust corrections are
  from Schminovich et al. (2005), at $z$$\sim$3-6 from Bouwens et
  al.\ (2009b), and at $z\gtrsim7$ they are assumed to be zero.  The
  solid circles at $z\sim7$ and $z\sim8$ are the present
  determinations.  For context, we have also included the SFR density
  estimates at $z$$\sim$4-6 from Bouwens et al.\ (2007), at
  $z$$\sim$2-3 from Reddy \& Steidel (2009: green crosses), and at
  $z$$\lesssim$2 from Schiminovich et al.\ (2005: black squares).  The
  estimates at $z\sim10$ (Bouwens et al.\ 2011a) are given with the
  blue circles and upper limit.  See Table~\ref{tab:sfrdens} for a
  tabulation of the $z\gtrsim4$ luminosity densities and SFR densities
  presented here.
\label{fig:dustsfz}}
\end{figure*}

\subsection{Shape of the UV LF at $z$$\sim$7 and $z$$\sim$8}

One potentially important open question about the $UV$ LFs at $z\geq7$
concerns their overall shape (e.g., see discussion in \S5.5 of Bouwens
et al.\ 2008).  While we have generally attempted to represent the LFs
here with Schechter functions (i.e., a power-law with an exponential
cut-off at the bright end: see \S5.2 and \S5.3), it is possible that
the LFs may have a shape closer to that of a power-law, e.g.,
resembling the halo mass function (e.g., Douglas et al.\ 2009).  This
is simply because at the high redshifts we are probing the typical
halo masses would be much lower ($<$10$^{12}$ $M_{\odot}$) and
therefore many of the physical mechanisms likely to be important for
truncating the LFs at $z\sim0$ (e.g., transition to hot flows:
Birnboim \& Dekel 2003; Rees \& Ostriker 1977; Binney 1977; Silk 1977
or AGN feedback: Croton et al.\ 2006) may not be as relevant at
$z\gtrsim8$.  Of course, at sufficiently high redshifts (i.e.,
$z\gtrsim7$), we would eventually expect some truncation in the LF at
high masses (and luminosities) due to a similar exponential cut-off in
the halo mass function (e.g., Trenti et al.\ 2010).

\begin{figure}
\epsscale{1.20}
\plotone{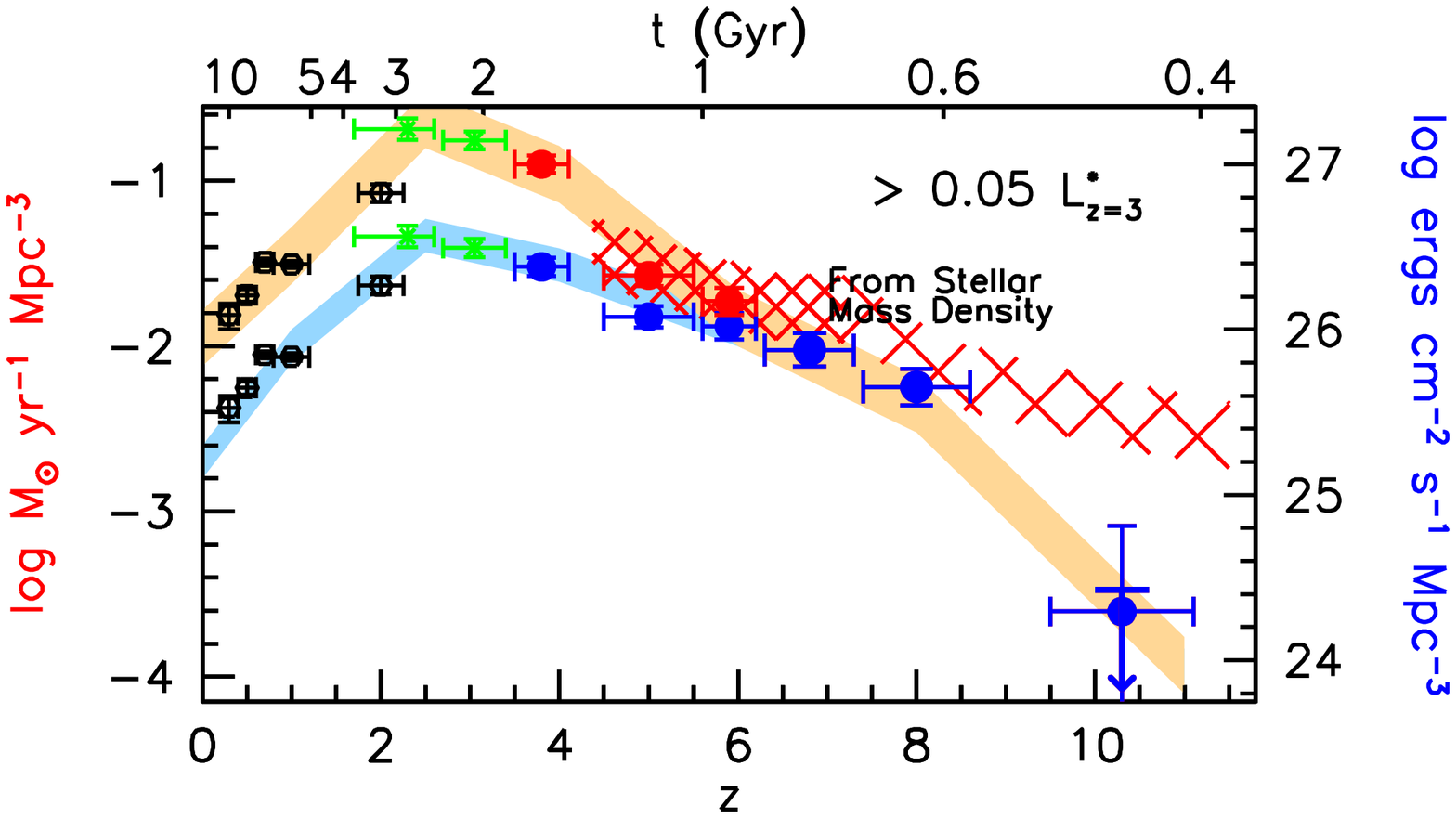}
\caption{Comparison of the SFR density determinations at $z\gtrsim4$
  with that implied by recent determinations of the stellar mass
  density (\S7.4).  The red hatched region gives the SFR densities
  implied by current compilations of the stellar mass density at
  $z\gtrsim4$ (Labb\'{e} et al.\ 2010a,b; Stark et al.\ 2009; Gonzalez
  et al.\ 2010) where the width of this region indicates a roughly
  $\pm$1$\sigma$ uncertainty.  The change in density of the
  cross-hatching at $z>7$ is intended to indicate that the
  mass-density-derived SFR may be more subject to systematic
  uncertainty from limited measurements and the extrapolation required
  from the measurements at $z\lesssim7$.  At $z\gtrsim8$, the stellar
  mass density in galaxies brightward of $-18$ AB mag is assumed to
  scale with redshift as $(1+z)^{-4.4}$ (as found in Labb\'{e} et
  al.\ 2010b).  Interestingly, at $z<7$ the mass-density-derived SFR
  is only about 40\% larger than the SFR density estimated from the
  LFs; such a small difference is quite encouraging.  This suggests
  that (1) the IMF at high redshift ($z\sim4$-7) may not be that
  different from what it is at later times and (2) that stellar mass
  density estimates at $z>4$ are reasonably
  accurate.\label{fig:sfzmass}}
\end{figure}

\begin{deluxetable}{lcccc}
\tablewidth{0pt}
\tabletypesize{\footnotesize}
\tablecaption{$UV$ Luminosity Densities and Star Formation Rate Densities to $-17.7$ AB mag.\tablenotemark{a}\label{tab:sfrdens}}
\tablehead{
\colhead{} & \colhead{} & \colhead{$\textrm{log}_{10} \mathcal{L}$} & \multicolumn{2}{c}{$\textrm{log}_{10}$ SFR density} \\
\colhead{Dropout} & \colhead{} & \colhead{(ergs s$^{-1}$} & \multicolumn{2}{c}{($M_{\odot}$ Mpc$^{-3}$ yr$^{-1}$)} \\
\colhead{Sample} & \colhead{$<z>$} & \colhead{Hz$^{-1}$ Mpc$^{-3}$)} & \colhead{Uncorrected} & \colhead{Corrected\tablenotemark{b}}}
\startdata
$z$ & 6.8 & 25.88$\pm$0.10 & $-2.02\pm0.10$ & $-2.02\pm0.10$ \\
$Y$ & 8.0 & 25.65$\pm$0.11 & $-2.25\pm0.11$ & $-2.25\pm0.11$ \\
$J$\tablenotemark{d} & 10.3 & 24.29$_{-0.76}^{+0.51}$ & $-$3.61$_{-0.76}^{+0.51}$ & $-$3.61$_{-0.76}^{+0.51}$\\
$J$\tablenotemark{d} & 10.3 & $<$24.42\tablenotemark{c} & $<-$3.48\tablenotemark{c} & $<-$3.48\tablenotemark{c}\\

\multicolumn{5}{c}{-----------------------------------------------------------------} \\
$B$ & 3.8 & 26.38$\pm$0.05 & $-1.52\pm$0.05 & $-0.90\pm0.05$ \\
$V$ & 5.0 & 26.08$\pm$0.06 & $-1.82\pm$0.06 & $-1.57\pm0.06$ \\
$i$ & 5.9 & 26.02$\pm$0.08 & $-1.88\pm$0.08 & $-1.73\pm0.08$ \\
\enddata
\tablenotetext{a}{Integrated down to 0.05 $L_{z=3}^{*}$.  Based upon
  LF parameters in Table~\ref{tab:lfparm} (see \S7.3).  The SFR
  density estimates assume $\gtrsim100$ Myr constant SFR and a
  Salpeter IMF (e.g., Madau et al.\ 1998).  Conversion to a Kroupa
  (2001) IMF would result in a factor of $\sim$1.7 (0.23 dex) decrease
  in the SFR density estimates given here.}
\tablenotetext{b}{The dust correction is taken to be 0 for galaxies at
  $z$$\sim$7-8, given the very blue $\beta$'s (see also Bouwens et
  al. 2009; Oesch et al.\ 2010a; Bouwens et al.\ 2010a).}
\tablenotetext{c}{Upper limits here are $1\sigma$ (68\%
  confidence).}
\tablenotetext{d}{$z\sim10$ determinations and limits are from
  Bouwens et al.\ (2011a) and assume 0.8 $z\sim10$ candidates in the
  first case and no $z\sim10$ candidates (i.e., an upper limit) in the
  second case.}
\end{deluxetable}

Overall, we do not find any evidence that the $UV$ LF at $z\sim7$ or
$z\sim8$ has a fundamentally different shape than the LF has at lower
redshift.  Both LFs seem to have roughly a power-law-like shape at the
faint end and a modest cut-off at the bright end -- consistent with
that of a Schechter function (though the evidence for such a cut-off
in the $z\sim8$ LF is weak).  Ascertaining whether the LF really has a
Schechter-like form or has a tail extending to bright magnitudes
(e.g., Capak et al.\ 2011) will require significantly more statistics,
particularly at bright magnitudes, to better establish the shape of
the LF there.  Fortunately, we should soon have such statistics as a
result of observations being obtained over the GOODS fields as part of
the CANDELS multi-cycle treasury program, the wide-area SEDS/CANDELS
fields, and two pure parallel programs (Trenti et al.\ 2011; Yan et
al.\ 2011).

\subsection{Luminosity Densities and SFR Densities}

With the completion of the wide-area ERS and ultra-deep HUDF09
observations, we have been able to significantly improve our
constraints on the $UV$ LFs at $z\sim7$ and $z\sim8$.  These new LF
determinations allow us to update our luminosity density and SFR
density estimates at $z$$\sim$7-8 from that already given in Bouwens
et al.\ (2010b) based on the initial HUDF09 data on the HUDF.  These
densities continue to be of great interest for discussions about the
role of galaxies in reionization as well as the build-up of stellar
mass.

For our luminosity density (and SFR density) estimates, we simply
integrate up our stepwise LF determinations to the limiting depth of
our LF determinations, i.e., to $-17.7$ AB mag (0.05 $L_{z=3}^{*}$).
The resultant luminosity densities are tabulated in
Table~\ref{tab:sfrdens}.  They are also shown in
Figure~\ref{fig:dustsfz} along with luminosity density determinations
at lower redshift.

The luminosity density determinations are converted into SFR densities
using the Madau et al.\ (1998) conversion factor -- assuming a
Salpeter IMF (with a stellar mass function ranging from 0.1
$M_{\odot}$ and 125 $M_{\odot}$).  As often noted (e.g., Verma et
al.\ 2007; Stanway et al.\ 2005; Bouwens et al.\ 2009), this
conversion factor assumes that the SFRs in galaxies are relatively
constant for $\gtrsim100$ Myr prior to observation.  Consequently, use
of this conversion would result in an underestimate of the SFR (by
factors of $\sim$2) for galaxies with particularly young ages.

Given the very blue $UV$-continuum slopes $\beta$ of $z\gtrsim7$
galaxies (e.g., Bouwens et al.\ 2010a,b; Oesch et al.\ 2010a;
Finkelstein et al.\ 2010; Bunker et al.\ 2010; see also Bouwens et
al.\ 2009), the dust extinction in $z\gtrsim7$ galaxies is likely
small, and so no correction for dust is made in computing the SFR
densities at $z\gtrsim7$.  At $z$$\lesssim$6, we adopt the dust
corrections given in Schiminovich et al.\ (2005), Reddy \& Steidel
(2009), and Bouwens et al.\ (2009).  Our SFR density estimates are
included in Table~\ref{tab:sfrdens} and on Figure~\ref{fig:dustsfz}.

The $UV$ luminosity densities we find to $-17.7$ AB mag at $z\sim7$
and $z\sim8$ are just 32\% and 18\%, respectively, of that found to
the same limiting luminosity at $z\sim4$.  The SFR densities we find
to these limits are just 6\% and 4\%, respectively, of that at
$z\sim4$.  This provides some measure of how substantial galaxy
build-up has been in the $\sim$1 Gyr from $z$$\sim$7-8 to $z\sim4$
when the universe was nearing its peak SFR density at $z\sim2$-3.

\subsection{SFR Densities from the Stellar Mass Density}

It is interesting to compare the present SFR density determinations
with that inferred from recent estimates of stellar mass density at
$z>4$ (Labb\'{e} et al. 2010a,b; Stark et al.\ 2009; Gonzalez et
al.\ 2010).  The comparison provides us with a very approximate test
of the shape of the IMF at high redshift as well as allowing us to
test the basic consistency of these very fundamental measurements.

In estimating the SFR densities from the stellar mass densities (Stark
et al.\ 2009; Labb{\'e} et al.\ 2010a, 2010b; Gonzalez et al.\ 2010),
there are a few issues we need to consider.  One issue is the stellar
mass lost due to SNe recycling (Bruzual \& Charlot 2003).  To correct
for this, we have multiplied the star formation rate densities we
infer from the stellar the stellar mass densities by a factor of
$1/0.72\sim1.39$ (appropriate for the Salpeter IMF assumed in the
Labb{\'e} et al.\ 2010b mass estimates and a $\sim$300 Myr typical
age).

A second issue we need to consider is that new galaxies will be
entering magnitude-limited samples at all redshifts simply due to the
growth of galaxies with cosmic time (e.g., galaxies not present in a
$z\sim7$ sample would grow enough from $z\sim7$ to $z\sim6$ to enter
magnitude-limited selections at $z\sim6$).  Therefore, it is not quite
possible to determine the SFR density by subtracting the stellar mass
density determination in one redshift interval from that immediately
below it.  To account for those galaxies entering magnitude-limited
samples due to luminosity growth, galaxies are assumed to brighten by
0.33 mag per unit redshift (the approximate $M^*$ evolution: see
\S7.5).  This results in a typical 20\% correction to the SFR density.
Based on these corrections and considerations and the stellar mass
density at $z\sim 8$ (Labb{\'e} et al. 2010b), we derive the SFR
density down to $z\sim4$.  The results are shown in
Figure~\ref{fig:sfzmass} (\textit{hatched red region}).
\footnote{The stellar mass density estimated at $z\sim8$ (Labb\'{e} et
  al. 2010b) -- although uncertain -- suggests a certain SFR density
  at $z\gtrsim8$.  Of course, since the stellar mass is an integral
  quantity (and there are a large range of plausible SF histories that
  yield the observed stellar mass density), a redshift dependence
  needs to be assumed.  We adopt the same scaling of the stellar mass
  density with redshift at $z\gtrsim8$, i.e.,$(1+z)^{-4.4}$, as
  Labb\'{e} et al.\ (2010b) found to fit the observational results at
  $z$$\sim$4-8.}  Interestingly, the SFR densities inferred from the
stellar mass density are only $\sim$40\% higher than that inferred
from the $UV$ LFs at $z<7$ (where the stellar masses are more well
determined).\footnote{The difference increases slightly at earlier
  times where the galaxy stellar mass function is more poorly
  determined and extrapolations are used.} This suggests that (1) the
IMF for SF galaxies at high redshift ($z\sim4$-7) may not be that
different from what it is at later times ($z\lesssim 3$: but see
discussion in Chary 2008) and (2) the stellar mass densities at $z>4$
inferred from the Spitzer IRAC photometry are reasonably accurate.
This latter conclusion is important since nebular emission line fluxes
have been suggested to dominate the measured light in $z\geq7$
galaxies with IRAC (Schaerer \& de Barros 2010: but see Finlator et
al.\ 2011).  These comparisons suggest that nebular emission is not as
dominant as has been suggested (Schaerer \& de Barros 2010).

\subsection{Extrapolations to $z\geq8$: A LF-Fitting Formula}

The present LF determinations provide us with reasonably accurate
measures of luminosity density and SFR density at $z\sim7$ and
$z\sim8$ -- which are of considerable importance for assessing the
impact of galaxies in the build-up of mass and metals in the universe,
as well as understanding their role in reionization.  However, knowing
the LF results at $z$$\sim$7-8 only provide us with part of the
information we want.  To address some of the most outstanding of
today's questions (i.e., whether galaxies reionize the universe, how
stellar mass and metals are built up versus cosmic time), we really
need to know what the $UV$ LFs look like at even higher redshift.

For these questions, perhaps the best approach is simply to
extrapolate the present LF results to even higher redshifts.
Certainly, given the very uniform rate of evolution of the $UV$ LF
from $z\sim8$ to $z\sim4$, it seems quite plausible that such an
extrapolation would be reasonably accurate.  As in our previous work
(\S5.3 of Bouwens et al.\ 2008), we frame the evolution of the LF in
terms of a Schechter function and assume that each Schechter parameter
$M^*$, $\alpha$, and $\phi^*$ can be expressed as a linear function of
redshift.  We then identify linear coefficients that maximize the
likelihood of reproducing the observed contours in
Figure~\ref{fig:schlf} at $z\sim4$, $z\sim5$, $z\sim6$, $z\sim7$, and
$z\sim8$.  The result is
\begin{eqnarray*}
M_{UV} ^{*} =& (-21.02\pm0.09) + (0.33\pm0.06) (z - 3.8)\\
\phi^* =& (1.14\pm0.20) 10^{(0.003\pm0.055)(z-3.8)}10^{-3} \textrm{Mpc}^{-3}\\
\alpha =& (-1.73\pm0.05) + (-0.01\pm0.04)(z-3.8)
\label{eq:empfit}
\end{eqnarray*}
We see that the evolution in $M_{UV}^*$ is significant at the
5$\sigma$ level.  However, the evolution in $\phi^*$ and $\alpha$ is
still not significant.  The fitting formula above are similar to those
found in Bouwens et al.\ (2008) based upon a large $z\sim4$-6 ACS
sample and $z\sim7$ NICMOS sample.

\subsection{Implications for $z\sim10$ $J_{125}$-dropout Searches}

In the previous section \S7.5, we combined the present constraints on
the $UV$ LFs at $z\sim7$ and $z\sim8$ with that available at
$z$$\sim$4-6 to derive fitting formula for the evolution of the $UV$
LF with redshift.  These formula arguably provide us with our best
means for extrapolating the $UV$ LF to even higher redshifts and
therefore estimating the form of the $UV$ LF at $z\sim10$.  This
enterprise has important implications for $z\sim10$ galaxy searches in
present and future WFC3/IR observations.

What surface density of $z\sim10$ $J_{125}$-dropout candidates do we
expect to find in current ultra-deep WFC3/IR observations?  We can
estimate this by taking advantage of the fitting formula we derive in
the previous section and the detailed simulations carried out by
Bouwens et al.\ (2011a) to estimate $J_{125}$-dropout selection
efficiencies.  We predict 1.0 and 2.3 $J_{125}$-dropouts arcmin$^{-2}$
to an $H_{160}$ magnitude of 28.5 and 29.5 AB mag, respectively.
These numbers are similar albeit somewhat larger than that found by
Bouwens et al.\ (2011a) over the HUDF to $H_{160}\sim29.5$.
Certainly, the fact that a few $J_{125}$-dropout galaxies are expected
in the HUDF search suggests that the $z\sim10$ galaxy candidate
identified by Bouwens et al.\ (2011a) is plausible.  Of course, deeper
observations are clearly needed for more definitive statements.  We do
not address the Yan et al.\ (2010) $J_{125}$-dropout numbers here,
given the challenges in interpreting those numbers (both as a result
of very low $3\sigma$ detection thresholds used in that study and the
likely high contamination rates expected from the proximity of the
many Yan et al.\ 2010 candidates to bright foreground galaxies: see
discussion in Bouwens et al.\ 2011a).

\section{Summary}

The ultra-deep (14 arcmin$^2$) HUDF09 and deep wide-area (40
arcmin$^2$) ERS WFC3/IR observations are a significant resource for
identifying large samples of star-forming galaxies at $z\gtrsim7$.
The depth of these observations in three distinct wavelength channels
combined with their availability over regions of the sky with
ultra-deep optical data make such identifications possible with
well-established selection procedures like the Lyman Break Galaxy
technique.

We utilize LBG selection criteria over these fields to identify 73
candidate $z\sim7$ galaxies and 59 $z\sim8$ candidate galaxies
(Table~\ref{tab:dropsamp}).  These candidates extend over a $\sim$3
mag magnitude range in luminosity from $H_{160,AB}$ $\sim$26 to
$\sim$29.4 mag -- thus reaching to $-17.7$ AB mag (0.05
$L_{z=3}^{*}$).  The use of LBG criteria allows us to identify likely
high-redshift sources from the extremely deep WFC3/IR data and then to
exploit fully the deep, high spatial resolution ACS optical data to
remove contaminants in a very robust way.  Our sample of 132 galaxies
in the reionization epoch is the largest currently available.  

To determine the contamination rate we thoroughly assess the impact of
lower redshift sources, photometric scatter, AGN, spurious sources,
low mass stars, and transients (e.g., SNe) on our selection (\S3.5).
The contamination rate that we find is just $\sim$7\% for the HUDF
sample and $\sim$30\% for the brighter, but less crucial, ERS sample
(where the shallower optical data provide weaker constraints).

\textit{Through simulations, we have found that very strict optical
  non-detection criteria are essential to obtaining contamination-free
  $z\geq7$ samples.}  We set a requirement that an accepted source
must {\it not} show a $2\sigma$-detection in any optical filter ({\it
  nor} a $1.5\sigma$-detection in $>$1 optical filter).  However,
through extensive simulations we have found that this alone is not
rigorous enough; it is likely that $\sim$30-40\% of our LBG samples
over the HUDF09-1, HUDF09-2, and ERS fields would have consisted of
contaminants if that had been the only optical constraint. What is
much more effective is to combine the information in the entire
optical dataset to derive a $\chi_{opt}^2$ value for each source.  In
addition, we recommend validating the selection by looking at the
$\chi_{opt}^2$ distribution for the entire sample (\S4.2:
Figure~\ref{fig:chisq}).  The use of this technique results in much
smaller contamination and is the reason why we can reach as faint as
29 AB mag and still reach an overall contamination rate of $\sim$11\%
($\sim$7\% in the HUDF09 selection).  This is discussed extensively in
\S3.5 and Appendix D, while our concerns about contamination in some
other recent studies are discussed in \S4.1.2 and \S4.1.3.

The high surface density of galaxies in our samples has direct
implications for the LFs of galaxies in the first 800 Myr of the
universe and the build-up of galaxies with cosmic time.  We derive
selection volume for our LBG-selected galaxy samples through careful
simulations, and use these selection volumes to derive luminosity
functions with well-understood errors.  The techniques we use to
derive the LFs are insensitive to large-scale structure biases and
also explicitly account for other observational biases.  We combine
our LF determinations with other wide-area search results (see
Table~\ref{tab:osearch}) to derive more robust constraints on the
overall shape of the LF.  We then compare these LFs with those found
at $z$$\sim$4-6 to quantify the evolution of the $UV$ LFs in the first
1.6 Gyr of the universe.

Here are our primary findings:
\begin{itemize}
\item{The faint-end slope $\alpha$ for the $UV$ LF at $z\sim7$ is very
  steep, with values of $-2.01\pm0.21$.  The current faint-end slope
  determination is consistent with previous estimates at $z\sim7$
  (Ouchi et al.\ 2009; Oesch et al.\ 2010a) and is a logical extension
  of that found at lower redshift where $\alpha$$\sim$$-1.7$ from
  $z\sim6$ to $z\sim1$ (Oesch et al.\ 2010c; Bouwens et al.\ 2007;
  Reddy \& Steidel 2009; Oesch et al.\ 2007; Yoshida et al.\ 2006;
  Beckwith et al.\ 2006; Yan \& Windhorst 2004).  The faint-end slope
  $\alpha$ at $z\sim8$ is more uncertain, but also consistent with
  being very steep: $\alpha=-1.91\pm 0.32$.  We have verified that the
  present faint-end slope determinations are not significantly
  affected by large-scale structure (see \S6.2).  Such steep slopes do
  not necessarily lead to a divergent integrated luminosity density
  since it is expected that the galaxy LF will almost certainly
  truncate at some physically-established low-mass or luminosity
  limit.  These results provide increasingly strong evidence that very
  low luminosity galaxies provide the dominant contribution to the
  luminosity density at $z\geq6$.  Strikingly, $\gtrsim$75\% of the
  luminosity density is from galaxies fainter than $-$18 AB mag (if
  $\alpha\leq-1.7$).  Interestingly, such large contributions are
  expected from theory (e.g., Salvaterra et al.\ 2010; Trenti et
  al.\ 2010; Barkana \& Loeb 2000).}
\item{The characteristic luminosity $M^*$ and the normalization
  $\phi^*$ we find at $z\sim7$ are $-20.14\pm0.26$ and
  $0.86_{-0.39}^{+0.70}\times10^{-3}$ Mpc$^{-3}$, respectively.  These
  values are consistent with other recent determinations of the $UV$
  LF at $z\sim7$ (Ouchi et al.\ 2009; Oesch et al.\ 2010a; Castellano
  et al.\ 2010a) and are also consistent with lower redshift trends
  (i.e., Bouwens et al.\ 2007, 2008).  If we represent the $z\sim8$ LF
  with a Schechter function, the characteristic luminosity $M^*$ and
  normalization $\phi^*$ that we determine is $-20.10\pm0.52$ and
  $0.59_{-0.37}^{+1.01}\times10^{-3}$ Mpc$^{-3}$, respectively.  This
  suggests that the evolution in the $UV$ LF at $z\gtrsim4$ occurs not
  only through changes in $M^*$ (as preferred from the $z\sim7$ to
  $z\sim4$ results: \S7.1), but also through changes in $\phi^*$.  The
  result however is quite sensitive to the surface density of very
  luminous galaxies [in this case the two bright $z\sim8$ galaxies in
    the HUDF09-2 field which would appear to represent an overdensity:
    \S5.3].  The uncertainties are still large at present.}
\item{The $z\sim7$ LF appears to be much better represented by a
  Schechter function than by a power law and shows evidence for an
  exponential cut-off at the bright end (\S7.2).  The $z\sim8$ LF also
  seems to be better represented by a Schechter function -- though the
  evidence is weaker.  It is unclear what physical explanation would
  impart such a cut-off in the LF (particularly if $M^*$ becomes
  fainter at higher redshift: \S7.5), but it is possible this may be
  due (at least partially) to a similar exponential cut-off in the
  halo mass function (e.g., Trenti et al. 2010).}
\item{By integrating the $UV$ LFs derived here at $z$$\sim$7 and 8, we
  estimate the $UV$ luminosity density and SFR density at $z\sim7$ and
  $z\sim8$ to a limiting luminosity of $-17.7$ AB mag (0.05
  $L_{z=3}^{*}$).  Consistent with the very blue UV-continuum slopes
  $\beta$ observed for $z$$\sim$7-8 galaxies (e.g., Bouwens et
  al.\ 2010a,b; Oesch et al.\ 2010a; Finkelstein et al.\ 2010; Bunker
  et al.\ 2010), we assume that the dust extinction is zero. The $UV$
  luminosity densities at $z$$\sim$7 and $z$$\sim$8 are $\sim$32\% and
  $\sim$18\% of the value at $z\sim4$, while the SFR densities
  estimated at $z$$\sim$7 and $z$$\sim$8 to these same luminosity
  limits are $\sim$6\% and $\sim$4\% of the value at $z\sim4$ (see
  Figure~\ref{fig:dustsfz} and Table~\ref{tab:sfrdens}: \S7.3).}
\item{The SFR densities we estimate from published stellar mass
  density determinations at $z\sim4$-7 (e.g., Labb{\'e} et al.\ 2010b;
  Stark et al.\ 2009; Gonzalez et al.\ 2010) are only $\sim$40\%
  higher than that derived directly from the LFs.  See \S7.4 and
  Figure~\ref{fig:sfzmass}. This suggests that (1) the initial mass
  function for star formation at very high redshift may not be that
  different from what it is at later times and (2) the stellar mass
  densities at $z>4$ inferred from the IRAC photometry are reasonably
  accurate.  The latter conclusion is important since it has been
  suggested that nebular emission line fluxes may dominate the
  measured light in $z\geq7$ galaxies with Spitzer IRAC (Schaerer \&
  de Barros 2010: but see also Finlator et al.\ 2011).  The present
  comparison suggests otherwise.}
\item{We use the current contraints on the LF at $z$$\sim$7-8 to
  update the Bouwens et al.\ (2008) fitting formula describing the
  evolution of $M^*$, $\phi^*$, and $\alpha$ as a function of
  redshift (\S7.5).  The revised formula are $M_{UV} ^{*} = (-21.02\pm0.09) +
  (0.33\pm0.06) (z - 3.8)$, $\phi^* = (1.14\pm0.20)
  10^{(0.003\pm0.055)(z-3.8)}10^{-3} \textrm{Mpc}^{-3}$, and $\alpha =
  (-1.73\pm0.05) + (-0.01\pm0.04)(z-3.8)$ (\S7.5).}
\item{Given the interest in the existence of massive black holes at
  $z>6$ we have considered whether our $z\sim7$ and $z\sim8$ samples
  include any obvious evidence in the WFC3/IR images for AGNs.  It
  appears that the frequency of obvious AGN (unresolved ``stellar''
  sources that are comparable to, or brighter than, the underlying
  galaxy) is quite small (i.e., $\lesssim$2\%) in these $z\sim7$-8
  samples (\S3.5.3).}
\item{These new LF results allow us to estimate the expected number of
  $z\sim10$ galaxies from a search of the current HUDF09 data.  We
  predict 1.0 and 2.3 $J_{125}$-dropouts arcmin$^{-2}$ to an
  $H_{160}$-band magnitude of 28.5 and 29.5 AB mag, respectively.
  These numbers are consistent with the single $z\sim10$ candidate
  found by Bouwens et al. (2011a) over the HUDF to $H\sim 29.5$
  (\S7.6).  See Oesch et al.\ (2011) for an updated discussion using a
  larger set of observations.}
\end{itemize}
The present WFC3/IR observations allowed us to select 132 galaxies at
$z$$\sim$7 and $z$$\sim$8, over a 3 mag range in luminosity, reaching
as faint as 29 AB mag in the near-IR.  These observations have
provided us with our best insight yet into the build-up and evolution
of galaxies at early times -- providing us with a clear measurement of
rate of evolution in the $UV$ LF from $z\sim8$ to $z\sim4$ in the
first 1.6 Gyr of the universe.  Surveying the results, it is
remarkable how uniform the evolution of the LF is with cosmic time.
These new observations also provide us with further insight to the
contributions of galaxies to reionization (see Bouwens et al.\ 2011b for
new estimates based on these deep LFs).

Further progress should be made over the next few years in the field
of high-redshift LFs from the wide-area HST Multi-Cycle Treasury (MCT)
CANDELS program.  The CANDELS program will substantially improve our
constraints on the prevalence of bright galaxies at $z\gtrsim7$ and
the overall shape of the LF.  However, the crucial observations needed
for ultimately making progress on characterizing the role of galaxies
in reionizing the universe in the period from z$\sim$11 to z$\sim$6
and on the nature of the dominant population of low-luminosity
galaxies at z$\sim$7-10 are further ultra-deep observations with
WFC3/IR.  Such observations would significantly improve our
determinations of the faint-end slope $\alpha$, the $UV$-continuum
slopes $\beta$ of very low luminosity galaxies (e.g., Bouwens et
al.\ 2010a), and the prevalence of $z\sim10$ galaxies (e.g., Bouwens
et al.\ 2011a).

\acknowledgements

We would like to thank the STScI director Matt Mountain for making the
early release science (ERS) observations possible by granting DD time
to observe the CDF-South GOODS field with the newly-installed WFC3/IR
camera aboard HST.  We acknowledge useful conversations with Marco
Castellano, Simon Lilly, Casey Papovich, Naveen Reddy, and Brant
Robertson.  We deeply appreciate all those at NASA, its contractors,
STScI and throughout the community who have worked so diligently to
make Hubble the remarkable observatory that it is today.  The
servicing missions, like the recent SM4, have rejuvenated Hubble and
made it an extraordinarily productive scientific facility time and
time again.  We greatly appreciate the support of policymakers, and
all those at NASA in the flight and servicing programs who contributed
to the repeated successes of the HST servicing missions.  We
acknowledge the support of NASA grant NAG5-7697 and NASA grant
HST-GO-11563.  PO acknowledges support from the Swiss National Science
Foundation.

\appendix

\section{A.  Sizes of Our Model Galaxy Population and a Comparison with the Observations}

\begin{figure}
\epsscale{1.2}
\plotone{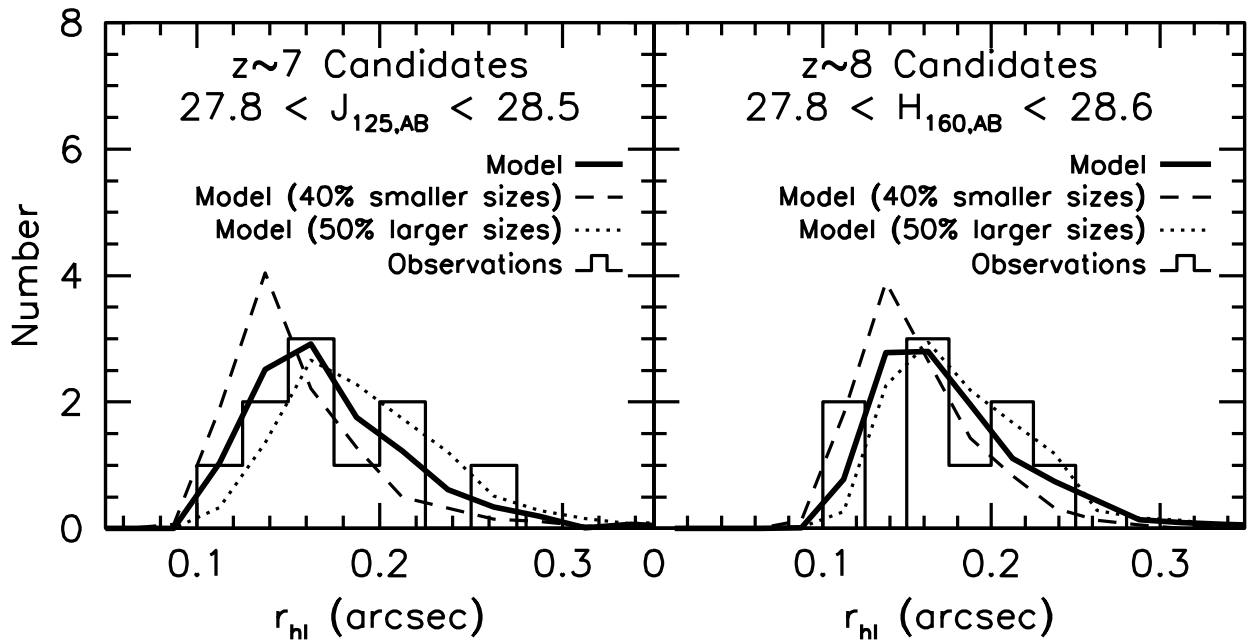}
\caption{Comparison of the half-light radii of $z\sim7$-8 galaxies
  observed over the HUDF09 field (\textit{histogram}) with those from
  our fiducial galaxy model (\textit{thick solid line}: which are used
  to estimate the selection volumes, see Appendix C).  The left panel
  is for our $z\sim7$ selection and the right panel is our for
  $z\sim8$ selection.  Galaxies are selected from our simulations in
  exactly the same way as from the observations.  The magnitude
  intervals plotted are chosen to be sufficiently brightward of the
  faint limit to allow for the selection of galaxies over a range of
  sizes, while still including a significant fraction of our HUDF09
  $z\sim7$-8 sample.  Also shown is the half-light radius distribution
  for a galaxy population with sizes that are 40\% smaller and 50\%
  larger than our fiducial model (\textit{dashed lines} and
  \textit{dotted lines}, respectively).  The observed half-light
  radius distribution is in excellent agreement with the sizes from
  our fiducial model.  The fit to the other size models is less good
  (the disagreement ranges from $\sim$1-2$\sigma$), but the
  uncertainties are large.  Deeper observations will be required to
  establish the size distribution more accurately to very low
  luminosities.  We find no evidence that our model size distribution
  is too large, and therefore that our selection volumes are too low,
  as suggested by Grazian et al.\ (2011: see Appendix
  A).\label{fig:sizes}}
\end{figure}

An essential input to determinations of the luminosity functions are
estimates of the selection volume.  For determining these selection
volumes, one of the most important inputs is the size (or surface
brightness) distribution of the model galaxy population (e.g., Bouwens
et al.\ 2003a,b; Bouwens et al.\ 2004; Bouwens et al.\ 2006; Oesch et
al.\ 2007).  Use of a model galaxy population with sizes that are too
large results in selection volumes that are too low, while use of a
model galaxy population with sizes that are too small results in
selection volumes that are too high.

Fortunately, we do not expect this to be an issue because of our
approach.  We expect that the sizes (or surface brightnesses) of our
model galaxy population will match that seen in the observations since
we base the sizes of our model galaxy population on the sizes already
measured for $z\sim7$-8 galaxies in the HUDF (Oesch et al.\ 2010b) and
on the size evolution observed from $z\sim8$ to $z\sim4$.
Nonetheless, it is useful to verify that there is good agreement
between our model galaxy population and observations using our current
$z\sim7$-8 samples and the full two-year HUDF09 data.

We provide such a check in Figure~\ref{fig:sizes} -- comparing the
sizes of $z\sim7$ galaxies seen in the observations and from our
simulations.  The simulations are based on $z\sim4$ galaxies from the
HUDF and the size-redshift evolution seen from $z\sim7$ to $z\sim4$.
Sources from the simulations are selected in an identical manner.  For
this comparison, we only utilize galaxies between $J_{125,AB}\sim27.8$
and $J_{125,AB}\sim28.5$ for our $z\sim7$ selection and
$H_{160,AB}\sim27.8$ and $H_{160,AB}\sim28.6$ for our $z\sim8$
selection -- since these magnitude intervals allow for the selections
of galaxies over a range of sizes and thus usefully probe the size
distribution.  Faintward of this, only the smallest galaxies are
selected, and so comparisons between the observed and model size
distributions do not allow us to effectively test our model (see
discussion in \S3.7 and Figure 5 of Bouwens et al.\ 2006; Appendix A
of Bouwens et al.\ 2008; Figure 3 of Grazian et al.\ 2011).  Also
shown in this figure is the half-light radius distribution for a
galaxy population with sizes that are 40\% smaller and 50\% larger
than our fiducial model.

As Figure~\ref{fig:sizes} illustrates, the observed size distribution
is in excellent agreement with that predicted by our fiducial model.
The fit to the other size models is less good, particularly the model
with the smaller size distribution (the disagreement ranges from
$\sim$1-2$\sigma$), but the uncertainties are large.  The poor fit of
the observations to the models with smaller sizes suggest that this
issue is unlikely to bias us towards shallower slopes. Deeper
observations will be required to establish the size distribution more
accurately to very low luminosities.  We find no evidence that our
model size distribution is too large, and therefore that our selection
volumes are too low, as suggested by Grazian et al.\ (2011).

Incidentally, it would appear that Grazian et al.\ (2011)'s claim is
based on a mistaken understanding of our methodology (shown in their
Figures 2 and 4).  They suggest that we model all galaxies using
spectroscopically confirmed (and therefore predominantly bright)
$z\sim4$ galaxies from the HUDF.  Regretably, this is a
misunderstanding of our approach and leads them to the wrong
conclusion about our results.  We explicitly consider luminosity when
modeling source sizes and structures, i.e., the structure of luminous
$z\sim7$ galaxies are modeled with luminous $z\sim4$ galaxies (from
the HUDF) and the structure of lower luminosity $z\sim7$ galaxies are
modeled with lower luminosity $z\sim4$ galaxies (from the HUDF).
Since lower luminosity $z\sim4$ galaxies are typically much smaller
than more luminous galaxies, we model the sizes of faint galaxies at
$z\sim7$-8 with much smaller galaxies than has been suggested.
Moreover, as shown in Figure~\ref{fig:sizes}, these sizes appear to be
in excellent agreement with the observations.

\section{B.  Expected Contamination Rates}

\begin{deluxetable}{ccccc}
\tablewidth{12cm}
\tabletypesize{\footnotesize}
\tablecaption{Estimated Number of Contaminants in our $z\sim7$
  and $z\sim8$ samples.\tablenotemark{a}\label{tab:cont}}
\tablehead{
\colhead{Mag Range\tablenotemark{b}} & \colhead{HUDF\tablenotemark{c}} & \colhead{HUDF09-1} & \colhead{HUDF09-2\tablenotemark{d}} & \colhead{ERS}}
\startdata
\multicolumn{5}{c}{$z\sim7$ Samples}\\
25.0-25.5 & 0.00 & 0.00 & 0.00 & 0.01\\
25.5-26.0 & 0.00 & 0.00 & 0.00 & 0.24\\
26.0-26.5 & 0.00 & 0.02 & 0.03 & 0.66\\
26.5-27.0 & 0.00 & 0.09 & 0.06 & 1.05\\
27.0-27.5 & 0.02 & 0.15 & 0.15 & 0.85\\
27.5-28.0 & 0.13 & 0.41 & 0.33 & 0.12\\
28.0-28.5 & 0.20 & 0.86 & 0.63 & $-$--\\
28.5-29.0 & 0.36 & 0.58 & 0.71 & $-$--\\
29.0-29.5 & 0.27 & $-$-- & $-$-- & $-$--\\
\multicolumn{5}{c}{$z\sim8$ Samples}\\
25.0-25.5 & 0.00 & 0.00 & 0.00 & 0.01\\
25.5-26.0 & 0.00 & 0.00 & 0.00 & 0.19\\
26.0-26.5 & 0.00 & 0.00 & 0.00 & 0.47\\
26.5-27.0 & 0.00 & 0.00 & 0.00 & 0.70\\
27.0-27.5 & 0.00 & 0.06 & 0.02 & 0.86\\
27.5-28.0 & 0.02 & 0.25 & 0.15 & $-$--\\
28.0-28.5 & 0.17 & 0.49 & 0.39 & $-$--\\
28.5-29.0 & 0.50 & 0.45 & 0.38 & $-$--\\
29.0-29.5 & 0.41 & 0.01 & 0.06 & $-$--\\
\enddata
\tablenotetext{a}{For our LBG selections over our three HUDF09
  fields, our estimated contamination rates are made assuming that the
  faint photometric samples have the same color distributions as found
  in the high S/N HUDF09 $26.5<H_{160,AB}<28.0$ subsample.  For our ERS
  selections, our contamination estimates are based upon the high S/N
  HUDF09 $26<H_{160,AB}<27.0$ subsample.  Noise is then added to the
  photometry, and our $z\sim7$ or $z\sim8$ selection
  criteria are then applied.  Any source that is selected by our
  LBG criteria (after adding noise), but detected in the optical
  in the original high S/N observations is counted towards the total
  contamination level.  See \S3.5 for details.}
\tablenotetext{b}{$J_{125}$ and $H_{160}$ magnitudes used for our
  $z\sim7$ and $z\sim8$ selections,
  respectively.}
\tablenotetext{c}{It is striking how much lower the estimated
  contamination levels are in our HUDF selections than in our
  HUDF09-1/ HUDF09-2 selections.  This difference is a direct result
  of the much greater depth of the optical data over the HUDF (by
  $\sim$1 mag) than the HUDF09-1/HUDF09-2 selections.}
\tablenotetext{d}{As a result of the ACS parallel observations
  acquired as part of the HUDF09 WFC3/IR program, the effective
  contamination rate over the HUDF09-2 field decreases by a factor of
  $\sim$3 for those regions with deeper ACS data (the cyan ``P1''
  region shown on Figure~\ref{fig:obsdata}).  The effect of these
  additional data is reflected in the numbers we quote in this table.}
\end{deluxetable}

In \S3.5, we consider many different sources of contamination for our
$z\sim7$ and $z\sim8$ selections, including (1) low-mass stars (T
dwarfs), (2) transient sources (SNe), (3) AGN, (4) low redshift
galaxies, (5) photometric scatter, and (6) spurious sources.  After
careful consideration of the source characteristics and simulations,
we find that the only meaningful source of contamination for our
$z\gtrsim7$ samples are lower-redshift galaxies photometrically
scattering into our selection.  In Table~\ref{tab:cont}, we give the
expected contamination rates for the $z\sim7$ and $z\sim8$ selections
as a function of magnitude for all four fields considered here.

\section{C.  Effective Selection Volumes}

To derive the rest-frame $UV$ LFs from the observed surface densities
of $z\sim7$-8 candidates, we require a good estimate of the effective
selection volume for our selections, as a function of magnitude.  To
obtain these estimates, we construct mock catalogs for our search
fields.  Pixel-by-pixel images are generated for each of the sources
in our catalogs, these sources are added to our observational data,
and finally source catalogs are constructed from the simulated data in
the same way as on the observations.  Dropouts are then selected using
the criteria described earlier in this section.  In generating
two-dimensional images for each source in our catalogs, we randomly
select a similar luminosity $z\sim4$ $B$-dropout from the Bouwens et
al.\ (2007) HUDF sample and then scale its size (physical) as
$(1+z)^{-1}$ to match the observed size evolution to $z\sim8$ (Oesch
et al.\ 2010b; see also Ferguson et al.\ 2004; Bouwens et al.\ 2004,
2006).  In setting the $UV$ colors of the sources, we assume that the
$UV$-continuum slopes $\beta$ are distributed as found by Bouwens et
al.\ (2010a): i.e., with a mean value of $-2.0$ at $L_{z=3}^{*}$
($-21$ AB mag), a mean value of $-2.7$ at $0.1L_{z=3}^{*}$ ($-18.5$ AB
mag), and with a mean $\beta$ between these two extremes for
luminosities in the range 0.1-1$L_{z=3}^{*}$ (e.g., Bouwens et
al.\ 2009; Bouwens et al.\ 2010a; Finkelstein et al.\ 2010; Bunker et
al.\ 2010; R.J. Bouwens et al.\ 2011, in prep).  These simulations are
performed for our $z\sim7$-8 selection over the ERS observations and
for all three of the HUDF09 search fields.

\section{D.  Use of the $\chi_{opt} ^2$ distribution to control for contamination}

\begin{figure}
\epsscale{1.15}
\plotone{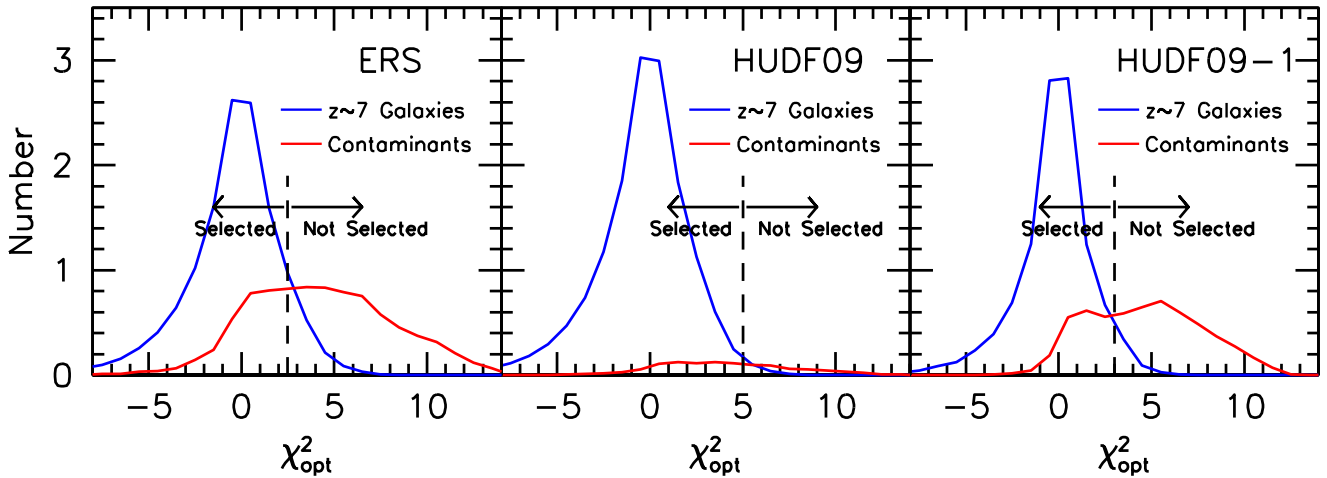}
\caption{Apparent detection significance $\chi_{opt}^2$ (Appendix D:
  \textit{blue lines}) expected for $z\sim7$ galaxy candidates from
  simulations for three different data sets considered here: ERS
  (\textit{left}), HUDF09 (\textit{center}), and HUDF09-1
  (\textit{right}).  Since $z\sim7$ galaxies should not be detected in
  the optical, these distributions (\textit{blue lines}) are, as
  expected, approximately centered on 0, with $\sim$50\% above and
  $\sim$50\% below 0 (due to noise).  Also presented here (\textit{red
    lines}) are the distributions expected for lower-redshift
  contaminants in our fields (after excluding the obvious contaminants
  detected at $>$2$\sigma$ in $\geq$1 bands).  The latter
  distributions (\textit{red lines}: for the low-redshift
  contaminants) are estimated through the photometric scatter
  simulations described in \S3.5.5.  A dramatic demonstration of the
  added value from our full $\chi_{opt} ^2$ approach can be seen in
  these figures.  Note the large number of low-redshift galaxies we
  expect to satisfy our $z\sim7$ LBG criterion that are \textit{not}
  excluded by a $2\sigma$ optical non-detection criterion.  This is
  effectively illustrated by the area under the red line.  For fields
  with optical data of limited depths (e.g., the ERS or HUDF09-1
  fields), this number can be comparable to the number of bona-fide
  $z\sim7$ galaxies.  To effectively exclude these low redshift
  contaminants, we derive $\chi_{opt} ^2$ values for sources in our
  selections and then only select sources with small values of
  $\chi_{opt} ^2$, i.e., $\chi_{opt}^2 < \chi_{lim}^2$.  For the
  HUDF09 and HUDF09-1 fields, the limiting values of $\chi_{opt}^2$
  shown in this diagram are for sources selected $\sim$0.7 mag
  brighter than the selection limit ($\sim$28.5 AB mag but this
  depends on the field).  The dashed vertical lines show these
  $\chi_{opt} ^2$ limits.  The sources which exceed the $\chi_{opt}^2$
  limits (but otherwise meet our selection criteria) are given in
  Tables~\ref{tab:hudf09miss7}, ~\ref{tab:hudf09miss8} and
  \ref{tab:possiblecand}.\label{fig:chisq}}
\end{figure}

The best way to control for contamination in the $z\geq7$ samples is
to take advantage of the optical imaging observations over these
samples.  Typically, this is done by imposing an optical non-detection
($<2\sigma$) criterion in one or more bands or in a stack of the
optical data -- since high-redshift sources should not show a
detection at optical wavelengths.  Unfortunately, for the present
$z\gtrsim7$ selections, the optical observations available over our
WFC3/IR fields are sufficiently shallow (in some cases) that it is not
clear that the above procedure will be effective in excluding all low
redshift contaminants from our selections (or even the majority).
This concern is particularly germane for the ERS, HUDF09-1, and
HUDF09-2 fields -- where the optical data are not appreciably deeper
than the near-IR data.

\subsection{D.1.  Definition of $\chi_{opt}^2$}

To make maximal use of the optical imaging data that exist over the
HUDF09 and GOODS fields, we clearly need to use a more sophisticated
procedure.  To this end, we define the quantity $\chi_{opt} ^2$ that
effectively combines information in all the optical imaging data.  We
define $\chi_{opt}^{2}$ as
\begin{equation}
\chi_{opt}^2 = \Sigma_{i} \textrm{SGN}(f_{i}) (f_{i}/\sigma_{i})^2
\end{equation}
where $f_{i}$ is the flux in band $i$ in our smaller scalable
apertures, $\sigma_i$ is the uncertainty in this flux, and
SGN($f_{i}$) is equal to 1 if $f_{i}>0$ and $-1$ if $f_{i}<0$.
Included in the computed $\chi_{opt} ^2$ values are the $B_{435}$,
$V_{606}$, $i_{775}$ band observations for the $z\sim7$
$z_{850}$-dropout selection and the $B_{435}$, $V_{606}$, $i_{775}$,
and $z_{850}$ band observations for the $z\sim8$ $Y_{105}$-dropout
selection.

\subsection{D.2.  $\chi_{opt}^2$ Distribution for $z\gtrsim7$ Galaxies and Low-Redshift Contaminants}

How shall we use the measured $\chi_{opt}^2$ values for sources in our
selection to exclude low-redshift contaminants?  To answer this
question, we must know the $\chi_{opt} ^2$ distribution expected for
bona-fide $z\geq7$ galaxies and also that distribution expected to
come from possible low-redshift contaminants.  The $\chi_{opt} ^2$
distribution for $z\sim7$ candidates is calculated assuming no optical
flux in $z\sim7$ candidates and ideal noise properties.  Meanwhile,
the $\chi_{opt} ^2$ distribution for the low-redshift contaminants is
calculated in the same way as we calculate contamination from
photometric scatter in \S3.5, i.e., starting with the color
distribution for an intermediate magnitude subsample of real sources
in our data and then adding noise to match the fluxes of faint sources
in our search fields.  While discussing the use of $\chi_{opt}^2$ to
control for contamination in our $z\gtrsim6$ selections, we will
assume that we have already removed sources that are detected at
$\geq2\sigma$ in individual images.

We show the model $\chi_{opt} ^2$ distributions in
Figure~\ref{fig:chisq} for the $z\sim7$ selections we perform over the
ERS, HUDF09, and HUDF09-1 fields (which represent a good sampling of
the selections considered here).  For the HUDF09 and HUDF09-1 fields,
these distributions are shown $\sim$0.7 mag from the approximate
selection limit for each field.  As Figure~\ref{fig:chisq} shows, the
$\chi_{opt}^2$ values for $z\gtrsim7$ galaxies are centered on 0 and
have lower values than for low-redshift galaxies.  However, due to the
effects of noise, both $z\gtrsim7$ galaxies and low-redshift
contaminants are spread over a wide range in $\chi_{opt} ^2$ values.

Indeed, it is striking how low the $\chi_{opt}^2$ values are expected
to become for some low-redshift galaxy contaminants that would appear
in our selections.  At relatively low $\chi_{opt}^2$ values of 3, for
example, the total number of contaminants in our ERS or HUDF09-1
selections \textit{(red lines)} is expected to exceed the number of
bona-fide $z\geq7$ galaxies (\textit{blue lines}:
Figure~\ref{fig:chisq}).  A few contaminants are even expected to have
observed $\chi_{opt}^2$ values that are negative, particularly in
those fields with the shallowest optical observations (i.e., the ERS
or HUDF09-1 data).  The prevalence of contaminants to even very low
values of $\chi_{opt} ^2$ underlines the importance of adopting very
conservative optical non-detection criteria in our selection.

\subsection{D.3.  Use of the measured $\chi_{opt}^2$ for selecting $z\gtrsim7$ galaxies}

We only select galaxies to a value of $\chi_{opt}^2$ where we expect
the number of $z\sim7$ galaxies to be greater than the number of
contaminants.  This critical value $\chi_{lim} ^2$ depends on both the
depth of the optical data and the magnitude of the source -- since we
would expect our fainter selections to admit more contamination than
our brighter selections.  We have found that $\chi_{lim} ^2$ values of
5, 3, and 5, respectively, worked well for our HUDF09, HUDF09-1, and
HUDF09-2 data sets, respectively, at brighter
($J_{125},H_{160}\lesssim28.5$) magnitudes; however, at fainter
magnitudes ($J_{125},H_{160}\sim29.2$), $\chi_{lim} ^2$ needed to be
2.5, 1.5, and 2.5, respectively, for the same fields.  For sources
with magnitudes between 28.5 and 29.2, the upper limits on $\chi_{opt}
^2$ are an interpolation between these extremes.  For the small
fraction ($\sim$30\%) of sources in the HUDF09-2 field without
additional optical/ACS coverage from our HUDF09 program (those not
within the P1 field shown in Figure 1), it was appropriate to adopt
the same $\chi_{opt}^2$ limits as for the HUDF09-1 field.

We derive $\chi_{opt}^2$ values for our $z\geq7$ candidates in three
different apertures: $0.35''$-diameter circular apertures, small
$0.18''$-diameter circular apertures, and Kron-style scalable
apertures (Kron factor of 1.2).  The use of three different apertures
allows us to be much more comprehensive in our identification of
possible optical light in sources (which may be spatially very compact
or extended) and therefore rejection of these sources as low-z
contaminants.  No smoothing of the ACS optical images is performed,
prior to making these formal flux measurements.  This ensures that we
obtain the highest possible S/N for these flux measurements --
effectively taking advantage of the sharper PSF in the ACS images.

\subsection{D.4.  Validation of the technique}

The $\chi_{opt} ^2$ distributions observed for our samples also
provide a convenient way of ascertaining the actual number of
bona-fide $z\geq7$ galaxies in our search fields and therefore the
contamination rate (by inference).  Looking at Figure~\ref{fig:chisq},
we see that an exceedingly small number of contaminants are expected
to have $\chi_{opt} ^2$ $<0$ from the simulations, and so essentially
all of sources satisfying our LBG criteria with $\chi_{opt} ^2$ $<0$
are bona-fide $z$$\sim$7-8 galaxies.  However, we also know the
$\chi_{opt} ^2$ $<0$ requirement introduces an incompleteness of
$\sim$50\% (simply because $\sim$50\% of the sources will have
positive values of $\chi_{opt}^2$ and $\sim$50\% negative values --
given our definition of $\chi_{opt}^2$).  Therefore the actual number
of bona-fide $z\geq7$ candidates is approximately double the number of
candidates with $\chi_{opt} ^2$ $<0$.  Moreover, since we quantify the
$\chi_{opt} ^2$ distribution for each search field in three different
apertures (0.35$''$-diameter circular, 0.18$''$-diameter circular,
scalable Kron), this provides us with three different estimates of the
actual number of $z\geq7$ candidates.

Reassuringly, the number of high-redshift candidates in our samples is
in good agreement with the numbers derived from the above
argumentation.  Multiplying the number of $z\sim7$ candidates with
$\chi_{opt}^2<0$ by 2, we expect 28.7, 15.3, 15.3, and 8.7 $z\sim7$
candidates in the HUDF09, HUDF09-1, HUDF09-2, and ERS fields,
respectively, vs. the 29, 17, 14, and 13 candidates in our observed
samples.  Similarly, we expect 28, 15.3, 10.7, and 6 $z\sim8$
candidates, respectively, in the above fields vs. the 24, 14, 15, and
6 candidates in our observed samples.

\subsection{D.5.  Limitations of high-redshift selections not making use of quantities like $\chi_{opt} ^2$}

It is instructive to contrast the present criteria for rejecting
low-redshift contaminants with other frequently-used methods.  Perhaps
the most common of these is one relying just upon the $2\sigma$
detections of the sources in individual optical bands for removing
low-redshift contaminants from high-redshift selections (e.g., Bouwens
et al.\ 2007; Wilkins et al.\ 2011).  The difficulty with this
approach is that it relies upon the optical data being very deep in
each individual band (which is rarely the case).

As an example of the limitations of this approach we found that had we
only used a $2\sigma$-optical rejection criterion in selecting
$z\sim7$ and $z\sim8$ galaxies, a large fraction ($\sim$30-40\%) of
our $z\sim7$-8 samples over the HUDF09-1, HUDF09-2, and ERS fields
would have consisted of contaminants.  Such large contamination rates
significantly reduce the value of the samples for quantitative work.

The red lines in Figure~\ref{fig:chisq} show the $\chi_{opt}^2$
distribution expected for low-redshift contaminants not detected in
any optical band at $>$2$\sigma$.  The area under the red curves in
Figure~\ref{fig:chisq} is comparable to that under the blue curves for
data sets like HUDF09-1 or the ERS fields.  Clearly, contamination
from low-redshift sources can be quite substantial and have a
significant impact in fields with shallower optical data (e.g., recall
the discussion of contamination in the Wilkins et al.\ 2010, 2011
sample in \S4.1.2 and \S4.1.3).  This is perhaps an even more
important concern than uncertainties about the size distribution at
faint magnitudes (see, e.g., Bouwens et al.\ 2004; Grazian et
al.\ 2011).  Careful use of measures like $\chi_{opt} ^2$ is required
to limit contamination (or correct for it statistically in a robust
way).

\section{E.  $z$$\sim$7-8 Samples}

In this section, we provide detailed catalogs and postage stamps for
our $z\sim7$ and $z\sim8$ samples from the three ultra-deep WFC3/IR
HUDF09 fields and ERS fields.

\begin{figure}
\epsscale{1.2}
\plotone{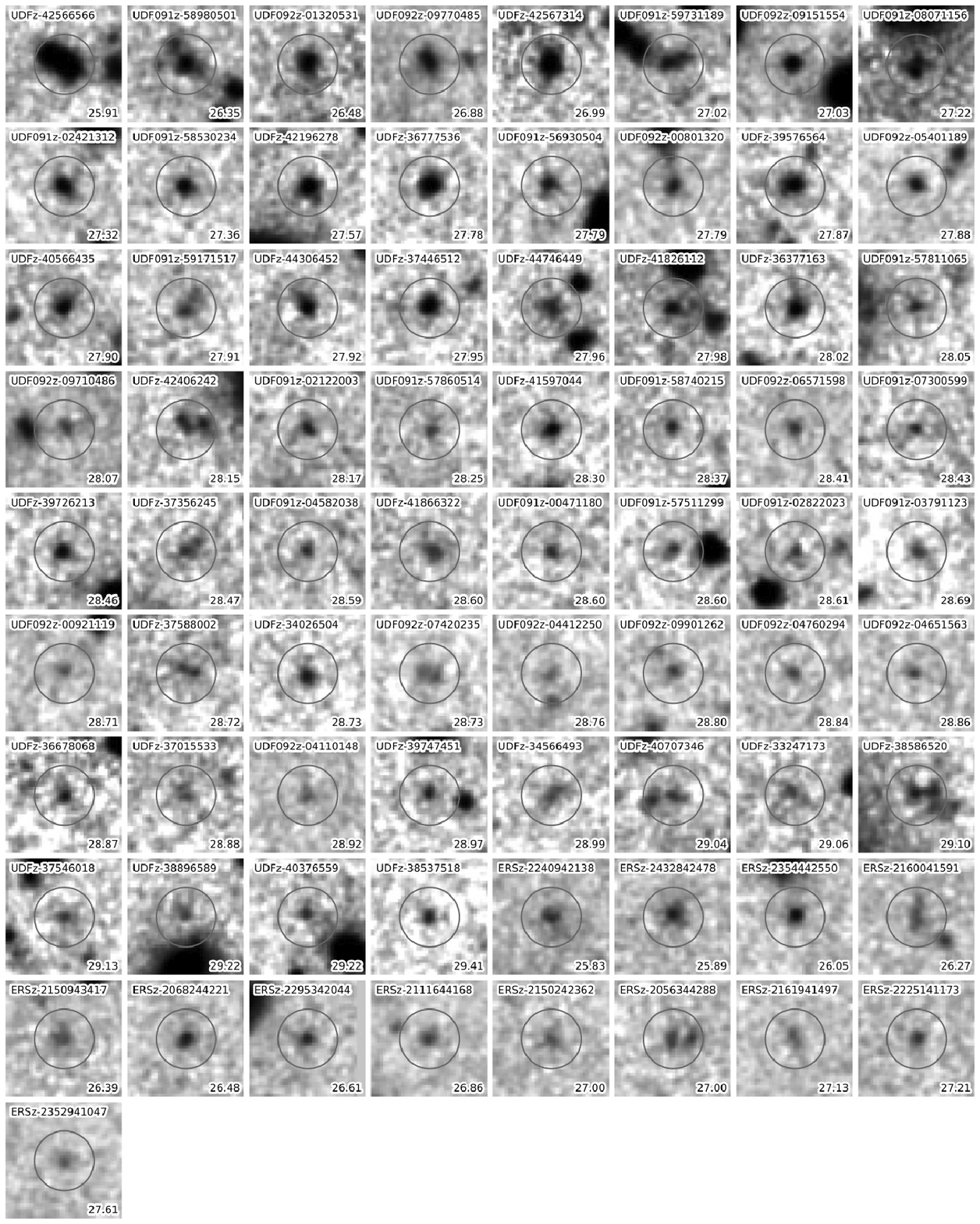}
\caption{Coadded $Y_{105}/Y_{098}+J_{125}+H_{160}$ images
  ($2.34''\times2.34''$) of the 73 $z\sim7$ candidates identified over
  our HUDF09 and ERS search fields.  Each candidate is annotated with
  its source ID and $J_{125}$-band magnitude.  The candidates are
  ordered in terms of their $J_{125}$-band magnitude, with those from
  the HUDF09 program given first and the ERS observations
  second.\label{fig:zstamp}}
\end{figure}

\begin{figure}
\epsscale{1.2}
\plotone{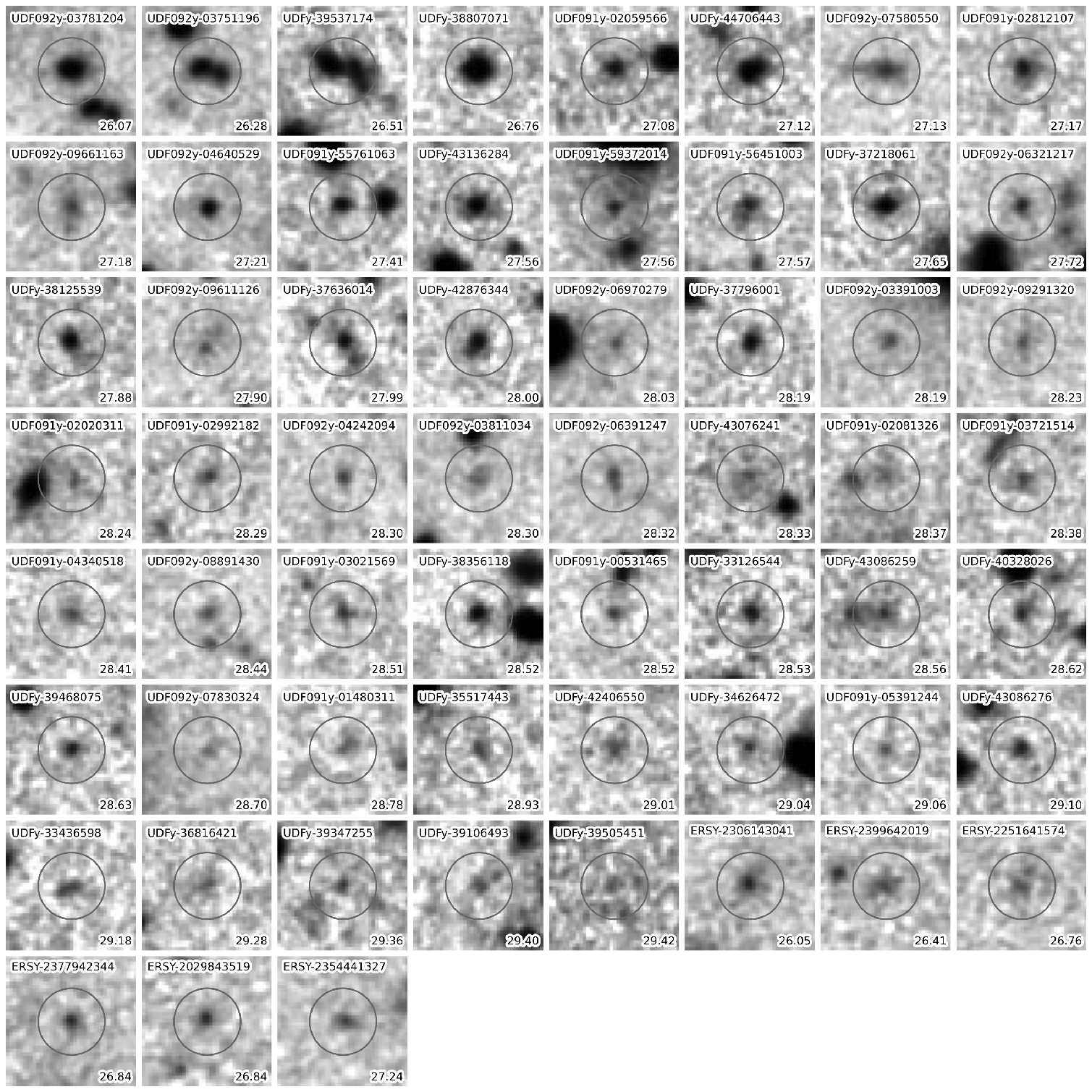}
\caption{Coadded $J_{125}+H_{160}$ images ($2.34''\times2.34''$) of
  the 59 $z\sim8$ identified over our HUDF09 and ERS search fields.
  Each candidate is annotated with its source ID and $H_{160}$-band
  magnitude.  The candidates are ordered in terms of their
  $H_{160}$-band magnitude, with those from the HUDF09 program given
  first and the ERS observations second.\label{fig:ystamp}}
\end{figure}

A description of the HUDF09 $z\sim7$ and $z\sim8$ selections is given
in \S3.2.  A similar description of the ERS $z\sim7$ and $z\sim8$
selections is given in \S3.4.  The coordinates, total magnitudes,
colors, and sizes measured for our $z\sim7$ candidates over our
ultra-deep HUDF09 and ERS fields are provided in
Tables~\ref{tab:z09candlist}-\ref{tab:zcandlist}.  Small postage stamp
cutouts of the $z\sim7$ candidates are given in
Figure~\ref{fig:zstamp}.  The properties of our $z\sim8$ candidates in
the three ultra-deep HUDF09 and ERS fields are given in
Tables~\ref{tab:y09candlist}-\ref{tab:z98candlist}.  Postage stamps of
the $z\sim8$ candidates are given in Figure~\ref{fig:ystamp}.  The
selection of these candidates is described in \S3.2 and \S3.4.

Sources that just missed our HUDF09 or ERS $z\gtrsim7$ selections as a
result of flux in the optical bands ($\chi_{opt}^2$) or colors that
just missed our selection criteria are given in
Tables~\ref{tab:hudf09miss7}-\ref{tab:possiblecand} (see Appendix
D).

\begin{deluxetable*}{ccccccccc}
\tablewidth{17cm}
\tablecolumns{9} 
\tablecaption{$z\sim7$ candidates identified in the ultra-deep HUDF09 WFC3/IR observations over the HUDF.\label{tab:z09candlist}} 
\tablehead{ \colhead{Object ID} &
\colhead{R.A.} & \colhead{Dec} & \colhead{$J_{125}$} & \colhead{$z_{850}-Y_{105}$\tablenotemark{a}} & \colhead{$Y_{105}-J_{125}$\tablenotemark{a}} & \colhead{$J_{125}-H_{160}$} & \colhead{$r_{hl}$\tablenotemark{b}} & \colhead{Ref\tablenotemark{c}}}
\startdata
UDFz-42566566 & 03:32:42.56 & $-$27:46:56.6 & 25.9$\pm$0.0 & 1.6$\pm$0.1 & 0.2$\pm$0.0 & $-$0.1$\pm$0.0 & 0.26$''$ & 1,2,3,4,6,8,9,10,11,12\\
UDFz-42567314 & 03:32:42.56 & $-$27:47:31.4 & 27.0$\pm$0.1 & 1.7$\pm$0.5 & 0.3$\pm$0.1 & $-$0.1$\pm$0.1 & 0.19$''$ & 1,3,4,6,8,9,10,11\\
UDFz-42196278 & 03:32:42.19 & $-$27:46:27.8 & 27.6$\pm$0.1 & 0.9$\pm$0.2 & 0.0$\pm$0.1 & $-$0.1$\pm$0.1 & 0.17$''$ & 8,11,12\\
UDFz-36777536 & 03:32:36.77 & $-$27:47:53.6 & 27.8$\pm$0.1 & 1.0$\pm$0.2 & $-$0.1$\pm$0.1 & $-$0.1$\pm$0.1 & 0.14$''$ & 6,8,12\\
UDFz-44306452 & 03:32:44.30 & $-$27:46:45.2 & 27.9$\pm$0.1 & 0.8$\pm$0.3 & 0.1$\pm$0.2 & $-$0.3$\pm$0.1 & 0.18$''$ & 8,11,12\\
UDFz-37446512 & 03:32:37.44 & $-$27:46:51.2 & 27.9$\pm$0.1 & 1.1$\pm$0.3 & $-$0.1$\pm$0.1 & $-$0.2$\pm$0.1 & 0.14$''$ & 6,8,10,11,12\\
UDFz-39576564 & 03:32:39.57 & $-$27:46:56.4 & 27.9$\pm$0.1 & 1.2$\pm$0.3 & 0.0$\pm$0.1 & $-$0.1$\pm$0.1 & 0.16$''$ & 6,8,10,11,12\\
UDFz-40566435 & 03:32:40.56 & $-$27:46:43.5 & 27.9$\pm$0.1 & 1.3$\pm$0.5 & $-$0.0$\pm$0.1 & $-$0.2$\pm$0.1 & 0.19$''$ & 6,8,9,10,11,12\\
UDFz-44746449 & 03:32:44.74 & $-$27:46:44.9 & 28.0$\pm$0.1 & $>$1.6 & 0.1$\pm$0.2 & $-$0.0$\pm$0.2 & 0.27$''$ & 8,10,11\\
UDFz-36377163 & 03:32:36.37 & $-$27:47:16.3 & 28.0$\pm$0.1 & 1.1$\pm$0.3 & $-$0.1$\pm$0.1 & $-$0.4$\pm$0.2 & 0.15$''$ & 6,8,9,10,11,12\\
UDFz-41826112 & 03:32:41.82 & $-$27:46:11.2 & 28.0$\pm$0.1 & 1.1$\pm$0.6 & 0.0$\pm$0.2 & 0.0$\pm$0.2 & 0.23$''$ & 10,11,12\\
UDFz-42406242 & 03:32:42.40 & $-$27:46:24.2 & 28.1$\pm$0.1 & $>$2.0 & $-$0.1$\pm$0.2 & $-$0.2$\pm$0.2 & 0.24$''$ & 10,11\\
UDFz-41597044 & 03:32:41.59 & $-$27:47:04.4 & 28.3$\pm$0.1 & 1.1$\pm$0.4 & $-$0.1$\pm$0.2 & $-$0.3$\pm$0.2 & 0.15$''$ & 12\\
UDFz-39726213 & 03:32:39.72 & $-$27:46:21.3 & 28.5$\pm$0.1 & $>$1.9 & 0.2$\pm$0.2 & $-$0.3$\pm$0.2 & 0.12$''$ & 6,8,9,10,11,12\\
UDFz-37356245 & 03:32:37.35 & $-$27:46:24.5 & 28.5$\pm$0.2 & 1.1$\pm$0.7 & $-$0.2$\pm$0.2 & $-$0.4$\pm$0.3 & 0.22$''$ & \\
UDFz-41866322 & 03:32:41.86 & $-$27:46:32.2 & 28.6$\pm$0.2 & $>$1.4 & 0.1$\pm$0.3 & $-$0.4$\pm$0.3 & 0.19$''$ & \\
UDFz-34026504 & 03:32:34.02 & $-$27:46:50.4 & 28.7$\pm$0.1 & $>$1.6 & 0.3$\pm$0.2 & $-$0.6$\pm$0.3 & 0.12$''$ & \\
UDFz-37588002 & 03:32:37.58 & $-$27:48:00.2 & 28.7$\pm$0.2 & $>$0.9 & 0.2$\pm$0.4 & 0.3$\pm$0.3 & 0.24$''$ & 10\\
UDFz-36678068 & 03:32:36.67 & $-$27:48:06.8 & 28.9$\pm$0.2 & $>$1.4 & 0.2$\pm$0.3 & $-$0.3$\pm$0.3 & 0.14$''$ & \\
UDFz-37015533 & 03:32:37.01 & $-$27:45:53.3 & 28.9$\pm$0.2 & 0.7$\pm$0.6 & 0.0$\pm$0.3 & $-$1.0$\pm$0.8 & 0.17$''$ & \\
UDFz-34566493 & 03:32:34.56 & $-$27:46:49.3 & 29.0$\pm$0.2 & $>$1.0 & 0.2$\pm$0.4 & 0.1$\pm$0.3 & 0.19$''$ & \\
UDFz-39747451 & 03:32:39.74 & $-$27:47:45.1 & 29.0$\pm$0.2 & $>$1.2 & 0.1$\pm$0.3 & $-$0.1$\pm$0.3 & 0.14$''$ & \\
UDFz-40707346 & 03:32:40.70 & $-$27:47:34.6 & 29.0$\pm$0.2 & $>$1.4 & $-$0.2$\pm$0.3 & $-$0.3$\pm$0.4 & 0.16$''$ & \\
UDFz-37546018 & 03:32:37.54 & $-$27:46:01.8 & 29.1$\pm$0.2 & $>$0.9 & 0.2$\pm$0.4 & $-$0.1$\pm$0.4 & 0.17$''$ & \\
UDFz-33247173 & 03:32:33.24 & $-$27:47:17.3 & 29.1$\pm$0.2 & $>$1.0 & 0.3$\pm$0.4 & $-$0.2$\pm$0.4 & 0.17$''$ & \\
UDFz-38586520 & 03:32:38.58 & $-$27:46:52.0 & 29.1$\pm$0.2 & 1.4$\pm$0.8 & 0.1$\pm$0.3 & 0.0$\pm$0.3 & 0.14$''$ & \\
UDFz-40376559 & 03:32:40.37 & $-$27:46:55.9 & 29.2$\pm$0.2 & $>$1.3 & $-$0.2$\pm$0.4 & $-$0.3$\pm$0.4 & 0.15$''$ & \\
UDFz-38896589 & 03:32:38.89 & $-$27:46:58.9 & 29.2$\pm$0.3 & $>$0.8 & 0.2$\pm$0.5 & $-$0.1$\pm$0.4 & 0.16$''$ & \\
UDFz-38537518 & 03:32:38.53 & $-$27:47:51.8 & 29.4$\pm$0.2 & $>$1.5 & 0.1$\pm$0.3 & $-$0.6$\pm$0.3 & 0.09$''$ & 6\\
\enddata
\tablenotetext{a}{Lower limits on the measured colors are
the $1\sigma$ limits.  Magnitudes are AB.}

\tablenotetext{b}{Half-light radius calculated using SExtractor from the square root of $\chi^2$ image.  The
  median difference between the half-light radii estimates in the
  first-year observations and that estimated from the full two-year
  observations are $\sim$0.02$''$.}
\tablenotetext{c}{References: [1] Bouwens et al.\ (2004), [2] Bouwens \& Illingworth (2006), [3] Labb\'{e} et al.\ (2006), [4] Bouwens et al.\ (2008), [5] Oesch et al.\ (2009), [6] Oesch et al.\ (2010a), [7] Bouwens et al.\ (2010b), [8] McLure et al.\ (2010), [9] Bunker et al.\ (2010), [10] Yan et al.\ (2010), [11] Finkelstein et al.\ (2010), [12] McLure et al.\ (2011)}
\end{deluxetable*}

\begin{deluxetable*}{ccccccccc}
\tablewidth{15cm}
\tablecolumns{9} 
\tablecaption{$z\sim7$ candidates identified in the ultra-deep WFC3/IR observations over HUDF09-1.\label{tab:z051candlist}} 
\tablehead{ \colhead{Object ID} &
\colhead{R.A.} & \colhead{Dec} & \colhead{$J_{125}$} & \colhead{$z_{850}-Y_{105}$\tablenotemark{a}} & \colhead{$Y_{105}-J_{125}$\tablenotemark{a}} & \colhead{$J_{125}-H_{160}$} & \colhead{$r_{hl}$\tablenotemark{b}} & \colhead{Ref\tablenotemark{c}}}
\startdata
UDF091z-58980501 & 03:32:58.98 & $-$27:40:50.1 & 26.3$\pm$0.1 & 1.2$\pm$0.2 & 0.1$\pm$0.1 & $-$0.1$\pm$0.1 & 0.36$''$ & \\
UDF091z-59731189 & 03:32:59.73 & $-$27:41:18.9 & 27.0$\pm$0.1 & $>$2.6 & 0.2$\pm$0.1 & $-$0.0$\pm$0.1 & 0.24$''$ & \\
UDF091z-08071156 & 03:33:08.07 & $-$27:41:15.6 & 27.2$\pm$0.1 & 1.1$\pm$0.3 & $-$0.1$\pm$0.2 & 0.2$\pm$0.2 & 0.19$''$ & \\
UDF091z-02421312 & 03:33:02.42 & $-$27:41:31.2 & 27.3$\pm$0.1 & 2.5$\pm$0.7 & 0.1$\pm$0.1 & $-$0.0$\pm$0.1 & 0.16$''$ & 2\\
UDF091z-58530234 & 03:32:58.53 & $-$27:40:23.4 & 27.4$\pm$0.1 & $>$2.5 & 0.2$\pm$0.2 & $-$0.1$\pm$0.1 & 0.16$''$ & 2\\
UDF091z-56930504 & 03:32:56.93 & $-$27:40:50.4 & 27.8$\pm$0.1 & 1.4$\pm$0.4 & 0.1$\pm$0.2 & $-$0.5$\pm$0.3 & 0.16$''$ & \\
UDF091z-59171517 & 03:32:59.17 & $-$27:41:51.7 & 27.9$\pm$0.2 & 1.8$\pm$0.8 & 0.0$\pm$0.3 & 0.1$\pm$0.2 & 0.22$''$ & \\
UDF091z-57811065 & 03:32:57.81 & $-$27:41:06.5 & 28.1$\pm$0.2 & $>$1.6 & 0.2$\pm$0.3 & 0.0$\pm$0.3 & 0.19$''$ & \\
UDF091z-02122003 & 03:33:02.12 & $-$27:42:00.3 & 28.2$\pm$0.2 & $>$2.0 & $-$0.0$\pm$0.3 & 0.2$\pm$0.2 & 0.16$''$ & \\
UDF091z-57860514 & 03:32:57.86 & $-$27:40:51.4 & 28.2$\pm$0.2 & 1.0$\pm$0.6 & 0.4$\pm$0.4 & $-$0.9$\pm$0.5 & 0.16$''$ & \\
UDF091z-07300599 & 03:33:07.30 & $-$27:40:59.9 & 28.4$\pm$0.2 & $>$1.6 & 0.3$\pm$0.4 & $-$0.4$\pm$0.4 & 0.15$''$ & \\
UDF091z-58740215 & 03:32:58.74 & $-$27:40:21.5 & 28.4$\pm$0.2 & 1.1$\pm$0.5 & $-$0.0$\pm$0.3 & $-$0.1$\pm$0.3 & 0.15$''$ & \\
UDF091z-02822023 & 03:33:02.82 & $-$27:42:02.3 & 28.6$\pm$0.2 & $>$1.3 & 0.4$\pm$0.4 & $-$0.4$\pm$0.4 & 0.13$''$ & \\
UDF091z-57511299 & 03:32:57.51 & $-$27:41:29.9 & 28.6$\pm$0.2 & $>$1.5 & 0.3$\pm$0.4 & $-$0.2$\pm$0.4 & 0.13$''$ & \\
UDF091z-04582038 & 03:33:04.58 & $-$27:42:03.8 & 28.6$\pm$0.2 & 1.1$\pm$0.6 & 0.1$\pm$0.3 & $-$0.3$\pm$0.4 & 0.13$''$ & \\
UDF091z-00471180 & 03:33:00.47 & $-$27:41:18.0 & 28.6$\pm$0.2 & 1.3$\pm$0.7 & 0.1$\pm$0.3 & 0.0$\pm$0.3 & 0.12$''$ & \\
UDF091z-03791123 & 03:33:03.79 & $-$27:41:12.3 & 28.7$\pm$0.2 & $>$1.8 & $-$0.1$\pm$0.3 & $-$0.4$\pm$0.4 & 0.13$''$ & 1\\
\enddata
\tablenotetext{a}{Lower limits on the measured colors are
the $1\sigma$ limits.  Magnitudes are AB.}
\tablenotetext{b}{Half-light radius calculated using SExtractor.  See footnote b from Table~\ref{tab:z09candlist}.}
\tablenotetext{c}{References: [1] Bouwens et al.\ (2008), [2] Wilkins et al. (2011)}
\end{deluxetable*}

\begin{deluxetable*}{ccccccccc}
\tablewidth{15cm}
\tablecolumns{9} 
\tablecaption{$z\sim7$ candidates identified in the ultra-deep WFC3/IR observations over HUDF09-2.\label{tab:z052candlist}} 
\tablehead{ \colhead{Object ID} &
\colhead{R.A.} & \colhead{Dec} & \colhead{$J_{125}$} & \colhead{$z_{850}-Y_{105}$\tablenotemark{a}} & \colhead{$Y_{105}-J_{125}$\tablenotemark{a}} & \colhead{$J_{125}-H_{160}$\tablenotemark{b}} & \colhead{$r_{hl}$\tablenotemark{b}} & \colhead{Ref\tablenotemark{c}}}
\startdata
UDF092z-01320531 & 03:33:01.32 & $-$27:50:53.1 & 26.5$\pm$0.1 & 1.1$\pm$0.3 & 0.2$\pm$0.2 & $-$0.1$\pm$0.1 & 0.25$''$ & \\
UDF092z-09770485 & 03:33:09.77 & $-$27:50:48.5 & 26.9$\pm$0.1 & 1.4$\pm$0.5 & $-$0.0$\pm$0.1 & 0.3$\pm$0.1 & 0.26$''$ & 1\\
UDF092z-09151554 & 03:33:09.15 & $-$27:51:55.4 & 27.0$\pm$0.1 & $>$2.6 & $-$0.1$\pm$0.1 & $-$0.0$\pm$0.1 & 0.17$''$ & 1,2\\
UDF092z-00801320 & 03:33:00.80 & $-$27:51:32.0 & 27.8$\pm$0.1 & 0.8$\pm$0.3 & 0.1$\pm$0.2 & $-$0.0$\pm$0.2 & 0.16$''$ & \\
UDF092z-05401189 & 03:33:05.40 & $-$27:51:18.9 & 27.9$\pm$0.1 & 1.7$\pm$0.9 & 0.3$\pm$0.2 & $-$0.4$\pm$0.2 & 0.12$''$ & 1,2\\
UDF092z-09710486 & 03:33:09.71 & $-$27:50:48.6 & 28.1$\pm$0.1 & $>$1.5 & 0.1$\pm$0.3 & $-$0.1$\pm$0.2 & 0.18$''$ & \\
UDF092z-06571598 & 03:33:06.57 & $-$27:51:59.8 & 28.4$\pm$0.1 & $>$2.1 & $-$0.1$\pm$0.2 & $-$0.3$\pm$0.3 & 0.13$''$ & \\
UDF092z-00921119 & 03:33:00.92 & $-$27:51:11.9 & 28.7$\pm$0.2 & $>$1.4 & 0.0$\pm$0.4 & $-$0.2$\pm$0.4 & 0.15$''$ & \\
UDF092z-07420235 & 03:33:07.42 & $-$27:50:23.5 & 28.7$\pm$0.2 & 1.1$\pm$0.5 & $-$0.3$\pm$0.3 & 0.0$\pm$0.3 & 0.16$''$ & \\
UDF092z-09901262 & 03:33:09.90 & $-$27:51:26.2 & 28.8$\pm$0.2 & $>$1.3 & $-$0.1$\pm$0.3 & 0.2$\pm$0.3 & 0.12$''$ & \\
UDF092z-04412250 & 03:33:04.41 & $-$27:52:25.0 & 28.8$\pm$0.2 & 1.2$\pm$0.9 & 0.1$\pm$0.4 & $-$0.6$\pm$0.6 & 0.15$''$ & \\
UDF092z-04760294 & 03:33:04.76 & $-$27:50:29.4 & 28.8$\pm$0.2 & 1.7$\pm$0.7 & $-$0.1$\pm$0.3 & $-$0.3$\pm$0.4 & 0.11$''$ & \\
UDF092z-04651563 & 03:33:04.65 & $-$27:51:56.3 & 28.9$\pm$0.2 & $>$1.4 & 0.3$\pm$0.3 & $-$0.2$\pm$0.3 & 0.11$''$ & \\
UDF092z-04110148 & 03:33:04.11 & $-$27:50:14.8 & 28.9$\pm$0.2 & 1.2$\pm$0.9 & 0.1$\pm$0.4 & $-$0.0$\pm$0.4 & 0.13$''$ & \\
\enddata
\tablenotetext{a}{Lower limits on the measured colors are
the $1\sigma$ limits.  Magnitudes are AB.}
\tablenotetext{b}{Half-light radius calculated using SExtractor.  See footnote b from Table~\ref{tab:z09candlist}.}
\tablenotetext{c}{References: [1] Wilkins et al. (2011), [2] McLure et al.\ (2011)}
\end{deluxetable*}

\begin{deluxetable*}{ccccccccc}
\tablewidth{15cm}
\tablecolumns{9} 
\tablecaption{$z\sim7$ candidates identified in the deep wide-area ERS observations.\label{tab:zcandlist}} 
\tablehead{ \colhead{Object ID} &
\colhead{R.A.} & \colhead{Dec} & \colhead{$J_{125}$} & \colhead{$z_{850}-J_{125}$\tablenotemark{a}} & \colhead{$Y_{098}-J_{125}$\tablenotemark{a}} & \colhead{$J_{125}-H_{160}$} & \colhead{$r_{hl}$\tablenotemark{b}} & \colhead{Ref\tablenotemark{c}}}
\startdata
ERSz-2240942138 & 03:32:24.09 & $-$27:42:13.8 & 25.8$\pm$0.1 & 1.6$\pm$0.9 & 0.4$\pm$0.3 & $-$0.1$\pm$0.2 & 0.30$''$ & 4,5,6\\
ERSz-2432842478 & 03:32:43.28 & $-$27:42:47.8 & 25.9$\pm$0.1 & 1.2$\pm$0.4 & 0.7$\pm$0.3 & 0.1$\pm$0.1 & 0.24$''$ & 1\\
ERSz-2354442550 & 03:32:35.44 & $-$27:42:55.0 & 26.1$\pm$0.1 & 2.2$\pm$0.8 & 1.1$\pm$0.3 & $-$0.1$\pm$0.1 & 0.16$''$ & \\
ERSz-2160041591 & 03:32:16.00 & $-$27:41:59.1 & 26.3$\pm$0.1 & 1.6$\pm$0.7 & 0.3$\pm$0.3 & $-$0.3$\pm$0.2 & 0.26$''$ & 6\\
ERSz-2150943417 & 03:32:15.09 & $-$27:43:41.7 & 26.4$\pm$0.1 & $>$2.2 & 1.0$\pm$0.5 & $-$0.1$\pm$0.2 & 0.22$''$ & \\
ERSz-2068244221 & 03:32:06.82 & $-$27:44:22.1 & 26.5$\pm$0.1 & 1.1$\pm$0.4 & $-$0.0$\pm$0.2 & $-$0.4$\pm$0.2 & 0.15$''$ & 6\\
ERSz-2295342044 & 03:32:29.53 & $-$27:42:04.4 & 26.6$\pm$0.1 & 1.4$\pm$0.6 & 0.3$\pm$0.3 & $-$0.1$\pm$0.2 & 0.17$''$ & 3,4,5,6\\
ERSz-2111644168 & 03:32:11.16 & $-$27:44:16.8 & 26.9$\pm$0.1 & $>$2.0 & 1.1$\pm$0.5 & $-$0.3$\pm$0.2 & 0.14$''$ & \\
ERSz-2150242362 & 03:32:15.02 & $-$27:42:36.2 & 27.0$\pm$0.2 & $>$1.2 & 0.4$\pm$0.5 & 0.4$\pm$0.3 & 0.20$''$ & \\
ERSz-2056344288 & 03:32:05.63 & $-$27:44:28.8 & 27.0$\pm$0.2 & 1.4$\pm$0.7 & 0.2$\pm$0.4 & 0.3$\pm$0.2 & 0.16$''$ & \\
ERSz-2161941497 & 03:32:16.19 & $-$27:41:49.7 & 27.1$\pm$0.2 & $>$1.4 & 0.6$\pm$0.5 & $-$0.1$\pm$0.3 & 0.17$''$ & 6\\
ERSz-2225141173 & 03:32:22.51 & $-$27:41:17.3 & 27.2$\pm$0.2 & 1.1$\pm$0.5 & 0.1$\pm$0.3 & 0.1$\pm$0.3 & 0.13$''$ & \\
ERSz-2352941047 & 03:32:35.29 & $-$27:41:04.7 & 27.6$\pm$0.2 & 1.1$\pm$0.9 & 0.3$\pm$0.5 & 0.2$\pm$0.3 & 0.14$''$ & \\
\enddata 
\tablenotetext{a}{Lower limits on the measured colors are
the $1\sigma$ limits.  Magnitudes are AB.}
\tablenotetext{b}{Half-light radius calculated using SExtractor.  See footnote b from Table~\ref{tab:z09candlist}.}
\tablenotetext{c}{References: [1] Bouwens et al.\ (2010c), [2] Hickey et al.\ (2010),
[3] Castellano et al.\ (2010a), [4] Wilkins et al.\ (2010), [5] Wilkins et al.\ (2011), [6] McLure et al.\ (2011)}
\end{deluxetable*}

\begin{deluxetable*}{cccccccc}
\tablewidth{14cm}
\tablecolumns{8} 
\tablecaption{$z\sim8$ candidates identified in the ultra-deep WFC3/IR observations in the HUDF.\label{tab:y09candlist}} 
\tablehead{ \colhead{Object ID} &
\colhead{R.A.} & \colhead{Dec} & \colhead{$H_{160}$} & \colhead{$Y_{105}-J_{125}$\tablenotemark{a}} & \colhead{$J_{125}-H_{160}$} & \colhead{$r_{hl}$\tablenotemark{b}} & \colhead{Ref\tablenotemark{c}}}
\startdata
UDFy-39537174 & 03:32:39.53 & $-$27:47:17.4 & 26.5$\pm$0.0 & 0.7$\pm$0.1 & 0.1$\pm$0.0 & 0.32$''$ & 1,3,6,8,9,10,11,13\\
UDFy-38807071 & 03:32:38.80 & $-$27:47:07.1 & 26.8$\pm$0.0 & 0.5$\pm$0.1 & 0.1$\pm$0.0 & 0.15$''$ & 1,2,3,4,5,6,8,9,10,11,13\\
UDFy-44706443 & 03:32:44.70 & $-$27:46:44.3 & 27.1$\pm$0.1 & 0.9$\pm$0.2 & $-$0.1$\pm$0.1 & 0.19$''$ & 6,8,9,10,11,13\\
UDFy-43136284 & 03:32:43.13 & $-$27:46:28.4 & 27.6$\pm$0.1 & 0.5$\pm$0.2 & 0.1$\pm$0.1 & 0.18$''$ & 6,8,9,10,11,13\\
UDFy-37218061 & 03:32:37.21 & $-$27:48:06.1 & 27.6$\pm$0.1 & 0.6$\pm$0.2 & $-$0.0$\pm$0.1 & 0.17$''$ & 6,8,9,10,11,13\\
UDFy-38125539 & 03:32:38.12 & $-$27:45:53.9 & 27.9$\pm$0.1 & 1.9$\pm$0.7 & 0.1$\pm$0.1 & 0.17$''$ & 7,8,9,10,11,12,13\\
UDFy-42876344 & 03:32:42.87 & $-$27:46:34.4 & 28.0$\pm$0.1 & 0.8$\pm$0.2 & 0.0$\pm$0.1 & 0.16$''$ & 7,8,9,10,11,12,13\\
UDFy-37636014 & 03:32:37.63 & $-$27:46:01.4 & 28.0$\pm$0.1 & 1.8$\pm$0.6 & $-$0.1$\pm$0.1 & 0.22$''$ & 7,8,9,10,11\\
UDFy-37796001 & 03:32:37.79 & $-$27:46:00.1 & 28.2$\pm$0.1 & 1.4$\pm$0.4 & $-$0.1$\pm$0.1 & 0.15$''$ & 7,8,9,10,11,12\\
UDFy-43076241 & 03:32:43.07 & $-$27:46:24.1 & 28.3$\pm$0.2 & 1.4$\pm$0.8 & $-$0.1$\pm$0.3 & 0.28$''$ & 8,13\\
UDFy-38356118 & 03:32:38.35 & $-$27:46:11.8 & 28.5$\pm$0.1 & 0.5$\pm$0.2 & $-$0.2$\pm$0.1 & 0.12$''$ & 9,10,11,13\\
UDFy-33126544 & 03:32:33.12 & $-$27:46:54.4 & 28.5$\pm$0.2 & 0.7$\pm$0.4 & $-$0.1$\pm$0.2 & 0.14$''$ & 8,9,12,13\\
UDFy-39468075 & 03:32:39.46 & $-$27:48:07.5 & 28.6$\pm$0.1 & $>$1.7 & 0.5$\pm$0.2 & 0.12$''$ & \\
UDFy-43086259 & 03:32:43.08 & $-$27:46:25.9 & 28.6$\pm$0.2 & 0.8$\pm$0.5 & $-$0.1$\pm$0.2 & 0.20$''$ & \\
UDFy-40328026 & 03:32:40.32 & $-$27:48:02.6 & 28.6$\pm$0.2 & 1.2$\pm$0.6 & 0.0$\pm$0.2 & 0.17$''$ & 11\\
UDFy-35517443 & 03:32:35.51 & $-$27:47:44.3 & 28.9$\pm$0.2 & $>$1.4 & 0.2$\pm$0.3 & 0.15$''$ & \\
UDFy-34626472 & 03:32:34.62 & $-$27:46:47.2 & 29.0$\pm$0.2 & 0.9$\pm$0.8 & 0.2$\pm$0.3 & 0.14$''$ & \\
UDFy-42406550 & 03:32:42.40 & $-$27:46:55.0 & 29.0$\pm$0.2 & 1.1$\pm$0.7 & $-$0.0$\pm$0.3 & 0.16$''$ & \\
UDFy-43086276 & 03:32:43.08 & $-$27:46:27.6 & 29.1$\pm$0.2 & 1.2$\pm$0.5 & $-$0.4$\pm$0.2 & 0.12$''$ & 7,8,10,11,12\\
UDFy-33436598 & 03:32:33.43 & $-$27:46:59.8 & 29.2$\pm$0.2 & $>$1.6 & 0.2$\pm$0.3 & 0.13$''$ & 14\\
UDFy-36816421 & 03:32:36.81 & $-$27:46:42.1 & 29.3$\pm$0.3 & $>$1.3 & $-$0.2$\pm$0.4 & 0.18$''$ & \\
UDFy-39347255 & 03:32:39.34 & $-$27:47:25.5 & 29.4$\pm$0.2 & 0.8$\pm$0.7 & 0.2$\pm$0.4 & 0.11$''$ & \\
UDFy-39106493 & 03:32:39.10 & $-$27:46:49.3 & 29.4$\pm$0.3 & 0.9$\pm$0.5 & $-$0.7$\pm$0.3 & 0.19$''$ & \\
UDFy-39505451 & 03:32:39.50 & $-$27:45:45.1 & 29.4$\pm$0.4 & 1.0$\pm$1.0 & $-$0.2$\pm$0.5 & 0.15$''$ & \\
\enddata
\tablenotetext{a}{Lower limits on the measured colors are
the $1\sigma$ limits.  Magnitudes are AB.}
\tablenotetext{b}{Half-light radius calculated using SExtractor.  See footnote b from Table~\ref{tab:z09candlist}.}
\tablenotetext{c}{References: [1] Bouwens et al.\ (2004), [2] Bouwens \& Illingworth (2006), [3] Labb\'{e} et al.\ (2006), [4] Bouwens et al.\ (2008), [5] Oesch et al.\ (2009), [6] Oesch et al.\ (2010a), [7] Bouwens et al.\ (2010b), [8] McLure et al.\ (2010), [9] Bunker et al.\ (2010), [10] Yan et al.\ (2010), [11] Finkelstein et al.\ (2010), [12] Lorenzoni et al.\ (2011), [13] McLure et al.\ (2011), [14] Bouwens et al.\ (2011)}
\end{deluxetable*}

\begin{deluxetable*}{cccccccc}
\tablewidth{14cm}
\tablecolumns{8} 
\tablecaption{$z\sim8$ candidates identified in the ultra-deep WFC3/IR observations over HUDF09-1.\label{tab:y051candlist}} 
\tablehead{ \colhead{Object ID} &
\colhead{R.A.} & \colhead{Dec} & \colhead{$H_{160}$\tablenotemark{a}} & \colhead{$Y_{105}-J_{125}$\tablenotemark{b}} & \colhead{$J_{125}-H_{160}$\tablenotemark{a}} & \colhead{$r_{hl}$\tablenotemark{c}} & \colhead{Ref\tablenotemark{d}}}
\startdata
UDF091y-02059566 & 03:33:02.05 & $-$27:39:56.6 & 27.1$\pm$0.1 & 1.0$\pm$0.3 & 0.1$\pm$0.1 & 0.19$''$ & \\
UDF091y-02812107 & 03:33:02.81 & $-$27:42:10.7 & 27.2$\pm$0.1 & 0.7$\pm$0.3 & 0.2$\pm$0.1 & 0.19$''$ & \\
UDF091y-55761063 & 03:32:55.76 & $-$27:41:06.3 & 27.4$\pm$0.1 & 0.5$\pm$0.2 & 0.2$\pm$0.1 & 0.14$''$ & \\
UDF091y-56451003 & 03:32:56.45 & $-$27:41:00.3 & 27.6$\pm$0.1 & 0.6$\pm$0.2 & $-$0.1$\pm$0.2 & 0.19$''$ & \\
UDF091y-59372014 & 03:32:59.37 & $-$27:42:01.4 & 27.6$\pm$0.1 & 0.6$\pm$0.3 & 0.1$\pm$0.2 & 0.19$''$ & \\
UDF091y-02020311 & 03:33:02.02 & $-$27:40:31.1 & 28.2$\pm$0.3 & 1.4$\pm$1.0 & $-$0.0$\pm$0.4 & 0.20$''$ & \\
UDF091y-02992182 & 03:33:02.99 & $-$27:42:18.2 & 28.3$\pm$0.2 & 0.8$\pm$0.5 & $-$0.0$\pm$0.3 & 0.15$''$ & 1\\
UDF091y-04340518 & 03:33:04.34 & $-$27:40:51.8 & 28.4$\pm$0.2 & 0.8$\pm$0.6 & 0.2$\pm$0.3 & 0.14$''$ & \\
UDF091y-02081326 & 03:33:02.08 & $-$27:41:32.6 & 28.4$\pm$0.3 & $>$1.6 & $-$0.2$\pm$0.4 & 0.20$''$ & \\
UDF091y-03721514 & 03:33:03.72 & $-$27:41:51.4 & 28.4$\pm$0.3 & 1.0$\pm$0.5 & $-$0.4$\pm$0.3 & 0.18$''$ & \\
UDF091y-00531465 & 03:33:00.53 & $-$27:41:46.5 & 28.5$\pm$0.2 & 0.5$\pm$0.4 & 0.2$\pm$0.3 & 0.12$''$ & \\
UDF091y-03021569 & 03:33:03.02 & $-$27:41:56.9 & 28.5$\pm$0.2 & 1.3$\pm$0.6 & $-$0.4$\pm$0.3 & 0.14$''$ & 1\\
UDF091y-01480311 & 03:33:01.48 & $-$27:40:31.1 & 28.8$\pm$0.3 & 1.2$\pm$1.0 & 0.2$\pm$0.4 & 0.13$''$ & \\
UDF091y-05391244 & 03:33:05.39 & $-$27:41:24.4 & 29.1$\pm$0.4 & $>$1.4 & $-$0.3$\pm$0.4 & 0.12$''$ & \\
\enddata
\tablenotetext{a}{See footnote b in Table~\ref{tab:z051candlist}.}
\tablenotetext{b}{Lower limits on the measured colors are
the $1\sigma$ limits.  Magnitudes are AB.}
\tablenotetext{c}{Half-light radius calculated using SExtractor.  See footnote b from Table~\ref{tab:z09candlist}.}
\tablenotetext{d}{References: [1] Lorenzoni et al.\ (2011)}
\end{deluxetable*}

\begin{deluxetable*}{cccccccc}
\tablewidth{14cm}
\tablecolumns{8} 
\tablecaption{$z\sim8$ candidates identified in the ultra-deep WFC3/IR observations over HUDF09-2.\label{tab:y052candlist}} 
\tablehead{ \colhead{Object ID} &
\colhead{R.A.} & \colhead{Dec} & \colhead{$H_{160}$} & \colhead{$Y_{105}-J_{125}$\tablenotemark{a}} & \colhead{$J_{125}-H_{160}$} & \colhead{$r_{hl}$\tablenotemark{b}} & \colhead{Ref\tablenotemark{c}}}
\startdata
UDF092y-03781204\tablenotemark{d} & 03:33:03.78 & $-$27:51:20.4 & 26.1$\pm$0.0 & 0.6$\pm$0.1 & 0.1$\pm$0.0 & 0.20$''$ & 1,3\\
UDF092y-03751196\tablenotemark{d} & 03:33:03.75 & $-$27:51:19.6 & 26.3$\pm$0.0 & 0.7$\pm$0.1 & $-$0.0$\pm$0.1 & 0.23$''$ & 3\\
UDF092y-07580550 & 03:33:07.58 & $-$27:50:55.0 & 27.1$\pm$0.1 & $>$2.4 & 0.3$\pm$0.1 & 0.22$''$ & 3\\
UDF092y-04640529\tablenotemark{d} & 03:33:04.64 & $-$27:50:52.9 & 27.2$\pm$0.1 & 0.8$\pm$0.2 & $-$0.1$\pm$0.1 & 0.15$''$ & 3\\
UDF092y-09661163 & 03:33:09.66 & $-$27:51:16.3 & 27.2$\pm$0.1 & 1.3$\pm$0.6 & 0.5$\pm$0.2 & 0.24$''$ & 2\\
UDF092y-06321217 & 03:33:06.32 & $-$27:51:21.7 & 27.7$\pm$0.1 & 0.6$\pm$0.2 & 0.0$\pm$0.1 & 0.15$''$ & \\
UDF092y-09611126 & 03:33:09.61 & $-$27:51:12.6 & 27.9$\pm$0.2 & 0.8$\pm$0.5 & 0.1$\pm$0.2 & 0.22$''$ & \\
UDF092y-06970279 & 03:33:06.97 & $-$27:50:27.9 & 28.0$\pm$0.2 & $>$1.7 & $-$0.1$\pm$0.3 & 0.22$''$ & \\
UDF092y-09291320 & 03:33:09.29 & $-$27:51:32.0 & 28.2$\pm$0.2 & 0.8$\pm$0.5 & $-$0.2$\pm$0.3 & 0.20$''$ & \\
UDF092y-03391003 & 03:33:03.39 & $-$27:51:00.3 & 28.2$\pm$0.2 & 1.2$\pm$0.6 & 0.1$\pm$0.2 & 0.14$''$ & \\
UDF092y-04242094 & 03:33:04.24 & $-$27:52:09.4 & 28.3$\pm$0.1 & 1.0$\pm$0.6 & 0.5$\pm$0.2 & 0.11$''$ & \\
UDF092y-03811034 & 03:33:03.81 & $-$27:51:03.4 & 28.3$\pm$0.2 & 0.5$\pm$0.4 & 0.1$\pm$0.2 & 0.16$''$ & \\
UDF092y-06391247 & 03:33:06.39 & $-$27:51:24.7 & 28.3$\pm$0.2 & 0.7$\pm$0.3 & $-$0.5$\pm$0.2 & 0.16$''$ & \\
UDF092y-08891430 & 03:33:08.89 & $-$27:51:43.0 & 28.4$\pm$0.2 & 0.5$\pm$0.5 & 0.4$\pm$0.3 & 0.12$''$ & \\
UDF092y-01971379 & 03:33:01.97 & $-$27:51:37.9 & 28.4$\pm$0.2 & 0.8$\pm$0.6 & 0.0$\pm$0.3 & 0.20$''$ & \\
UDF092y-07830324 & 03:33:07.83 & $-$27:50:32.4 & 28.7$\pm$0.3 & $>$1.1 & 0.0$\pm$0.4 & 0.16$''$ & \\
\enddata
\tablenotetext{a}{Lower limits on the measured colors are
the $1\sigma$ limits.  Magnitudes are AB.}
\tablenotetext{b}{Half-light radius calculated using SExtractor.  See footnote b from Table~\ref{tab:z09candlist}.}
\tablenotetext{c}{References: [1] Wilkins et al. (2011), 
[2] Lorenzoni et al.\ (2011), [3] McLure et al.\ (2011)}
\tablenotetext{d}{All three sources here are within 30$''$ of each
  other (two being within $3''$ of each other), have consistent
  $Y_{105}-J_{125}$ colors (and thus redshifts), and are among the
  brightest $z\sim7.5$ galaxies in the field.  They would appear to be
  part of a possible overdensity.}
\end{deluxetable*}

\begin{deluxetable*}{ccccccccc}
\tablewidth{15cm}
\tablecolumns{9} 
\tablecaption{$z\sim8$ candidates identified in the deep wide-area ERS observations.\label{tab:z98candlist}} 
\tablehead{ \colhead{Object ID} &
\colhead{R.A.} & \colhead{Dec} & \colhead{$H_{160}$} & \colhead{$z_{850}-J_{125}$\tablenotemark{a}} & \colhead{$Y_{098}-J_{125}$\tablenotemark{a}} & \colhead{$J_{125}-H_{160}$} & \colhead{$r_{hl}$\tablenotemark{b}} & \colhead{Ref}}
\startdata
ERSY-2306143041\tablenotemark{c} & 03:32:30.61 & $-$27:43:04.1 & 26.0$\pm$0.1 & 2.0$\pm$0.8 & 1.4$\pm$0.7 & 0.3$\pm$0.2 & 0.23$''$ & \\
ERSY-2399642019 & 03:32:39.96 & $-$27:42:01.9 & 26.4$\pm$0.2 & $>$1.4 & $>$1.5 & 0.3$\pm$0.2 & 0.22$''$ & \\
ERSY-2251641574 & 03:32:25.16 & $-$27:41:57.4 & 26.8$\pm$0.2 & $>$1.7 & $>$1.5 & 0.0$\pm$0.3 & 0.21$''$ & \\
ERSY-2029843519 & 03:32:02.98 & $-$27:43:51.9 & 26.8$\pm$0.2 & $>$1.7 & $>$1.6 & 0.1$\pm$0.3 & 0.14$''$ & 1,2\\
ERSY-2377942344\tablenotemark{d} & 03:32:37.79 & $-$27:42:34.4 & 26.8$\pm$0.2 & $>$1.9 & $>$1.7 & 0.2$\pm$0.2 & 0.14$''$ & \\
ERSY-2354441327 & 03:32:35.44 & $-$27:41:32.7 & 27.2$\pm$0.2 & $>$1.6 & $>$1.7 & $-$0.1$\pm$0.3 & 0.14$''$ & 2\\

\enddata 
\tablenotetext{a}{Lower limits on the measured colors are
the $1\sigma$ limits.  Magnitudes are AB.}
\tablenotetext{b}{Half-light radius calculated using SExtractor.  See footnote b from Table~\ref{tab:z09candlist}.}
\tablenotetext{c}{References: [1] Lorenzoni et al. (2011), [2] McLure et al.\ (2011)}
\tablenotetext{c}{Source is formally ``detected'' at $\sim1\sigma$ in the $V_{606}$ band and $\sim1.3\sigma$ in the $z_{850}$ band, so is more likely to be at low-redshift than the other candidates}
\tablenotetext{d}{Source has $H_{160}-3.6\mu$ and $H_{160}-4.5\mu$ colors of $\sim$2 mag, which is quite red for a $z\sim8$ galaxy and may indicate it has a redshift $z\sim2-3$.}
\end{deluxetable*}

\begin{deluxetable*}{ccccccccc}
\tablewidth{17cm}
\tablecolumns{9} 
\tablecaption{Possible $z$$\sim$7 candidates over the HUDF09 fields that did not meet our selection criteria.\tablenotemark{a}\label{tab:hudf09miss7}} 
\tablehead{ \colhead{Object ID} &
\colhead{R.A.} & \colhead{Dec} & \colhead{$J_{125}$} & \colhead{$z_{850}-Y_{105}$\tablenotemark{b}} & \colhead{$Y_{105}-J_{125}$\tablenotemark{b}} & \colhead{$J_{125}-H_{160}$} & \colhead{$r_{hl}$\tablenotemark{c}} & \colhead{Ref\tablenotemark{d}}}
\startdata
UDF092z-01101160 & 03:33:01.10 & $-$27:51:16.0 & 25.9$\pm$0.0 & 0.8$\pm$0.1 & 0.2$\pm$0.1 & $-$0.0$\pm$0.0 & 0.26$''$ & 4\\
UDF091z-56701076 & 03:32:56.70 & $-$27:41:07.6 & 26.7$\pm$0.1 & 1.4$\pm$0.3 & 0.4$\pm$0.1 & $-$0.1$\pm$0.1 & 0.20$''$ & 3\\
UDF091z-59481461 & 03:32:59.48 & $-$27:41:46.1 & 27.6$\pm$0.1 & 1.1$\pm$0.5 & 0.1$\pm$0.2 & $-$0.6$\pm$0.3 & 0.25$''$ & \\
UDF091z-59270343 & 03:32:59.27 & $-$27:40:34.3 & 27.8$\pm$0.1 & $>$2.0 & 0.2$\pm$0.2 & $-$0.2$\pm$0.2 & 0.19$''$ & \\
UDFz-37296175 & 03:32:37.29 & $-$27:46:17.5 & 28.0$\pm$0.1 & $>$1.7 & 0.1$\pm$0.2 & 0.2$\pm$0.2 & 0.28$''$ & 2\\
UDF091z-58830381 & 03:32:58.83 & $-$27:40:38.1 & 28.1$\pm$0.1 & 1.0$\pm$0.4 & 0.1$\pm$0.2 & 0.1$\pm$0.2 & 0.15$''$ & \\
UDF092z-07961350 & 03:33:07.96 & $-$27:51:35.0 & 28.1$\pm$0.1 & 1.7$\pm$0.7 & 0.1$\pm$0.2 & 0.1$\pm$0.1 & 0.12$''$ & \\
UDFz-36916516 & 03:32:36.91 & $-$27:46:51.6 & 28.2$\pm$0.2 & $>$1.1 & 0.2$\pm$0.4 & 0.0$\pm$0.3 & 0.44$''$ & \\
UDFz-36467324 & 03:32:36.46 & $-$27:47:32.4 & 28.3$\pm$0.1 & 1.6$\pm$0.6 & $-$0.1$\pm$0.2 & 0.4$\pm$0.1 & 0.16$''$ & 1,2\\
UDF092z-08551377 & 03:33:08.55 & $-$27:51:37.7 & 28.3$\pm$0.2 & $>$1.2 & $-$0.1$\pm$0.3 & 0.2$\pm$0.3 & 0.22$''$ & \\
UDF091z-04781529 & 03:33:04.78 & $-$27:41:52.9 & 28.3$\pm$0.2 & $>$1.5 & 0.2$\pm$0.4 & $-$0.1$\pm$0.4 & 0.19$''$ & \\
UDF091z-59561338 & 03:32:59.56 & $-$27:41:33.8 & 28.3$\pm$0.2 & 1.8$\pm$0.6 & $-$0.3$\pm$0.3 & $-$0.4$\pm$0.4 & 0.19$''$ & \\
UDF091z-01020006 & 03:33:01.02 & $-$27:40:00.6 & 28.4$\pm$0.2 & 1.2$\pm$0.7 & $-$0.1$\pm$0.3 & $-$0.5$\pm$0.5 & 0.18$''$ & \\
UDFz-42247087 & 03:32:42.24 & $-$27:47:08.7 & 28.5$\pm$0.2 & 1.5$\pm$0.8 & $-$0.2$\pm$0.2 & 0.4$\pm$0.2 & 0.20$''$ & 2\\
UDF091z-59451483 & 03:32:59.45 & $-$27:41:48.3 & 28.6$\pm$0.2 & $>$1.4 & $-$0.1$\pm$0.4 & $-$0.1$\pm$0.4 & 0.16$''$ & \\
UDF091z-02071130 & 03:33:02.07 & $-$27:41:13.0 & 28.7$\pm$0.2 & 0.7$\pm$0.5 & $-$0.0$\pm$0.4 & 0.0$\pm$0.4 & 0.15$''$ & \\
UDF092z-10561177 & 03:33:10.56 & $-$27:51:17.7 & 28.7$\pm$0.2 & 0.9$\pm$0.6 & $-$0.1$\pm$0.3 & 0.1$\pm$0.3 & 0.15$''$ & \\
UDF092z-09690571 & 03:33:09.69 & $-$27:50:57.1 & 28.7$\pm$0.3 & $>$0.8 & 0.1$\pm$0.5 & $-$0.0$\pm$0.4 & 0.18$''$ & \\
UDFz-40506512 & 03:32:40.50 & $-$27:46:51.2 & 28.8$\pm$0.2 & $>$1.3 & 0.3$\pm$0.3 & $-$0.6$\pm$0.3 & 0.14$''$ & \\
UDF091z-59842044 & 03:32:59.84 & $-$27:42:04.4 & 28.8$\pm$0.2 & 1.4$\pm$0.9 & 0.2$\pm$0.4 & 0.1$\pm$0.3 & 0.14$''$ & \\
UDF091z-05751079 & 03:33:05.75 & $-$27:41:07.9 & 28.9$\pm$0.2 & $>$1.2 & 0.3$\pm$0.5 & $-$0.0$\pm$0.4 & 0.13$''$ & \\
UDF092z-07192310 & 03:33:07.19 & $-$27:52:31.0 & 28.9$\pm$0.3 & $>$0.9 & 0.2$\pm$0.5 & $-$0.0$\pm$0.4 & 0.15$''$ & \\
UDF092z-07582097 & 03:33:07.58 & $-$27:52:09.7 & 28.9$\pm$0.3 & 0.9$\pm$1.1 & 0.3$\pm$0.6 & 0.3$\pm$0.4 & 0.16$''$ & \\
UDF092z-06900317 & 03:33:06.90 & $-$27:50:31.7 & 29.0$\pm$0.2 & $>$1.4 & $-$0.2$\pm$0.4 & 0.4$\pm$0.3 & 0.14$''$ & \\
UDF092z-06960354 & 03:33:06.96 & $-$27:50:35.4 & 29.0$\pm$0.3 & $>$0.8 & 0.2$\pm$0.6 & 0.2$\pm$0.4 & 0.14$''$ & \\
UDF092z-04992391 & 03:33:04.99 & $-$27:52:39.1 & 29.0$\pm$0.3 & $>$0.9 & 0.1$\pm$0.5 & 0.3$\pm$0.4 & 0.15$''$ & \\
UDFz-41357078 & 03:32:41.35 & $-$27:47:07.8 & 29.2$\pm$0.2 & 0.9$\pm$0.9 & $-$0.1$\pm$0.4 & 0.1$\pm$0.3 & 0.16$''$ & \\
UDFz-37316420 & 03:32:37.31 & $-$27:46:42.0 & 29.2$\pm$0.2 & 1.5$\pm$0.6 & $-$0.5$\pm$0.3 & $-$0.9$\pm$0.7 & 0.15$''$ & \\
\enddata
\tablenotetext{a}{Each of the sources here satisfy the color criteria
  for our dropout selections, but show slight detections in the
  $\chi_{opt} ^2$ images generated from the $B_{435}$, $V_{606}$,
  $i_{775}$ data.  These sources therefore do not make it into our
  Lyman-Break galaxy samples.  See \S3.3 and Appendix D for a
  discussion of $\chi_{opt} ^2$ and our selection criteria.  While a
  few of these sources will correspond to $z\gtrsim7$ galaxies, the majority will not according to our simulations (Appendix D).  For a few
  sources, the $\chi_{opt}^2$ detection comes
  almost entirely from the $i_{775}$-band, suggesting that some might
  nevertheless have redshifts slightly above $z$$\sim$6.}
\tablenotetext{b}{Lower limits on the measured colors are
the $1\sigma$ limits.  Magnitudes are AB.}
\tablenotetext{c}{Half-light radius calculated using SExtractor}
\tablenotetext{d}{References: [1] McLure et al.\ (2010), [2]
  Finkelstein et al.\ (2010), [3] Wilkins et al. (2011), [4] McLure
  et al.\ (2011).}

\end{deluxetable*}

\begin{deluxetable*}{ccccccccc}
\tablewidth{17cm}
\tablecolumns{9} 
\tablecaption{Possible $z$$\sim$8 candidates over the HUDF09 fields that did not meet our selection criteria.\tablenotemark{a}\label{tab:hudf09miss8}} 
\tablehead{ \colhead{Object ID} &
\colhead{R.A.} & \colhead{Dec} & \colhead{$J_{125}$} & \colhead{$Y_{105}-J_{125}$\tablenotemark{b}} & \colhead{$J_{125}-H_{160}$} & \colhead{$r_{hl}$\tablenotemark{c}} & \colhead{Ref\tablenotemark{d}}}
\startdata
UDF092y-10571166 & 03:33:10.57 & $-$27:51:16.6 & 27.3$\pm$0.1 & 0.6$\pm$0.2 & 0.4$\pm$0.1 & 0.15$"$ & \\
UDF092y-07492060 & 03:33:07.49 & $-$27:52:06.0 & 27.5$\pm$0.1 & $>$1.9 & 0.5$\pm$0.2 & 0.20$"$ & \\
UDF091y-06330449 & 03:33:06.33 & $-$27:40:44.9 & 27.7$\pm$0.3 & 0.8$\pm$0.5 & $-$0.2$\pm$0.4 & 0.23$"$ & \\
UDF091y-59771436 & 03:32:59.77 & $-$27:41:43.6 & 28.0$\pm$0.2 & 0.5$\pm$0.5 & 0.3$\pm$0.3 & 0.17$"$ & \\
UDF092y-07091160 & 03:33:07.09 & $-$27:51:16.0 & 28.0$\pm$0.2 & 1.1$\pm$0.4 & $-$0.4$\pm$0.2 & 0.19$"$ & 2\\
UDFy-35177170 & 03:32:35.17 & $-$27:47:17.0 & 28.1$\pm$0.1 & 0.5$\pm$0.4 & 0.2$\pm$0.2 & 0.23$''$ & 1\\
UDF092y-05720119 & 03:33:05.72 & $-$27:50:11.9 & 28.1$\pm$0.2 & 0.5$\pm$0.5 & 0.3$\pm$0.3 & 0.20$"$ & \\
UDFy-43416360 & 03:32:43.41 & $-$27:46:36.0 & 28.2$\pm$0.1 & 1.0$\pm$0.5 & 0.2$\pm$0.2 & 0.19$''$ & 1\\
UDF092y-08721095 & 03:33:08.72 & $-$27:51:09.5 & 28.3$\pm$0.2 & 0.9$\pm$0.6 & 0.2$\pm$0.2 & 0.15$"$ & \\
UDF092y-03671041 & 03:33:03.67 & $-$27:51:04.1 & 28.3$\pm$0.2 & 1.0$\pm$0.5 & 0.1$\pm$0.2 & 0.16$"$ & \\
UDF092y-10621086 & 03:33:10.62 & $-$27:51:08.6 & 28.3$\pm$0.3 & 0.9$\pm$0.9 & 0.1$\pm$0.4 & 0.20$"$ & \\
UDF091y-59101018 & 03:32:59.10 & $-$27:41:01.8 & 28.3$\pm$0.3 & 1.4$\pm$0.9 & $-$0.2$\pm$0.4 & 0.20$"$ & \\
UDFy-41787344 & 03:32:41.78 & $-$27:47:34.4 & 28.4$\pm$0.1 & 0.5$\pm$0.3 & 0.0$\pm$0.2 & 0.19$''$ & \\
UDF092y-02731564 & 03:33:02.73 & $-$27:51:56.4 & 28.4$\pm$0.1 & 0.9$\pm$0.5 & 0.4$\pm$0.2 & 0.10$"$ & \\
UDF091y-59522015 & 03:32:59.52 & $-$27:42:01.5 & 28.5$\pm$0.3 & 0.6$\pm$0.6 & $-$0.0$\pm$0.4 & 0.18$"$ & \\
UDF092y-08680370 & 03:33:08.68 & $-$27:50:37.0 & 28.6$\pm$0.2 & 1.2$\pm$0.9 & 0.1$\pm$0.3 & 0.15$"$ & \\
UDF091y-00240470 & 03:33:00.24 & $-$27:40:47.0 & 28.6$\pm$0.3 & $>$1.5 & $-$0.2$\pm$0.4 & 0.19$"$ & \\
UDF092y-07090218 & 03:33:07.09 & $-$27:50:21.8 & 28.6$\pm$0.3 & 0.5$\pm$0.4 & $-$0.5$\pm$0.3 & 0.18$"$ & \\
UDF091y-00252004 & 03:33:00.25 & $-$27:42:00.4 & 28.6$\pm$0.3 & 0.6$\pm$0.4 & $-$0.3$\pm$0.3 & 0.13$"$ & \\
UDF092y-00841085 & 03:33:00.84 & $-$27:51:08.5 & 28.6$\pm$0.3 & 0.6$\pm$0.4 & $-$0.6$\pm$0.4 & 0.18$"$ & \\
UDF091y-06790538 & 03:33:06.79 & $-$27:40:53.8 & 28.6$\pm$0.4 & $>$1.4 & $-$0.4$\pm$0.5 & 0.19$"$ & \\
UDFy-38008154 & 03:32:38.00 & $-$27:48:15.4 & 28.7$\pm$0.2 & 0.6$\pm$0.3 & $-$0.3$\pm$0.2 & 0.19$''$ & \\
UDFy-36497453 & 03:32:36.49 & $-$27:47:45.3 & 28.7$\pm$0.2 & 1.2$\pm$1.0 & 0.3$\pm$0.3 & 0.19$''$ & \\
UDF091y-06211046 & 03:33:06.21 & $-$27:41:04.6 & 28.7$\pm$0.3 & 0.8$\pm$0.6 & $-$0.2$\pm$0.4 & 0.16$"$ & \\
UDF092y-09631474 & 03:33:09.63 & $-$27:51:47.4 & 28.7$\pm$0.3 & 1.1$\pm$1.0 & 0.2$\pm$0.4 & 0.14$"$ & \\
UDF092y-07941445 & 03:33:07.94 & $-$27:51:44.5 & 28.8$\pm$0.2 & 0.5$\pm$0.5 & 0.3$\pm$0.3 & 0.11$"$ & \\
UDFy-37398129 & 03:32:37.39 & $-$27:48:12.9 & 28.8$\pm$0.2 & 1.5$\pm$0.9 & $-$0.3$\pm$0.3 & 0.20$''$ & \\
UDF092y-01571044 & 03:33:01.57 & $-$27:51:04.4 & 28.8$\pm$0.3 & 0.7$\pm$0.8 & 0.2$\pm$0.4 & 0.14$"$ & \\
UDF091y-03941583 & 03:33:03.94 & $-$27:41:58.3 & 28.8$\pm$0.3 & 0.9$\pm$0.7 & 0.0$\pm$0.4 & 0.13$"$ & \\
UDF092y-05842094 & 03:33:05.84 & $-$27:52:09.4 & 28.8$\pm$0.4 & 1.0$\pm$0.9 & $-$0.2$\pm$0.4 & 0.18$"$ & \\
UDF092y-08061458 & 03:33:08.06 & $-$27:51:45.8 & 28.9$\pm$0.3 & $>$1.4 & $-$0.3$\pm$0.4 & 0.15$"$ & \\
UDFy-37517553 & 03:32:37.51 & $-$27:47:55.3 & 29.0$\pm$0.2 & $>$1.5 & 0.0$\pm$0.3 & 0.16$''$ & \\
UDF092y-07220444 & 03:33:07.22 & $-$27:50:44.4 & 29.0$\pm$0.3 & $>$1.1 & 0.0$\pm$0.4 & 0.12$"$ & \\
UDF092y-02271378 & 03:33:02.27 & $-$27:51:37.8 & 29.0$\pm$0.3 & 0.6$\pm$0.6 & $-$0.2$\pm$0.4 & 0.15$"$ & \\
UDF092y-10231298 & 03:33:10.23 & $-$27:51:29.8 & 29.0$\pm$0.3 & 0.8$\pm$0.6 & $-$0.3$\pm$0.4 & 0.12$"$ & \\
UDF092y-10121280 & 03:33:10.12 & $-$27:51:28.0 & 29.0$\pm$0.3 & 0.9$\pm$0.6 & $-$0.4$\pm$0.4 & 0.14$"$ & \\
UDFy-39405459 & 03:32:39.40 & $-$27:45:45.9 & 29.0$\pm$0.3 & 1.3$\pm$0.9 & $-$0.4$\pm$0.4 & 0.29$''$ & \\
UDF092y-09351539 & 03:33:09.35 & $-$27:51:53.9 & 29.0$\pm$0.4 & $>$1.2 & $-$0.1$\pm$0.5 & 0.14$"$ & \\
UDFy-43347121 & 03:32:43.34 & $-$27:47:12.1 & 29.1$\pm$0.3 & 1.1$\pm$0.8 & $-$0.1$\pm$0.3 & 0.17$''$ & \\
UDFy-38396038 & 03:32:38.39 & $-$27:46:03.8 & 29.2$\pm$0.2 & 1.0$\pm$0.7 & 0.1$\pm$0.3 & 0.14$''$ & \\
UDFy-39807470 & 03:32:39.80 & $-$27:47:47.0 & 29.3$\pm$0.2 & $>$1.0 & 0.3$\pm$0.4 & 0.15$''$ & \\
UDFy-34437343 & 03:32:34.43 & $-$27:47:34.3 & 29.3$\pm$0.4 & 0.8$\pm$0.8 & $-$0.4$\pm$0.5 & 0.20$''$ & \\
UDFy-37706328 & 03:32:37.70 & $-$27:46:32.8 & 29.4$\pm$0.3 & 0.9$\pm$0.7 & $-$0.2$\pm$0.4 & 0.15$''$ & \\
\enddata

\tablenotetext{a}{Each of the sources here satisfy the color criteria
  for our Lyman-Break galaxy selections, but show slight detections in
  the $\chi_{opt} ^2$ images generated from the $B_{435}$, $V_{606}$,
  $i_{775}$, $z_{850}$ data.  These sources therefore do not make it
  into our Lyman-Break galaxy samples.  See \S3.3 and Appendix D for a
  discussion of $\chi_{opt} ^2$ and our selection criteria.  While a
  few of these sources will correspond to $z\gtrsim7$ galaxies, the 
  majority will not according to our simulations (Appendix D).}

\tablenotetext{b}{Lower limits on the measured colors are
the $1\sigma$ limits.  Magnitudes are AB.}
\tablenotetext{c}{Half-light radius calculated using SExtractor}
\tablenotetext{d}{References: [1] Yan et al.\ (2010), [2] Wilkins et al. (2011)}
\end{deluxetable*}

\begin{deluxetable*}{ccccccccc}
\tablewidth{17cm}
\tablecolumns{9} 
\tablecaption{Possible $z$$\sim$7-8 candidates that did not meet our selection criteria.\tablenotemark{a}\label{tab:possiblecand}} 
\tablehead{ \colhead{Object ID} &
\colhead{R.A.} & \colhead{Dec} & \colhead{$J_{125}$} & \colhead{$z_{850}-J_{125}$\tablenotemark{b}} & \colhead{$Y_{098}-J_{125}$\tablenotemark{b}} & \colhead{$J_{125}-H_{160}$} & \colhead{$r_{hl}$\tablenotemark{c}} & \colhead{Ref\tablenotemark{d}}}
\startdata
\multicolumn{9}{c}{Possible $z\sim7$ candidates} \\
ERSz-2320541505\tablenotemark{f} & 03:32:32.05 & $-$27:41:50.5 & 25.9$\pm$0.1 & 1.9$\pm$0.7 & 1.2$\pm$0.4 & 0.3$\pm$0.1 & 0.28$''$ & \\
ERSz-2226543006\tablenotemark{g} & 03:32:22.65 & $-$27:43:00.6 & 26.1$\pm$0.1 & 1.1$\pm$0.3 & 0.1$\pm$0.2 & 0.4$\pm$0.1 & 0.20$''$ & 1,2,3,5\\
ERSz-2279041042\tablenotemark{e} & 03:32:27.90 & $-$27:41:04.2 & 26.2$\pm$0.1 & 2.7$\pm$0.8 & 0.7$\pm$0.2 & $-$0.6$\pm$0.1 & 0.11$''$ & 5\\
ERSz-2097545591\tablenotemark{f} & 03:32:09.75 & $-$27:45:59.1 & 26.2$\pm$0.2 & $>$1.8 & 0.7$\pm$0.5 & 0.0$\pm$0.3 & 0.24$''$ & \\
ERSz-2030845122\tablenotemark{e,g} & 03:32:03.08 & $-$27:45:12.2 & 26.3$\pm$0.1 & 1.1$\pm$0.4 & 1.1$\pm$0.4 & 0.2$\pm$0.1 & 0.13$''$ & \\
ERSz-2104045407\tablenotemark{g} & 03:32:10.40 & $-$27:45:40.7 & 26.5$\pm$0.1 & 0.8$\pm$0.4 & 0.4$\pm$0.3 & 0.2$\pm$0.2 & 0.23$''$ & \\
ERSz-1599542348\tablenotemark{f} & 03:31:59.95 & $-$27:42:34.8 & 26.5$\pm$0.2 & 0.9$\pm$0.6 & 0.7$\pm$0.6 & 0.3$\pm$0.2 & 0.27$''$ & \\
ERSz-2417341555\tablenotemark{g} & 03:32:41.73 & $-$27:41:55.5 & 26.6$\pm$0.2 & $>$1.2 & 0.5$\pm$0.5 & 0.4$\pm$0.3 & 0.31$''$ & \\
ERSz-2313142239\tablenotemark{f} & 03:32:31.31 & $-$27:42:23.9 & 26.6$\pm$0.2 & 1.0$\pm$0.5 & 0.8$\pm$0.5 & 0.2$\pm$0.2 & 0.21$''$ & \\
ERSz-2338641089\tablenotemark{f} & 03:32:33.86 & $-$27:41:08.9 & 26.6$\pm$0.2 & 1.4$\pm$0.8 & 0.6$\pm$0.4 & 0.1$\pm$0.2 & 0.27$''$ & \\
ERSz-2413843168\tablenotemark{g} & 03:32:41.38 & $-$27:43:16.8 & 26.8$\pm$0.1 & 0.9$\pm$0.3 & 0.3$\pm$0.2 & 0.0$\pm$0.2 & 0.13$''$ & \\
ERSz-2308840450\tablenotemark{h} & 03:32:30.88 & $-$27:40:45.0 & 26.8$\pm$0.2 & 1.1$\pm$0.6 & 0.8$\pm$0.6 & 0.1$\pm$0.3 & 0.22$''$ & \\
ERSz-2379242208\tablenotemark{i} & 03:32:37.92 & $-$27:42:20.8 & 27.0$\pm$0.2 & $>$1.4 & 1.1$\pm$0.9 & 0.3$\pm$0.3 & 0.24$''$ & \\
ERSz-2202443342\tablenotemark{g} & 03:32:20.24 & $-$27:43:34.2 & 27.0$\pm$0.2 & 0.8$\pm$0.5 & 0.3$\pm$0.3 & $-$0.1$\pm$0.3 & 0.19$''$ & \\
ERSz-2346843052\tablenotemark{g} & 03:32:34.68 & $-$27:43:05.2 & 27.0$\pm$0.2 & 1.7$\pm$0.9 & 0.5$\pm$0.4 & 0.1$\pm$0.3 & 0.19$''$ & \\
ERSz-2400642142\tablenotemark{i} & 03:32:40.06 & $-$27:42:14.2 & 27.2$\pm$0.2 & $>$1.5 & 1.0$\pm$0.7 & 0.2$\pm$0.3 & 0.17$''$ & \\
ERSz-2153943286\tablenotemark{g} & 03:32:15.39 & $-$27:43:28.6 & 27.2$\pm$0.2 & 0.9$\pm$0.5 & $-$0.0$\pm$0.3 & $-$0.3$\pm$0.3 & 0.13$''$ & \\
ERSz-2232742361\tablenotemark{f} & 03:32:23.27 & $-$27:42:36.1 & 27.2$\pm$0.2 & 0.9$\pm$0.5 & 0.2$\pm$0.4 & 0.0$\pm$0.3 & 0.16$''$ & \\
ERSz-2426643478\tablenotemark{g} & 03:32:42.66 & $-$27:43:47.8 & 27.2$\pm$0.2 & 1.2$\pm$0.8 & $>$1.2 & 0.3$\pm$0.3 & 0.15$''$ & \\
ERSz-2229344099\tablenotemark{g} & 03:32:22.93 & $-$27:44:09.9 & 27.3$\pm$0.2 & 1.0$\pm$0.5 & 0.1$\pm$0.4 & 0.3$\pm$0.3 & 0.14$''$ & 5\\
ERSz-2267841314\tablenotemark{f,i} & 03:32:26.78 & $-$27:41:31.4 & 27.4$\pm$0.2 & $>$1.6 & 0.8$\pm$0.6 & $-$0.0$\pm$0.3 & 0.16$''$ & \\
ERSz-2270041428\tablenotemark{g} & 03:32:27.00 & $-$27:41:42.8 & 27.5$\pm$0.2 & 0.8$\pm$0.5 & 0.1$\pm$0.3 & $-$0.2$\pm$0.3 & 0.15$''$ & \\
ERSz-2202541197\tablenotemark{f} & 03:32:20.25 & $-$27:41:19.7 & 27.5$\pm$0.2 & 1.3$\pm$0.9 & 0.3$\pm$0.5 & 0.3$\pm$0.3 & 0.15$''$ & \\
\multicolumn{9}{c}{Possible $z\sim8$ candidates} \\
ERSY-2146242369\tablenotemark{l} & 03:32:14.62 & $-$27:42:36.9 & 26.0$\pm$0.1 & $>$1.7 & 1.3$\pm$0.7 & 0.5$\pm$0.2 & 0.25$''$\\
ERSY-2416741275\tablenotemark{f,m} & 03:32:41.67 & $-$27:41:27.5 & 26.1$\pm$0.1 & $>$1.4 & $>$1.4 & 0.6$\pm$0.2 & 0.21$''$\\
ERSY-2165043527\tablenotemark{k} & 03:32:16.50 & $-$27:43:52.7 & 26.2$\pm$0.1 & $>$2.2 & 1.3$\pm$0.5 & 0.0$\pm$0.2 & 0.21$''$\\
ERSY-2376440061\tablenotemark{n} & 03:32:37.64 & $-$27:40:06.1 & 27.0$\pm$0.3 & 1.3$\pm$0.9 & $>$1.3 & 0.1$\pm$0.3 & 0.41$''$\\
ERSY-2367541341\tablenotemark{j} & 03:32:36.75 & $-$27:41:34.1 & 27.1$\pm$0.2 & $>$1.4 & $>$1.4 & 0.3$\pm$0.3 & 0.17$''$\\
ERSY-2107343270\tablenotemark{f} & 03:32:10.73 & $-$27:43:27.0 & 28.2$\pm$0.3 & $>$1.1 & $>$1.4 & $-$0.3$\pm$0.4 & 0.12$''$
\enddata 
\tablenotetext{a}{Sources in this table either slightly miss our
  selection criteria or show slight detections in the $\chi_{opt} ^2$
  images generated from the $B_{435}$, $V_{606}$, $i_{775}$, $z_{850}$
  data.  These sources therefore do not make it into our Lyman-Break
  galaxy samples.  See \S3.3 and Appendix D for a discussion of
  $\chi_{opt} ^2$ and our selection criteria.}
\tablenotetext{b}{Lower limits on the measured colors are
the $1\sigma$ limits.  Magnitudes are AB.}
\tablenotetext{c}{Half-light radius calculated using SExtractor}
\tablenotetext{c}{References: [1] Bouwens et al.\ (2010c), [2] Hickey et al.\ (2010),
[3] Castellano et al.\ (2010a), [4] Wilkins et al.\ (2010), [5] Wilkins et al.\ (2011)}
\tablenotetext{e}{The size/structure of this source and its colors are consistent with the source being a T-dwarf.  The case is particularly compelling for ERS-2279041041}
\tablenotetext{f}{Show slight ($>1.5\sigma$) detections in one or more
  optical bands, strongly suggesting they do not correspond to
  bona-fide $z$$\sim$7-8 galaxies.  For ERSz-2097545591, the detection
  is only apparent in a small $0.2''$-diameter aperture.}
\tablenotetext{g}{If a star-forming galaxy, more likely at $z$$\sim$6-6.6}
\tablenotetext{h}{This source shows a red $Y-J\sim1$ color -- and
  hence a possible $z\sim7$ galaxy if the Lyman-Break falls between
  the $Y$ and $J$ bands -- but the source is also detected at
  $>$1.5$\sigma$ in the $i$ and $z$ bands.}
\tablenotetext{i}{Despite not satisfying our selection criteria, these
  sources may correspond to $z\gtrsim6.5$ galaxies.}
\tablenotetext{j}{Small $2\sigma$ detection evident in a small
  $0.2''$-diameter aperture in the $B_{435}$ band}
\tablenotetext{k}{$2\sigma$ detection in $i$ band and apparent
  $1\sigma$ detection in $z$ band inconsistent with source being at
  $z\sim8$}
\tablenotetext{l}{$2\sigma$ detection in $V_{606}$ band in
  $0.2''$-diameter aperture}
\tablenotetext{m}{Red $J-H\sim0.6$ color suggests this source is
  likely a low-redshift, intrinsic red galaxy}
\tablenotetext{n}{Slight detection in $i_{775}$, $z_{850}$ band in
  $0.2''$-diameter aperture}
\end{deluxetable*}

\end{document}